\documentclass[
 reprint,
 superscriptaddress,
 amsmath,amssymb,
 aps,
]{revtex4-2}

\usepackage{graphicx}
\usepackage{dcolumn}
\usepackage{bm}
\usepackage{color}
\usepackage{subfigure}
\usepackage{multirow}
\usepackage{here}
\usepackage[mathlines]{lineno}

\definecolor{green(html/cssgreen)}{rgb}{0.0, 0.5, 0.0}

\begin{document}

\title{Measurements of protons and charged pions emitted from $\nu_{\mu}$ charged\nobreakdash-current interactions on iron at a mean neutrino energy of 1.49\,GeV using a nuclear emulsion detector}


\author{H.~Oshima}\thanks{E-mail: hitoshi.oshima@sci.toho-u.ac.jp}
\affiliation{Department of Physics, Toho University, Funabashi 274--8510, Japan} 
\author{T.~Matsuo}\thanks{Present address: Nagoya University.}
\affiliation{Department of Physics, Toho University, Funabashi 274--8510, Japan} 
\author{A.~Ali}
\affiliation{Department of Physics, Kyoto University, Kyoto 606--8502, Japan} 
\author{S.~Aoki}
\affiliation{Graduate School of Human Development and Environment, Kobe University, Kobe 657--8501, Japan} 
\author{L.~Berns}
\affiliation{Department of Physics, Tokyo Institute of Technology, Tokyo 152--8551, Japan} 
\author{T.~Fukuda}
\affiliation{Department of Physics, Nagoya University, Nagoya 464--8602, Japan} 
\author{Y.~Hanaoka}
\affiliation{College of Industrial Technology, Nihon University, Narashino 275--8576, Japan} 
\author{Y.~Hayato}
\affiliation{Kamioka Observatory, Institute for Cosmic Ray Research, University of Tokyo, Kamioka 506--1205, Japan} 
\author{A.~Hiramoto}\thanks{Present address: Okayama University.}
\affiliation{Department of Physics, Kyoto University, Kyoto 606--8502, Japan} 
\author{A.~K.~Ichikawa}
\affiliation{Department of Physics, Tohoku University, Sendai 980--8578, Japan} 
\author{H.~Inamoto}
\affiliation{Department of Physics, Nagoya University, Nagoya 464--8602, Japan} 
\author{A.~Kasumi}
\affiliation{Department of Physics, Nagoya University, Nagoya 464--8602, Japan} 
\author{H.~Kawahara}
\affiliation{Department of Physics, Nagoya University, Nagoya 464--8602, Japan} 
\author{T.~Kikawa}
\affiliation{Department of Physics, Kyoto University, Kyoto 606--8502, Japan} 
\author{R.~Komatani}
\affiliation{Department of Physics, Nagoya University, Nagoya 464--8602, Japan} 
\author{M.~Komatsu}
\affiliation{Department of Physics, Nagoya University, Nagoya 464--8602, Japan} 
\author{K.~Kuretsubo}
\affiliation{Graduate School of Human Development and Environment, Kobe University, Kobe 657--8501, Japan} 
\author{T.~Marushima}
\affiliation{Graduate School of Human Development and Environment, Kobe University, Kobe 657--8501, Japan} 
\author{H.~Matsumoto}
\affiliation{Graduate School of Human Development and Environment, Kobe University, Kobe 657--8501, Japan} 
\author{S.~Mikado}
\affiliation{College of Industrial Technology, Nihon University, Narashino 275--8576, Japan} 
\author{A.~Minamino}
\affiliation{Faculty of Engineering, Yokohama National University, Yokohama 240--8501, Japan} 
\author{K.~Mizuno}
\affiliation{Department of Physics, Toho University, Funabashi 274--8510, Japan} 
\author{Y.~Morimoto}
\affiliation{Department of Physics, Toho University, Funabashi 274--8510, Japan} 
\author{K.~Morishima}
\affiliation{Department of Physics, Nagoya University, Nagoya 464--8602, Japan} 
\author{N.~Naganawa}
\affiliation{Department of Physics, Nagoya University, Nagoya 464--8602, Japan} 
\author{M.~Naiki}
\affiliation{Department of Physics, Nagoya University, Nagoya 464--8602, Japan} 
\author{M.~Nakamura}
\affiliation{Department of Physics, Nagoya University, Nagoya 464--8602, Japan} 
\author{Y.~Nakamura}
\affiliation{Department of Physics, Nagoya University, Nagoya 464--8602, Japan} 
\author{T.~Nakano}
\affiliation{Department of Physics, Nagoya University, Nagoya 464--8602, Japan} 
\author{T.~Nakaya}
\affiliation{Department of Physics, Kyoto University, Kyoto 606--8502, Japan} 
\author{A.~Nishio}
\affiliation{Department of Physics, Nagoya University, Nagoya 464--8602, Japan} 
\author{T.~Odagawa}
\affiliation{Department of Physics, Kyoto University, Kyoto 606--8502, Japan} 
\author{S.~Ogawa}
\affiliation{Department of Physics, Toho University, Funabashi 274--8510, Japan} 
\author{H.~Rokujo}
\affiliation{Department of Physics, Nagoya University, Nagoya 464--8602, Japan} 
\author{O.~Sato}
\affiliation{Department of Physics, Nagoya University, Nagoya 464--8602, Japan} 
\author{H.~Shibuya}
\affiliation{Department of Physics, Toho University, Funabashi 274--8510, Japan} 
\affiliation{Faculty of Engineering, Kanagawa University, Yokohama 221--8686, Japan} 
\author{K.~Sugimura}
\affiliation{Department of Physics, Nagoya University, Nagoya 464--8602, Japan} 
\author{L.~Suzui}
\affiliation{Department of Physics, Nagoya University, Nagoya 464--8602, Japan} 
\author{Y.~Suzuki}
\affiliation{Department of Physics, Nagoya University, Nagoya 464--8602, Japan} 
\author{H.~Takagi}
\affiliation{Department of Physics, Toho University, Funabashi 274--8510, Japan} 
\author{S.~Takahashi}
\affiliation{Graduate School of Human Development and Environment, Kobe University, Kobe 657--8501, Japan} 
\author{T.~Takao}
\affiliation{Department of Physics, Nagoya University, Nagoya 464--8602, Japan} 
\author{Y.~Tanihara}
\affiliation{Yokohama National University, Yokohama 240--8501, Japan} 
\author{M.~Watanabe}
\affiliation{Department of Physics, Nagoya University, Nagoya 464--8602, Japan} 
\author{K.~Yamada}
\affiliation{Graduate School of Human Development and Environment, Kobe University, Kobe 657--8501, Japan} 
\author{K.~Yasutome}
\affiliation{Department of Physics, Kyoto University, Kyoto 606--8502, Japan} 
\author{M.~Yokoyama}
\affiliation{Department of Physics, University of Tokyo, Tokyo 113--0033, Japan} 
\author{M.~Yoshimoto}
\affiliation{Nishina Center for Accelerator-Based Science, RIKEN, Wako 351--0198, Japan} 

\collaboration{The NINJA Collaboration}

\date{\today}

\begin{abstract}
This study conducted an analysis of muons, protons, and charged pions emitted from $\nu_{\mu}$~charged\nobreakdash-current interactions on iron using a nuclear emulsion detector.
The emulsion detector with a 65\,kg iron target was exposed to a neutrino beam corresponding to 4.0$\times$10$^{19}$ protons on target with a mean neutrino energy of 1.49\,GeV.
The measurements were performed at a momentum threshold of 200~(50)\,MeV/$c$ for protons~(pions), which are the lowest momentum thresholds attempted up to now.
The measured quantities are the multiplicities, emission angles, and momenta of the muons, protons, and charged pions.
In addition to these inclusive measurements, exclusive measurements such as the muon-proton emission-angle correlations of specific channels and the opening angle between the protons of CC0$\pi$2$p$ events were performed.
The data were compared to Monte~Carlo~(MC) predictions and some significant differences were observed.
The results of the study demonstrate the capability of detailed measurements of neutrino-nucleus interactions using a nuclear emulsion detector to improve neutrino interaction models.
\end{abstract}

\maketitle

\section{Introduction}\label{sec:introduction}
An important topic in particle physics is searching for $CP$ violations in the lepton sector~\cite{PhysRevD.74.072003,PhysRevLett.112.191801,Nature_T2K_2020,NOvA.2021nfi,HK_2015,arXiv_DUNE_2016}. 
Long\nobreakdash-baseline neutrino oscillation experiments searching for the $CP$ violation, such as T2K~\cite{T2K_2011} and NOvA~\cite{NOvA.2021nfi}, are performed at a neutrino energy around 1\,GeV.
The dominant modes of neutrino charged\nobreakdash-current (CC) interactions in this region are quasi\nobreakdash-elastic scattering (QE) and resonant pion production (RES).
In the interactions with free nucleons, the CCQE interaction includes one lepton and one nucleon, whereas CCRES includes one pion in addition to those particles.
Further, two\nobreakdash-particle\nobreakdash-two\nobreakdash-hole (2p2h) interactions are well known in electron scattering experiments~\cite{PhysRevC.68.014313,PhysRevC.71.044615,PhysRevLett.96.082501,PhysRevLett.98.132501,PhysRevLett.99.072501,subedi2008probing}, and the existence of a 2p2h interaction for neutrino interactions that emits one lepton and two nucleons is natural. In particular, the 2p2h interaction on low-{\it{A}} nuclei, such as carbon, is regarded as established~\cite{PhysRevD.93.112012,PhysRevD.94.093004,PhysRevC.80.065501,PhysRevD.88.113007,PhysRevLett.116.071802}.
The T2K far detector, Super\nobreakdash-Kamiokande~(SK)~\cite{SK_2003}, is a water Cherenkov detector that is insensitive to most protons owing to the momentum threshold for Cherenkov light production at 1.06\,GeV/$c$.
For a CCQE interaction, events with a single lepton and no other visible particles are selected as signal events.
Further, the incoming neutrino energy is reconstructed using only the outgoing lepton information and the known incoming neutrino direction for a CCQE interaction with a nucleon at rest.
However, this method cannot separate 2p2h from CCQE interactions, and the reconstructed neutrino energy is downward biased.
Furthermore, if a charged pion generated from the CCRES interaction is not detected, CCQE and CCRES cannot be distinguished, which results in the underestimation of the neutrino energy.
Moreover, the signals at SK are contaminated when the pions fall short of the Cherenkov threshold in water, although Michel-electron tagging can be used to veto such events.
In addition to each interaction mode, the pions and protons may experience final state interactions~(FSIs), such as re-scattering, absorption, particle production, or charge-exchange, inside the target nucleus, and distinguishing interaction modes from particles in the final state is challenging.
In neutrino oscillation experiments, the uncertainty on the neutrino energy reconstruction is an important source of systematic error since the oscillation probability is measured as a function of the neutrino energy.
Therefore, understanding the neutrino-nucleus interactions in the 1 GeV energy region helps to reduce the systematic error of oscillation experiments.
In particular, multiplicity and kinematic studies of protons and charged pions from neutrino interactions are required to validate reliable neutrino interaction models.
However, at present, measurements of these hadron kinematics are scarce, because these hadrons tend to exhibit low energies.
In particular, observing the low-momentum protons in scintillator-based tracking detector is challenging, wherein proton momentum thresholds are typically 400--700\,MeV/$c$~\cite{MicroBooNE_multiplicity_2019,T2K_proton_2018,MINERvA_CCQE_2013}. 

A series of neutrino-nucleus interaction measurements were conducted as part of the NINJA experiment~\cite{NINJA_Run4_ECC_2017,PhysRevD.102.072006,ninja_run6_xsec} using an emulsion detector in the near detector hall of the T2K experiment at J\nobreakdash-PARC.
The aim was to precisely measure charged particles from neutrino interactions in the 1\,GeV energy region.
The emulsion detector was made of emulsion films interleaved with a target material such as iron or water.
To understand neutrino-nucleus interactions, including nuclear effects, these interactions must be studied using various targets.
The emulsion detector is suitable for the precise measurement of charged particles with a low momentum threshold owing to its thin-layered structure.
Previous studies have reported the detector performance of the 2\,kg iron pilot measurement~\cite{NINJA_Run4_ECC_2017} and the results of the 3\,kg water pilot measurement~\cite{PhysRevD.102.072006}.
Further, the measurement of the flux\nobreakdash-averaged cross\nobreakdash-sections and muon kinematics of the $\nu_{\mu}$ CC interaction of the 65\,kg iron pilot measurement~\cite{ninja_run6_xsec} have also been reported.
The cross-section results of these studies are consistent with the T2K results~\cite{INGRID_CCinclusive_Xsec_2014,INGRID_CCinclusive_Xsec_2016,WAGSCI_2019} and Monte~Carlo~(MC) simulations.
In addition, the muon kinematics is also well reproduced by the MC simulation.
These results demonstrate the reliability of the detector used and validate the data analysis conducted.
In this study, the multiplicity and kinematic measurements of protons and charged pions emitted from $\nu_{\mu}$ CC interactions on iron in the 65\,kg iron pilot run were measured, and consequently, the first kinematic results of neutrino interactions on iron with slow proton kinematic information ($p \geq 200\,{\rm MeV}/c$) are presented.
The results represent the first step in a series of detailed neutrino-nucleus interaction studies and are expected to be important for building reliable neutrino interaction models.

\section{Detector configuration and data samples}\label{sec:detector_datasample}
The emulsion detector was located in the near detector hall of the T2K experiment at J\nobreakdash-PARC.
The high-intensity 30\,GeV proton beam from the J\nobreakdash-PARC accelerator struck a graphite target, producing charged pions and kaons that were focused and selected via a system of magnetic horns.
The hadrons decay in flight, producing an intense neutrino beam.
In addition, the neutrino and anti-neutrino beam modes can be switched by changing the polarity of the magnetic horns.
Further details on the J\nobreakdash-PARC neutrino beamline can be found in Ref.~\cite{Flux_JNUBEAM_2013}.

\begin{figure}[h]
		\includegraphics[width=8.6cm,pagebox=cropbox]{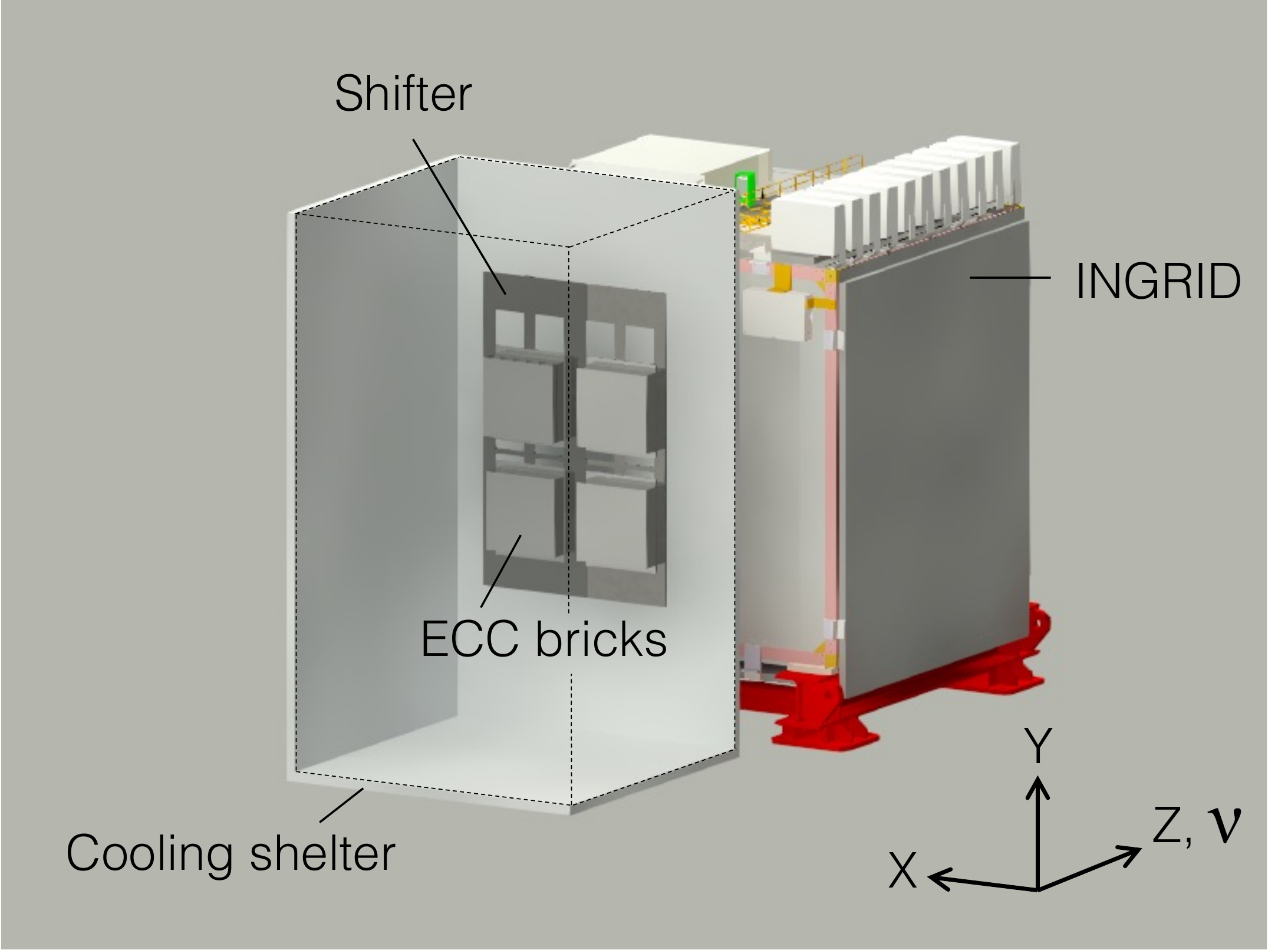}
		\caption{Schematic view of the detector. The ECC bricks and Shifter are enclosed in a cooling shelter, which is placed in front of an INGRID module.}
		\label{fig:Detector}
\end{figure}

Figure~\ref{fig:Detector} shows a schematic view of the detector, which is a hybrid apparatus composed of an iron-target emulsion cloud chamber~(ECC), an emulsion multistage shifter~(Shifter)~\cite{K.Kodama_2006,S.Takahashi_2010,ROKUJO2013127,NINJA_Run4_Shifter_2017,Mizutani_2019}, and an interactive neutrino grid~(INGRID)~\cite{INGRID_2010,INGRID_Otani_2010} at the center of the neutrino beam~(on\nobreakdash-axis).
The ECC bricks and Shifter are enclosed in a cooling shelter to maintain a temperature of approximately 10\,$^{\circ}$C and protect the emulsion films from sensitivity degradation and fading.
The X- and Y-axes are defined as the horizontal and vertical directions perpendicular to the beam direction~(Z-axis), respectively.

\begin{figure}[h]
		\includegraphics[width=7.0cm,pagebox=cropbox]{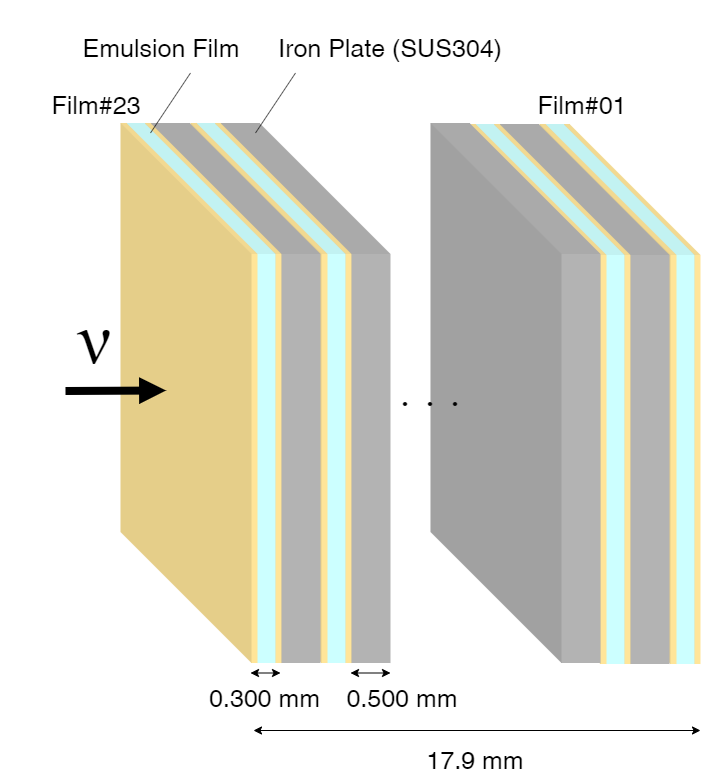}
		\caption{Structure of the ECC brick. Each ECC brick consists of 23 emulsion films interleaved with 22 iron plates.}
		\label{fig:ECC}
\end{figure}

The ECC comprises 12 basic units called bricks, each of which is composed of 23 emulsion films interleaved with 22 iron plates, as shown in Fig.~\ref{fig:ECC}.
The nuclear emulsion comprises AgBr crystals embedded in gelatine.
The emulsion film is a 180\,$\mu$m polystyrene sheet with a 60\,$\mu$m-thick nuclear emulsion layer on each face with an area of 25\,cm$\times$25\,cm.
The iron plate measures 25\,cm$\times$25\,cm$\times$0.05\,cm.
Charged particle trajectries leave latent images in the emulsion transformed into visible rows of grains during development.
The rows of grains are measured as tracks using an optical microscope.
An ECC brick is a three-dimensional tracking detector with a sub-$\mu$m spatial resolution.
Because the tracks in the ECC bricks were required to pass through at least one iron plate and two emulsion films, the momentum thresholds for protons and charged pions were 200\,MeV/$c$ and 50\,MeV/$c$, respectively.

\begin{figure}[h]
		\includegraphics[width=8.6cm,pagebox=cropbox]{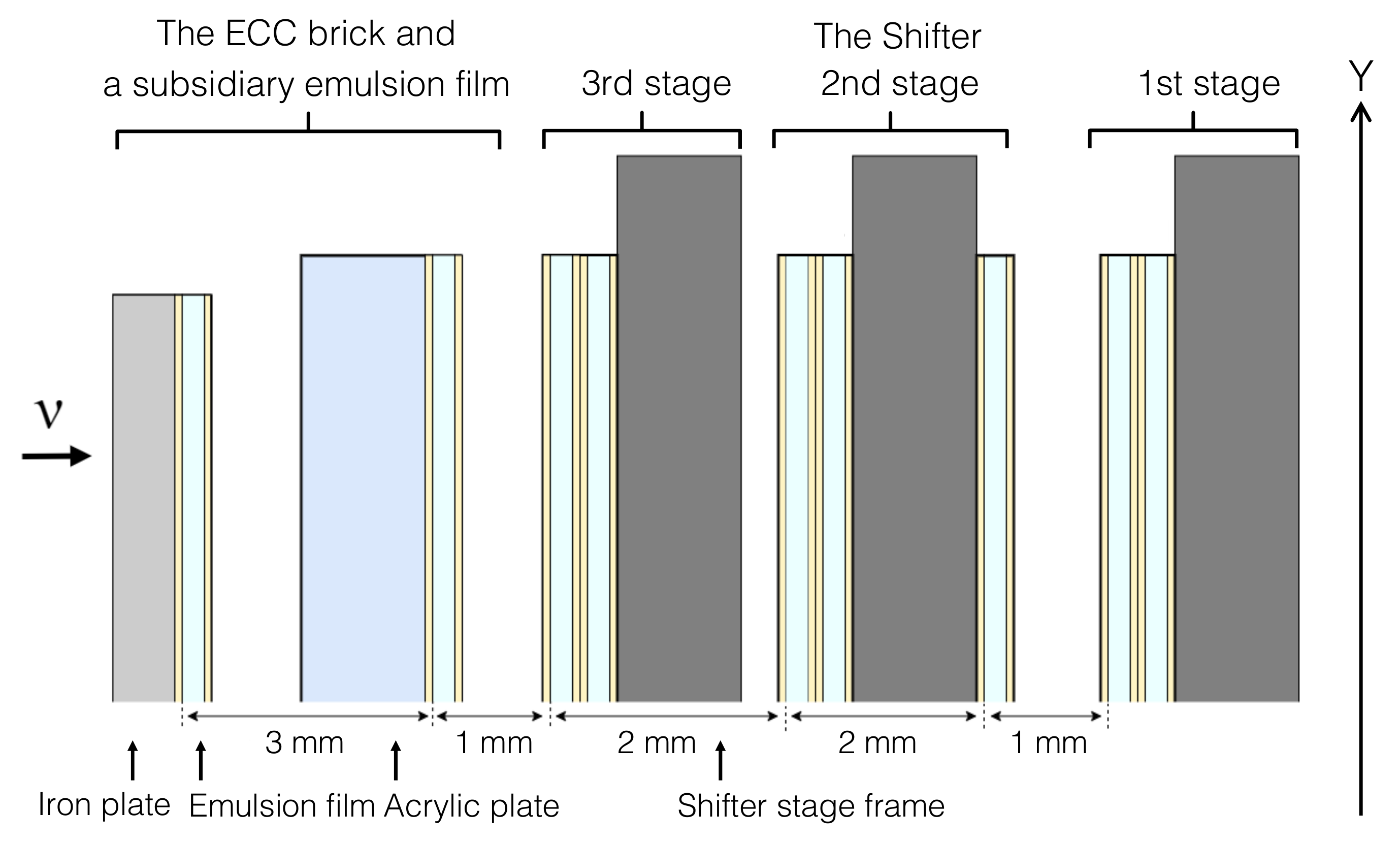}
		\caption{Structure of the Shifter. The Shifter is composed of seven emulsion films mounted on three stages. Each stage is driven at a different speed along the Y-direction.}
		\label{fig:Shifter}
\end{figure}

The Shifter is composed of seven emulsion films mounted on three stages, as shown in Fig.~\ref{fig:Shifter}.
A subsidiary emulsion film was placed between the ECC brick and Shifter to facilitate track connection.
Each stage is driven at a different speed along the Y-direction in order to add timing information to the tracks in the ECC bricks, which was consequently used to match the track with the corresponding muon track in INGRID.

\begin{figure}[h]
		\includegraphics[width=8.6cm,pagebox=cropbox]{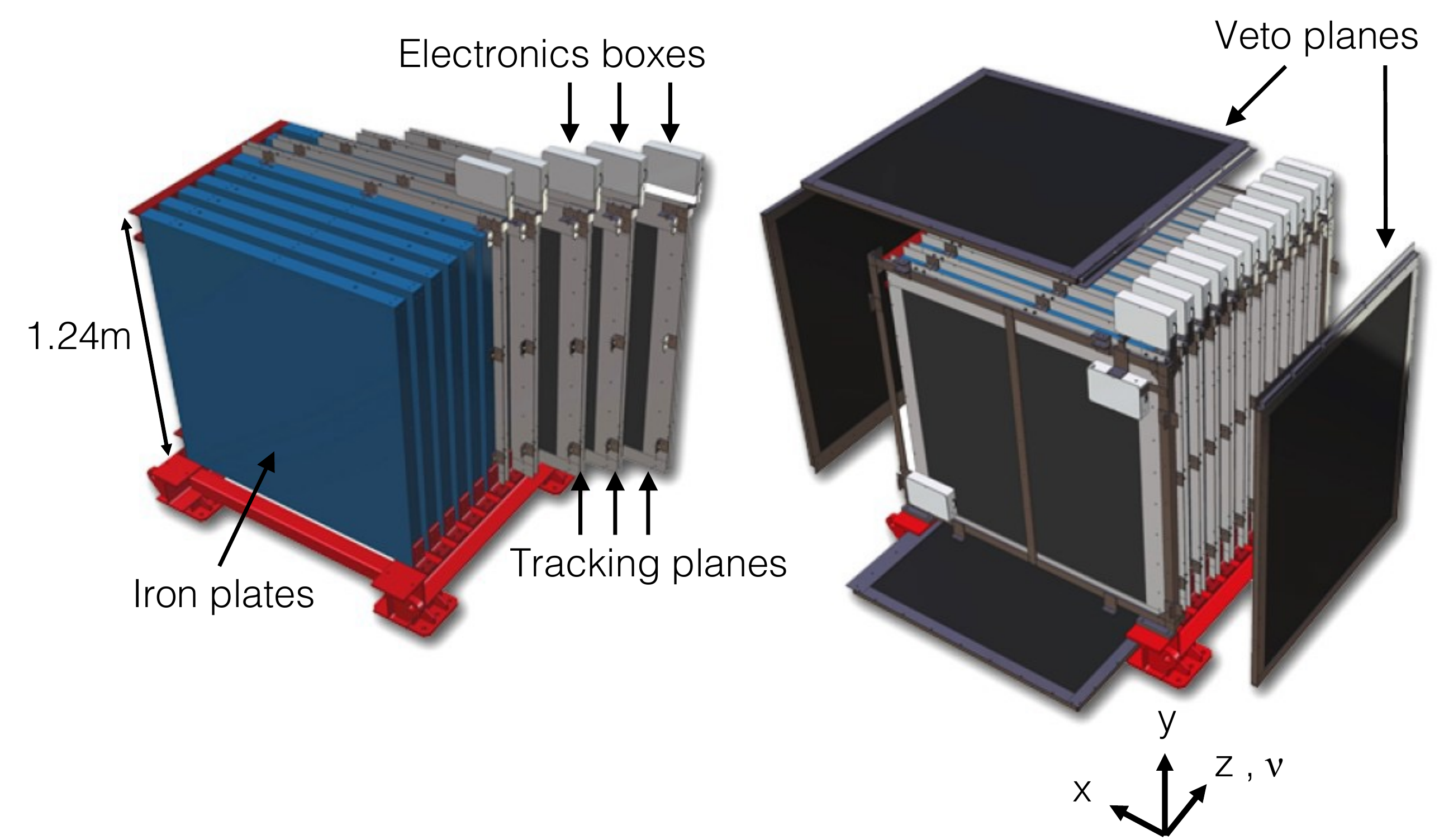}
		\caption{Exploded view of an INGRID module. INGRID module comprises 11 scintillator planes interleaved with 9 iron plates.}
		\label{fig:INGRID}
\end{figure}

INGRID is the on\nobreakdash-axis near detector for the T2K experiment.
INGRID module comprises 11 scintillator planes interleaved with 9 iron plates, as shown in Fig.~\ref{fig:INGRID}.
To select the $\nu_{\mu}$ CC interactions in the ECC bricks, INGRID was employed as a muon range detector in this study, and the set-up was exposed to the $\nu_{\mu}$ beam corresponding to 4.0$\times$10$^{19}$ protons on target~(POT) with a mean neutrino energy of 1.49\,GeV.
A total of 183 events were selected as $\nu_{\mu}$ CC interactions on iron.
Further details regarding the detector and the event reconstruction are described in Ref.~\cite{ninja_run6_xsec}.

\section{Monte Carlo simulation}\label{sec:MC}
An MC simulation was used to evaluate signal and background events, neutrino flux, and detection efficiency.
The MC simulation comprised three parts: JNUBEAM~\cite{Flux_JNUBEAM_2013} version 13av6.1, to predict the neutrino flux; NEUT~\cite{NEUT_2002,NEUT_2009,hayato2021neut} version 5.4.0, to model the neutrino-nucleus interactions; and GEANT4~\cite{Geant4_2003,1610988,ALLISON2016186} version 9.2.1, to simulate the detector response and the behavior of particles emitted from interactions in the detectors.
Further, these simulations were normalized by the POT and target mass.
Additional details regarding the MC simulation can be found in Ref.~\cite{ninja_run6_xsec}.
This section summarizes the neutrino interaction models.

The neutrino interactions are categorized into several modes as follows: CCQE and neutral current~(NC) elastic scattering, 2p2h interactions, CC and NC RES, coherent pion production~(COH\,$\pi$), and deep inelastic scattering~(DIS).
The one-particle-one-hole~(1p1h) model proposed by Nieves $et$ $al.$~\cite{1p1h_nieves_2012,rpa_nieves_2012} was used to simulate the CCQE.
A local Fermi gas model with random phase approximation corrections was used for the nuclear model.
The vector and axial masses in the form factors were set to 0.84 and 1.05\,GeV/$c^2$, respectively.
Nieves $et$ $al.$ modeled the 2p2h interaction~\cite{2p2h_nieves_2011}.
Further, the RES was simulated using the Rein-Sehgal model~\cite{res_rein_sehgal_1981}.
The vector and axial masses in the form factors were set to 0.84 and 0.95\,GeV/$c^2$, respectively.
In addition, the COH\,$\pi$ model described by Rein and Sehgal in Refs.~\cite{coh_rein_sehgal_1983,coh_rein_sehgal_2007} was used in this study. 
To describe the DIS, the parton distribution function GRV98 was applied with Bodek and Yang corrections~\cite{dis_pdf_1998,dis_pdf_2003,dis_pdf_2005}. 
In addition, NEUT modeled the FSIs for hadrons using a semi-classical intranuclear cascade model~\cite{NEUT_2009,hayato2021neut,pion_fsi_2014,pion_fsi_2019}.

\section{Detection and analysis method}\label{sec:analysis}
In this section, detection and analysis methods for protons and charged pions are described.
The detection method is described in Sec.~\ref{subsec:partner_track_search}, while the momentum measurement and particle identification are described in Sec.~\ref{subsec:momentum} and Sec.~\ref{subsec:PID}, respectively.

\subsection{Partner track search}\label{subsec:partner_track_search}
Induced muons from $\nu_{\mu}$ CC interactions in the ECC bricks were selected via track connection between the ECC bricks, Shifter, and INGRID.
The muon candidates were traced back from INGRID to the neutrino interaction vertices in the ECC bricks.
The muon candidate is required to start from an iron plate in the ECC brick and the vertex is inside of the fiducial volume~(FV).
The average fiducial scanning area of each film is 116\,mm$\times$78\,mm in X- and Y- directions, and the FV begins at the fourth film from the upstream face and at the second film from the downstream face of each ECC brick in Z-direction.
Detailed information on the detection and selection methods for $\nu_{\mu}$ CC interactions in ECC bricks is provided in Ref.~\cite{ninja_run6_xsec}.

The tracks attached to the muon tracks are charged hadrons from neutrino interactions and are referred to as ``partner tracks.''
The track pieces were recorded in emulsion layers and tracks connecting the positions of track pieces on both sides of the polystyrene sheet were used as track segments in this analysis.
Position and slope of each track segment were measured in three-dimensional space.
Further, the energy deposited in the emulsion layer was measured as track blackness, referred to as the volume pulse height~(VPH)~\cite{VPH_2006}.
Tracks of heavily ionizing particles~(HIPs), such as proton tracks, exhibit a large VPH, whereas those of minimum ionizing particles~(MIPs), such as the muon and pion tracks, exhibit a small VPH.
In this analysis, the partner tracks with VPH $<$ 150 and $\geq$ 150 were defined as thin tracks and black tracks, respectively.
Consequently, MIPs and HIPs can be separated using this boundary value which is corresponding to 3\,MeV\,g$^{-1}$\,cm$^{2}$ as the ionization loss in the emulsion film, although VPH exhibits an angle dependence.
In case of thin tracks, the slope- and position-related tolerances of track segment connections were defined as functions of the track slope, and were determined from the scattering angles of cosmic-muon~(MIP) data.
Regarding the black tracks, the track reconnection process was performed with larger tolerances than those of thin tracks.
The tolerances are defined as 3 $\sigma$ values of angular and positional differences by scattering of neutrino induced protons with momenta below 500\,MeV/$c$.
The connection efficiency of thin tracks exceeded 99\%, whereas that of black tracks exceeded 90\%.
Further, thin tracks were required to have at least three track segments because the accidental track rate for the tracks with three track segments or more is negligible, whereas black tracks were required to have at least two track segments in order to exclude nuclear fragments.

A track is called a partner track if its distance of closest approach to the muon starting segment (the most upstream segment of the muon track) is small.
The distance was required to be less than 50\,$\mu$m~(60\,$\mu$m) for the thin~(black) partner tracks.
The maximum distance was determined as the value at which the detection efficiency saturated using the MC simulation.
Further, in both cases, the distance along the neutrino beam direction between the starting segment and closest approach position were required to be less than 800\,$\mu$m.

\begin{figure}[h]
		\includegraphics[clip, width=0.95\columnwidth]{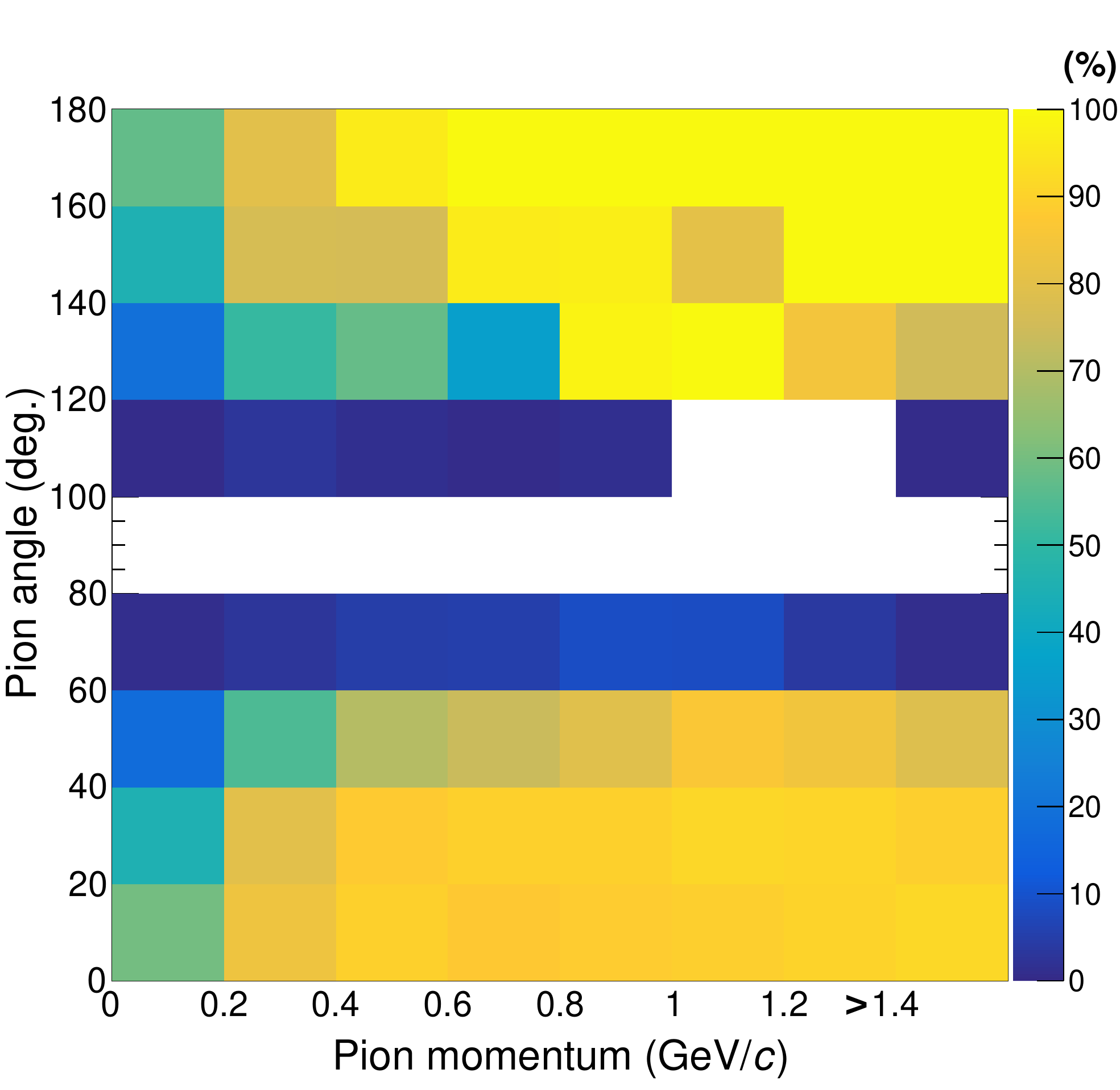}
		\caption{Pion detection efficiency estimated using the MC simulation. Track angle is the angle with respect to the beam direction.}
		\label{fig:eff_pion}
\end{figure}

\begin{figure}[h]
		\includegraphics[clip, width=0.95\columnwidth]{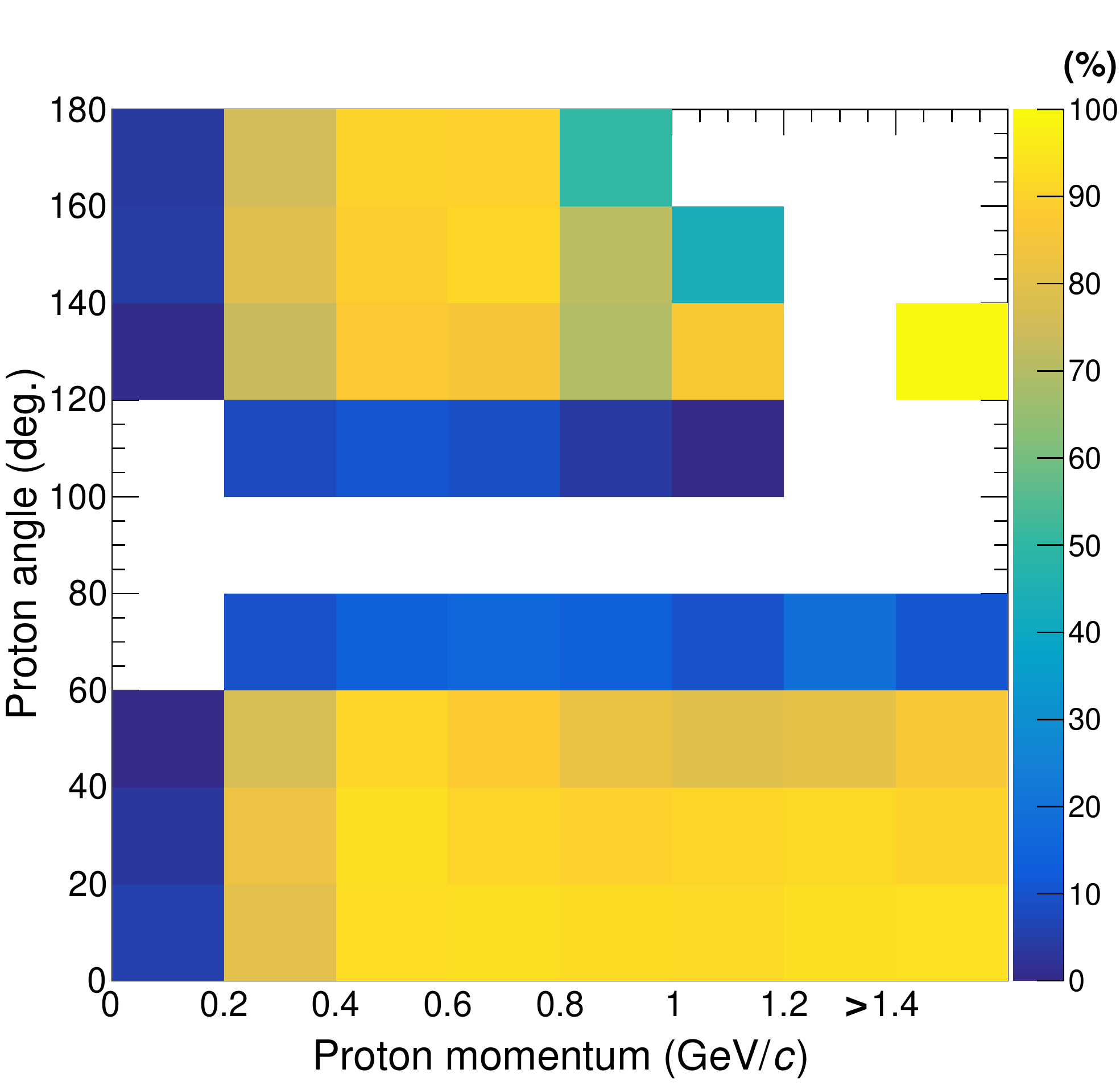}
		\caption{Proton detection efficiency estimated using the MC simulation. Track angle is the angle with respect to the beam direction. Blank bins in the region of momentum greater than 1\,GeV/$c$ and the backward direction denote that there are no events generated by the MC simulation.}
		\label{fig:eff_proton}
\end{figure}

Figures~\ref{fig:eff_pion} and \ref{fig:eff_proton} show the detection efficiencies of pions and protons emitted from $\nu_{\mu}$ CC interactions on iron estimated using the MC simulation.
Blank bins around 90$^{\circ}$ denote the region outside the acceptance.
Concerning the proton detection efficiency, blank bins in the region of momentum greater than 1\,GeV/$c$ and the backward direction denote that there are no events generated by the MC simulation.
It is expected that the detection efficiency for protons is generally greater than 70\% in the 200--400\,MeV/$c$ regions.

\subsection{Momentum measurement}\label{subsec:momentum}
\begin{figure}[h]
		\includegraphics[width=8.6cm,pagebox=cropbox]{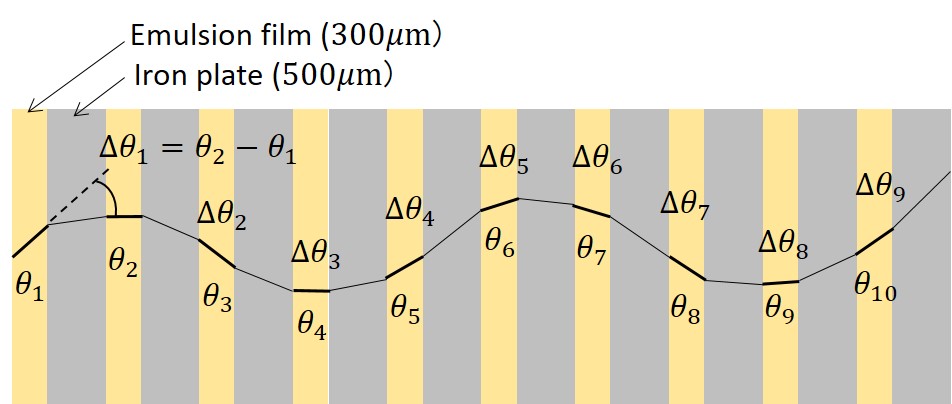}
		\caption{Schematic view of the angular method of MCS measurement. In the figure, $\theta_{i}$ is the track segment angle and $\Delta \theta_{i}$ is the angular difference in cell $i$.}
		\label{fig:AngularMethod}
\end{figure}

Two methods were employed for estimating the momentum of a charged particle in the ECC brick.
One involved measuring its multiple Coulomb scattering~(MCS) while the other derived its energy from the track range in the detector.
The MCS measurement can be performed by two methods: the angular method~\cite{MCS_angular_method_2012} and the coordinate method~\cite{MCS_angular_coordinate_method_2007}.
Figure~\ref{fig:AngularMethod} shows a schematic view of the angular method.
Here, a cell length defined as the length of one iron plate and one emulsion film was used.
The measured variance of the scattering angle distribution is
\begin{equation}
  \Delta \theta_{meas}^{2} = \Delta \theta^{2} + \Delta \theta_{err}^{2},
  \label{eq:angle_scattering}
\end{equation}
where $\Delta \theta$ is the angular scattering due to MCS and $\Delta \theta_{err}$ is the angular measurement uncertainty.
It is typically 2\,mrad owing to the track scanning, track reconstruction and film alignment.
For a particle of momentum $p$ and velocity $\beta$, $\Delta \theta$~\cite{MCS_1991} is given by
\begin{equation}
\Delta \theta=\frac{13.6 \, {\rm MeV}}{\beta cp}z\sqrt{\frac{x}{X_{0}}}\biggl[1+0.038 \, {\rm ln}\Bigl(\frac{x}{X_{0}}\Bigr)\biggr],
\label{eq:MCS_angle}
\end{equation}
where $c$ denotes the speed of light, $z$ denotes the charge number of the particle, $x$ denotes the thickness and $X_{0}$ denotes the radiation length of the material.
Using Eqs.~(\ref{eq:angle_scattering}) and (\ref{eq:MCS_angle}), $p\beta$ is obtained as the slope of an approximately linear function of the square of $\Delta \theta$ and material thickness.
The upper limit of the measurable momentum is about 1\,GeV/$c$.
For higher momenta, the angular scattering due to MCS becomes much smaller than the angular measurement uncertainty.

\begin{figure}[h]
		\includegraphics[width=8.6cm,pagebox=cropbox]{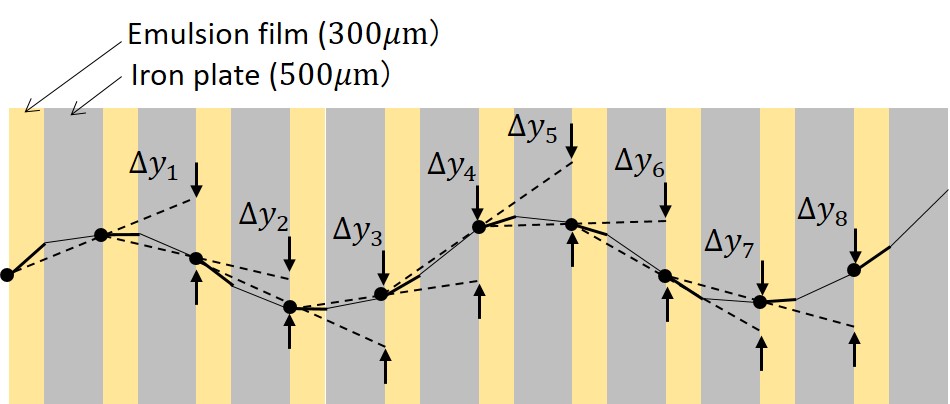}
		\caption{Schematic view of the coordinate method of MCS measurement. In the figure, $\Delta y_{i}$ is the positional displacement between track segments in cell $i$.}
		\label{fig:CoordinateMethod}
\end{figure}

In contrast, in the coordinate method, $p\beta$ is estimated by a position displacement measurement.
Figure~\ref{fig:CoordinateMethod} shows a schematic view of the coordinate method.
The measured variance of the positional scattering distribution is
\begin{equation}
\Delta y_{meas}^{2} = \Delta y^{2} + \Delta y_{err}^{2},
\label{eq:diff_position}
\end{equation}
where $\Delta y$ is the positional scattering due to MCS and $\Delta y_{err}$ is the positional measurement uncertainty.
The positional measurement uncertainty is typically 3\,${\rm \mu}$m for reasons similar to the angular uncertainty.
The relation between $p\beta$ and $\Delta y$~\cite{MCS_1991} is given by
\begin{equation}
\Delta y=\frac{1}{\sqrt{3}}\frac{13.6 \, {\rm MeV}}{\beta cp}zX_{0}\Bigl(\frac{x}{X_{0}}\Bigr)^{\frac{3}{2}}\biggl[[1+0.038 \, {\rm ln}\Bigl(\frac{x}{X_{0}}\Bigr)\biggr].
\label{eq:MCS_position}
\end{equation}
The positional displacement caused by MCS increases as the three-half power of the thickness of the material, whereas the angular difference increases as the square root of the thickness.
Therefore, the upper limit of the measurable momentum using the coordinate method is higher than that of the angular method, for a given material thickness.
It is approximately 5\,GeV/$c$.
The positional displacement was measured as the difference between the track and predicted positions, which is the extrapolated position using a slope angle reconstructed from the track positions on several films.
The coordinate method requires a larger number of track segments than the angular method because the slope angle is used in the coordinate method.
The coordinate method was applied to tracks with more than 15 segments in the ECC bricks, whereas the angular method was applied to the remaining tracks.

After particle identification~(PID), the momentum of a particle can also be measured from its range in the detector.
The momenta of muons stopping in the INGRID module were measured by the track range in the INGRID module and ECC bricks, while those of protons stopping in the ECC bricks were measured by the range in the ECC bricks.
The PID process is described in Sec.~\ref{subsec:PID}.
The momentum measurement uncertainty using this range is smaller than that obtained using the MCS measurement.
However, the charged pions in this energy region cause hadron interactions in the detector because their cross\nobreakdash-section is large.
Therefore, the range\nobreakdash-energy relation for the charged pion candidates was not used.

\begin{figure}[h]
		\includegraphics[clip, width=0.95\columnwidth]{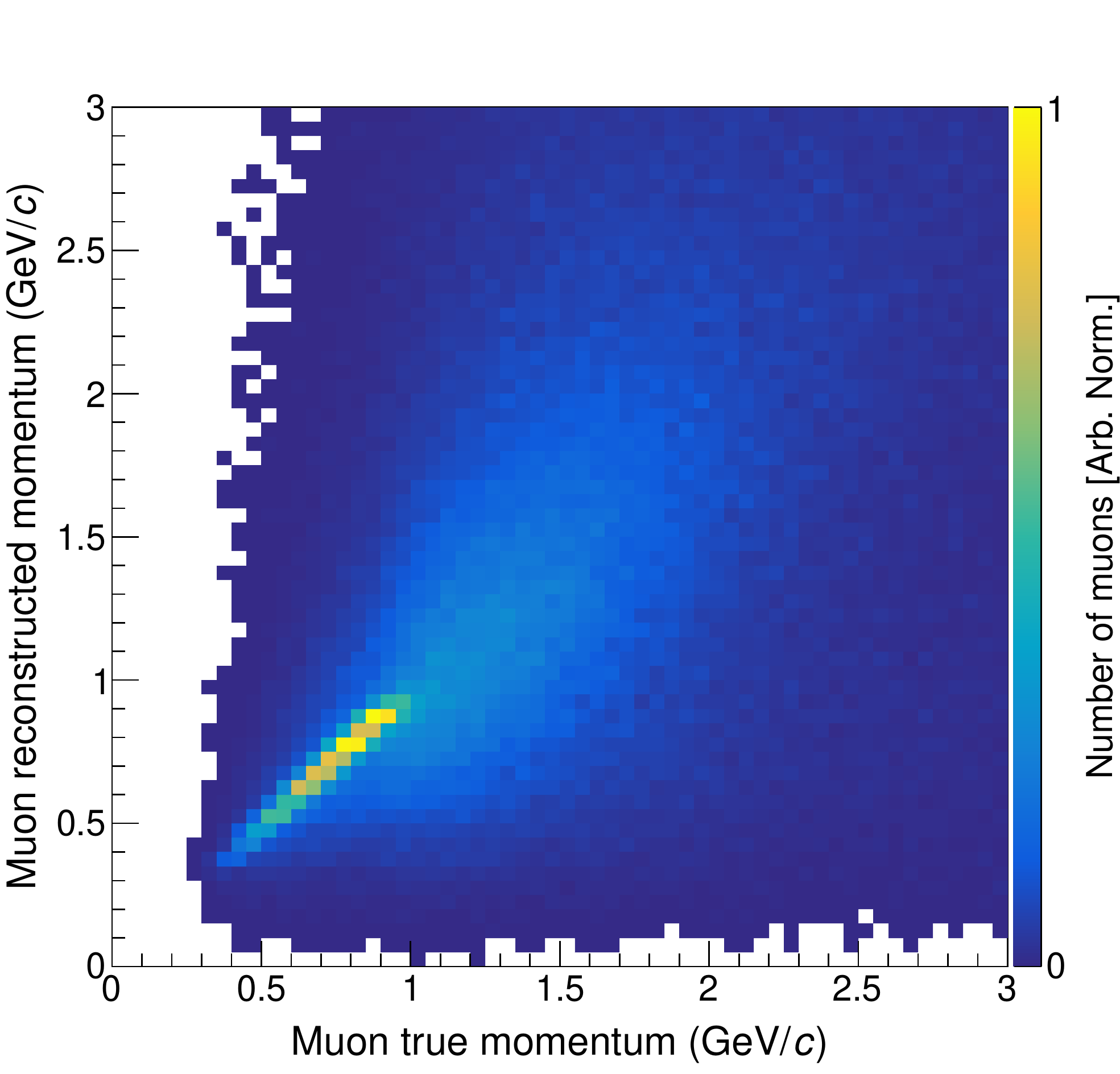}
		\caption{Relation between the reconstructed and true momenta of muons from the neutrino interactions in the MC simulation.}
		\label{fig:muon_momentum_recon_true_relation}
\end{figure}

\begin{figure}[h]
		\includegraphics[clip, width=0.95\columnwidth]{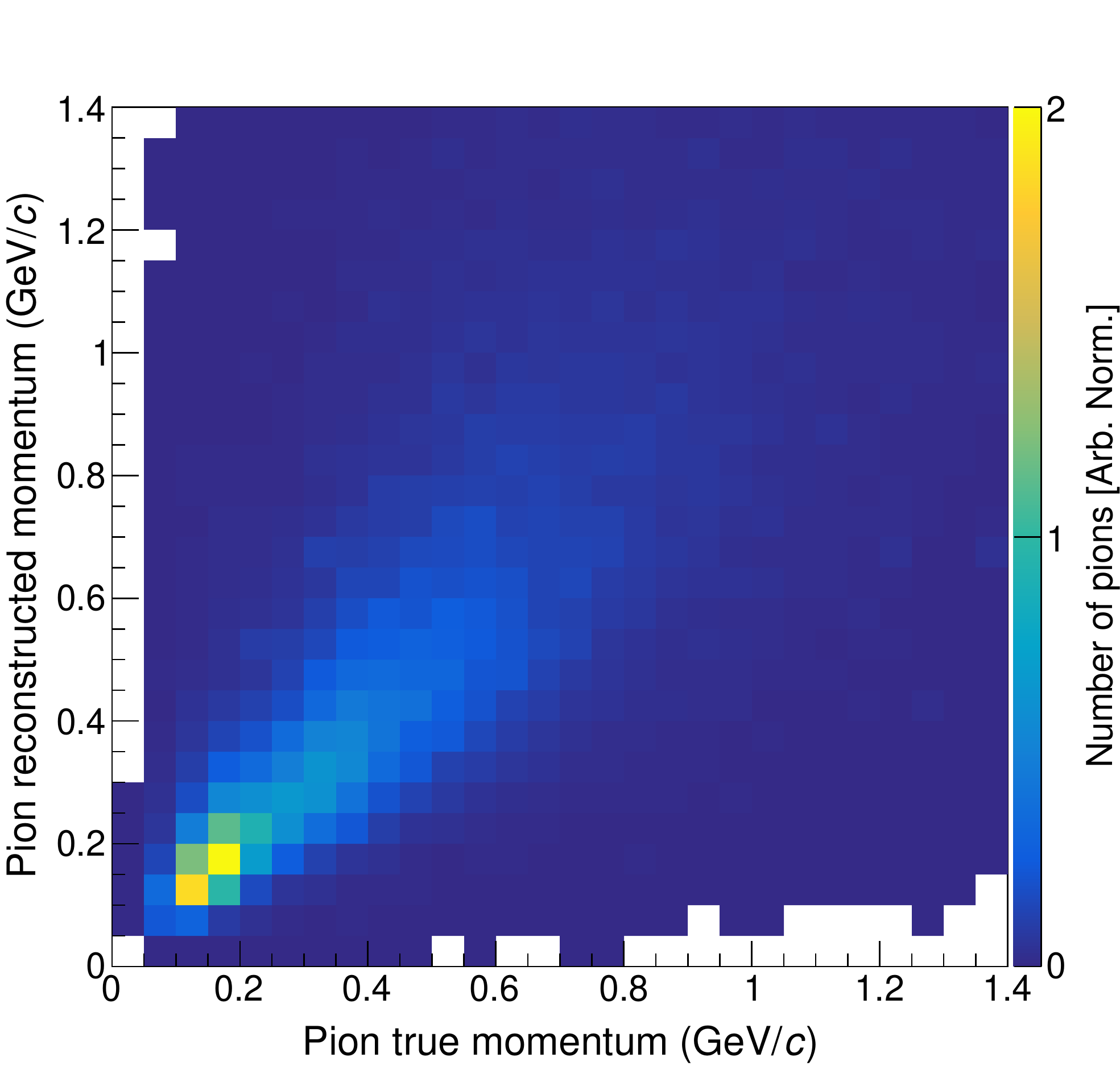}
		\caption{Relation between the reconstructed and true momenta of charged pions from the neutrino interactions in the MC simulation.}
		\label{fig:pion_momentum_recon_true_relation}
\end{figure}

\begin{figure}[h]
		\includegraphics[clip, width=0.95\columnwidth]{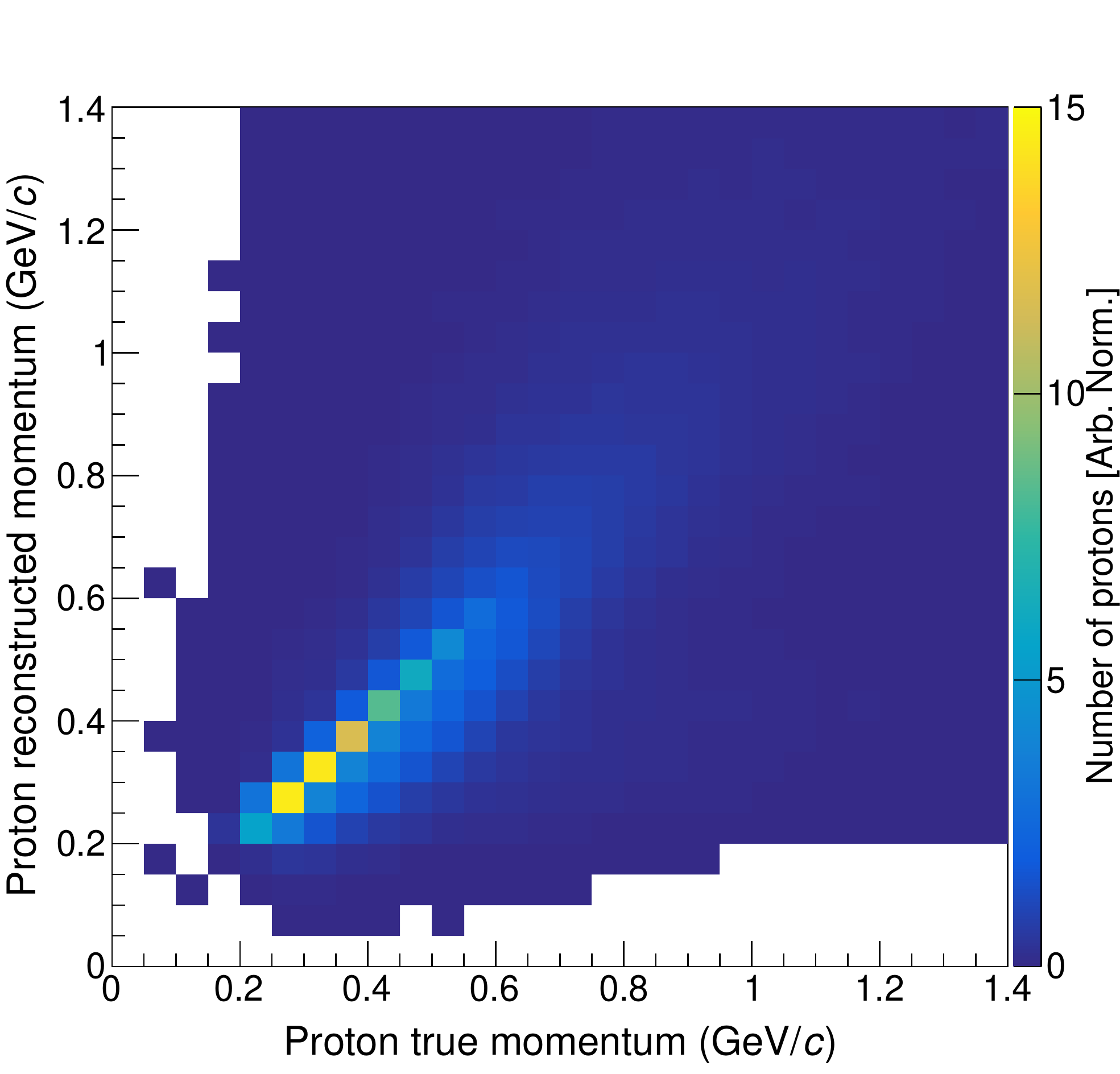}
		\caption{Relation between the reconstructed and true momenta of protons from the neutrino interactions in the MC simulation.}
		\label{fig:proton_momentum_recon_true_relation}
\end{figure}

The momentum-measurement resolutions for muons, pions, and protons were evaluated using the MC simulation.
Figure~\ref{fig:muon_momentum_recon_true_relation} shows the relation between the reconstructed and true momenta of muons from the neutrino interactions, while Figs.~\ref{fig:pion_momentum_recon_true_relation}~and~\ref{fig:proton_momentum_recon_true_relation} show that of charged pions and protons, respectively.
The mean values of the momentum measurement resolutions in this analysis are listed in Table~\ref{tab:momentum_accuracy}.
\begin{table}[h]
	\caption{Momentum measurement resolutions for muons, pions, and protons from neutrino interactions. Momentum estimations using the range-energy relation for pions were not applied because of the high probability of pion interactions in the detector.}
	\label{tab:momentum_accuracy}
	\begin{center}
		\begin{tabular}{cccc}
			\hline
			\hline
			& Muon & Pion & Proton \\
			\hline
			Angular method			& 43.0\%	& 29.6\%	& 36.0\% \\
			Coordinate method		& 25.9\%	& 25.2\%	& 30.7\% \\
			Range-energy relation		& 6.4\%		& -			& 3.8\% \\
			\hline
		\end{tabular}
	\end{center}
\end{table}

\subsection{Particle identification}\label{subsec:PID}
\begin{figure}[h]
		\subfigure{
			\includegraphics[clip, width=0.95\columnwidth]{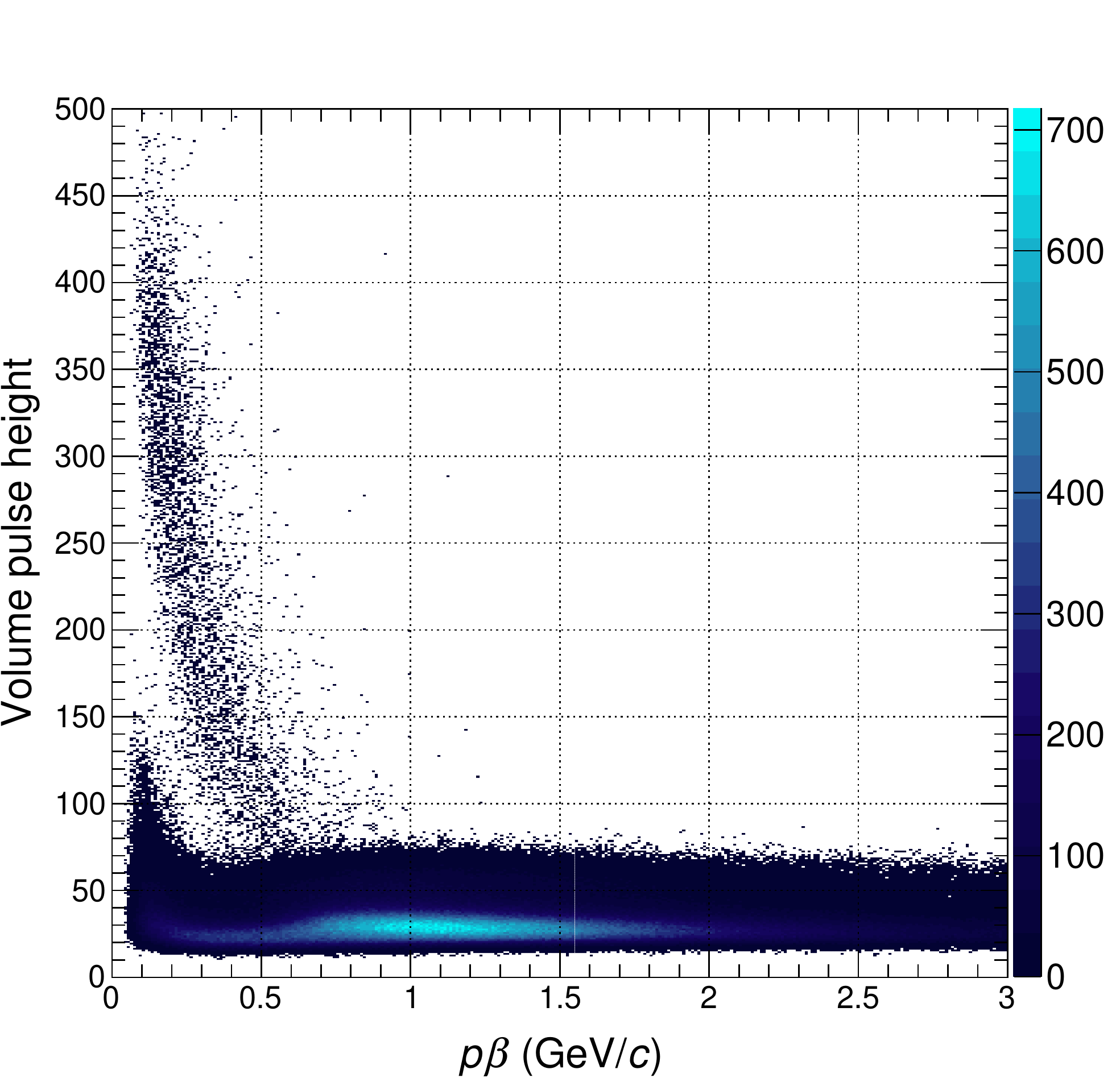}}
		\subfigure{
			\includegraphics[clip, width=0.95\columnwidth]{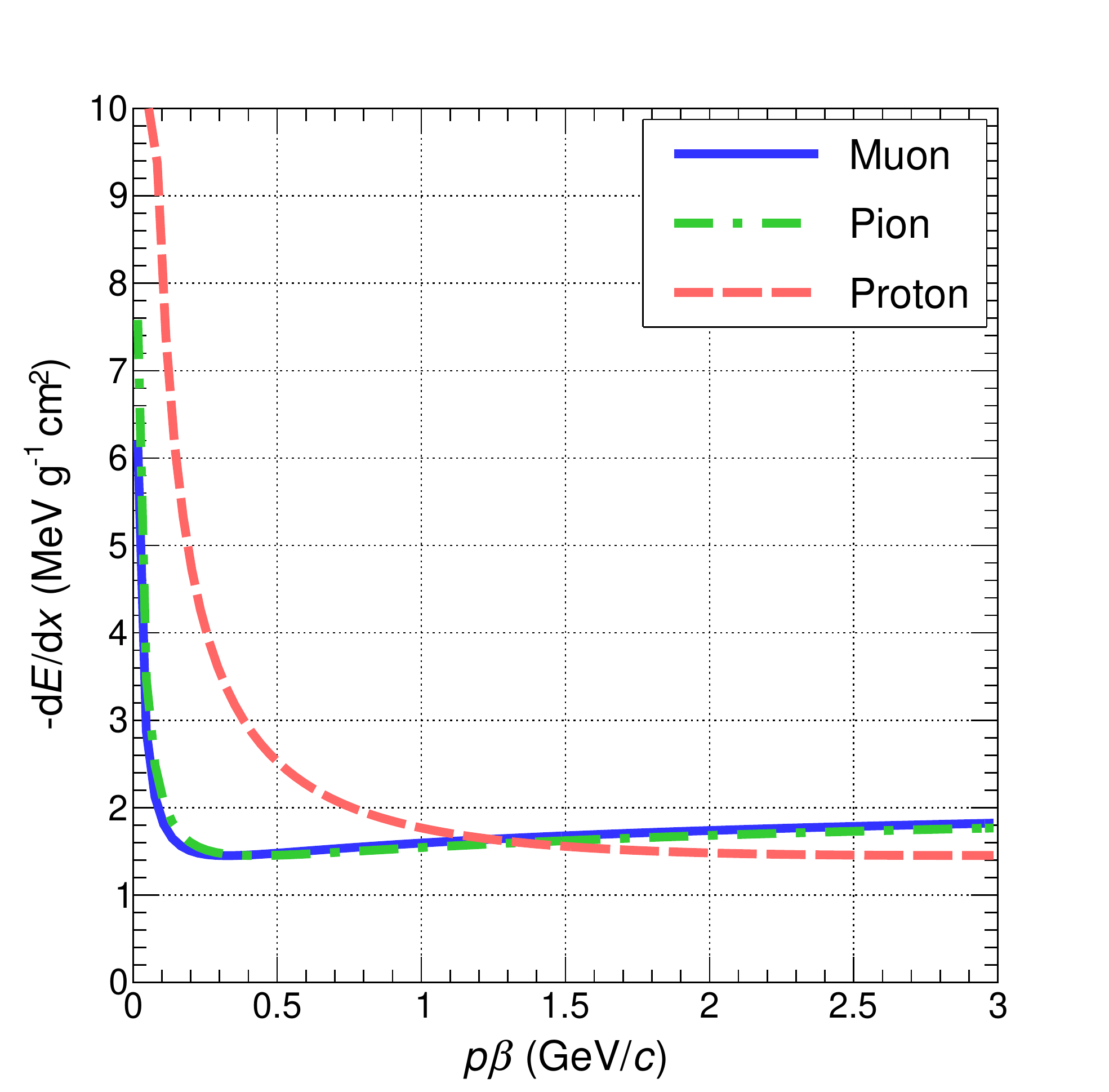}}
		\caption{Top figure shows correlations between VPH and $p\beta$ of the tracks in the ECC bricks, primarily cosmic-rays. The bottom figure shows the theoretical curves of the ionization loss in the emulsion film. The curves were calculated using the Bethe equation~\cite{Rossi:99081,Bethe:1930ku,RevModPhys.9.245,doi:10.1146/annurev.ns.13.120163.000245}.}
		\label{fig:VPH_dEdx_pbeta_correlation}
\end{figure}

\begin{figure}[h]
		\includegraphics[width=8.0cm]{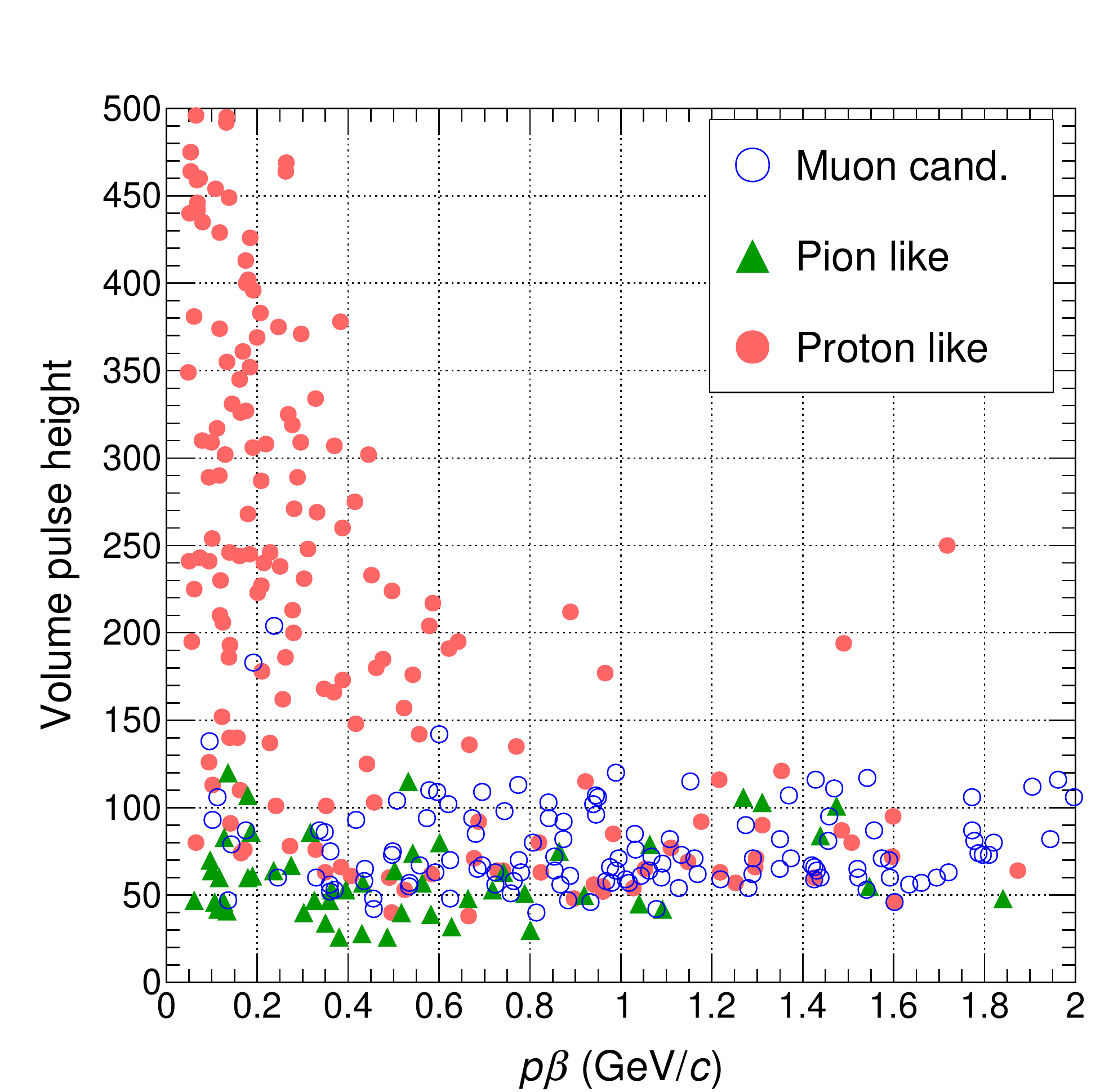}
		\caption{Correlation between VPH and $p\beta$ of the data is shown. The PID between protons and charged pions is performed using the VPH and $p\beta$ of the tracks in the ECC bricks, whereas the muon ID is performed using track matching among the ECC bricks, Shifter, and INGRID.}
		\label{fig:VPH_pbeta}
\end{figure}

\begin{figure}[h]
		\includegraphics[width=8.6cm,pagebox=cropbox]{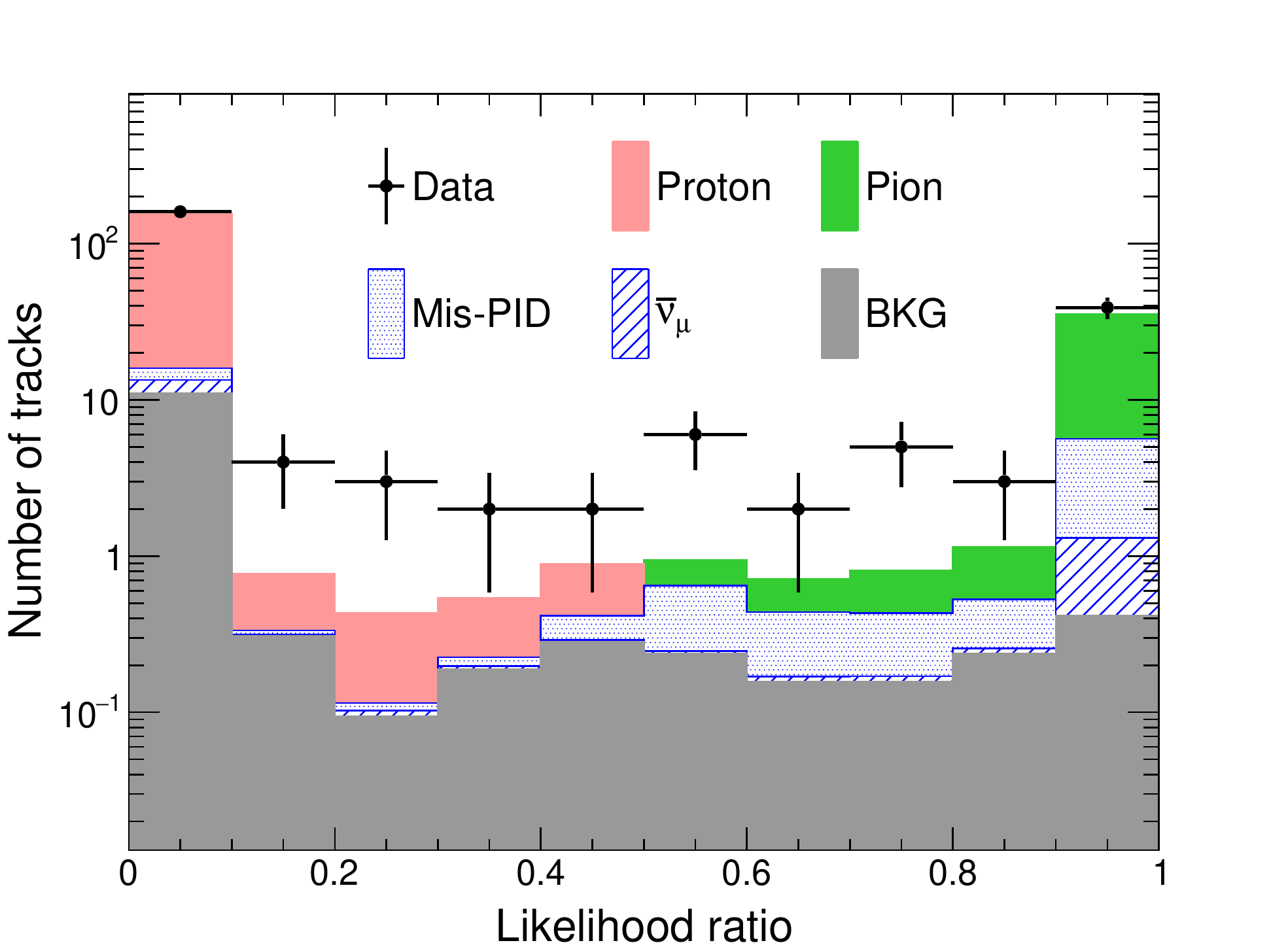}
	\caption{Likelihood ratio distributions of protons and charged pions. The marker represents the data, and the histogram represents the MC simulation. The background includes $\nu_{e}$ and $\bar{\nu}_{e}$ interactions in the ECC bricks, neutrino interactions in the upstream wall of the detector hall and INGRID modules, and misconnected events between the ECC bricks, Shifter, and INGRID. The difference between the data and the MC simulation in the range of 0.1 to 0.9 reflect the difference between the observed and predicted numbers of protons and pions, as discussed in Sec.~\ref{sec:results}.}
	\label{fig:Evaluation_LikelihoodRatio}
\end{figure}

\begin{figure}[h]
	\subfigure{
		\includegraphics[clip, width=0.95\columnwidth]{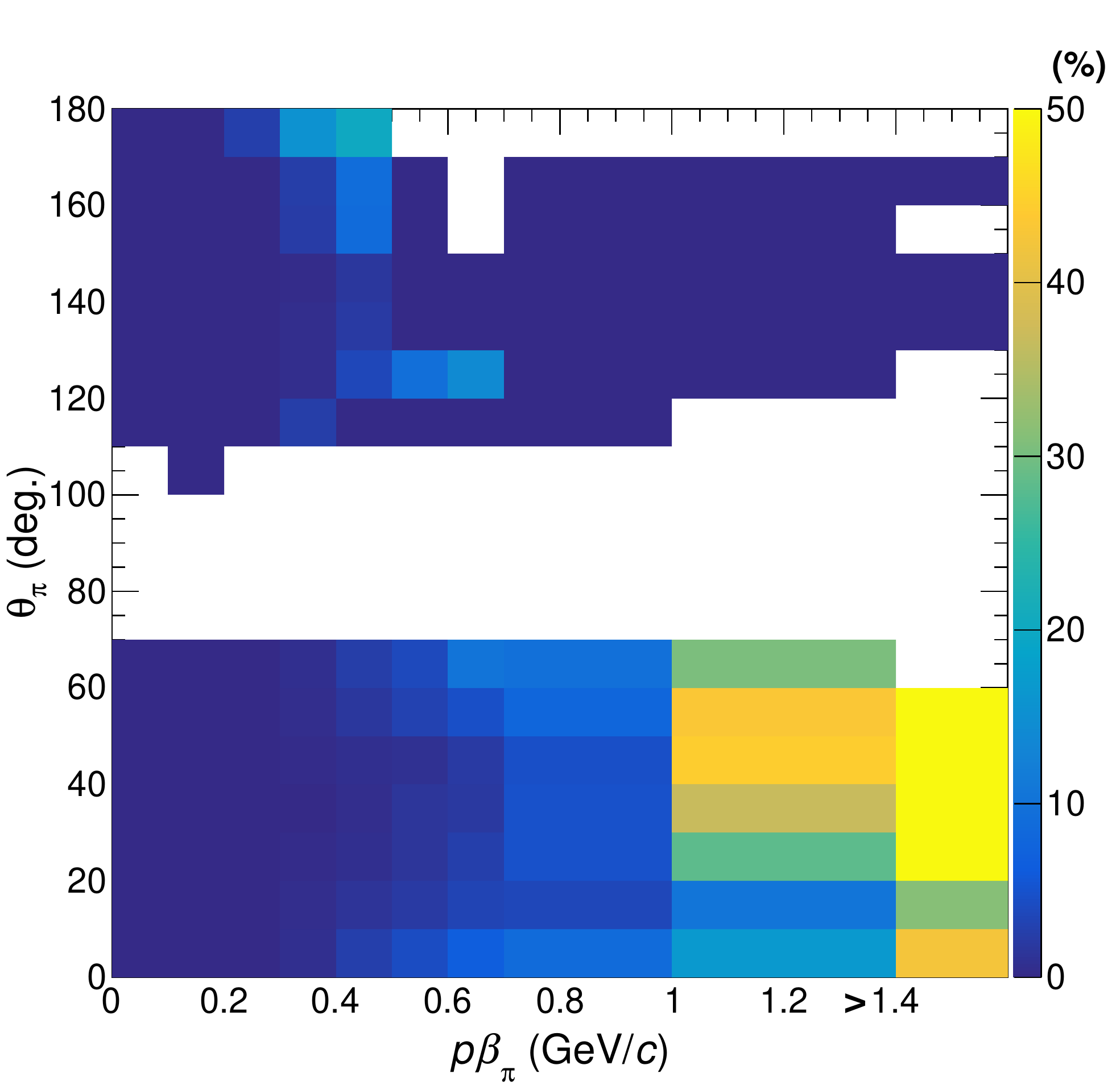}}
	\subfigure{
		\includegraphics[clip, width=0.95\columnwidth]{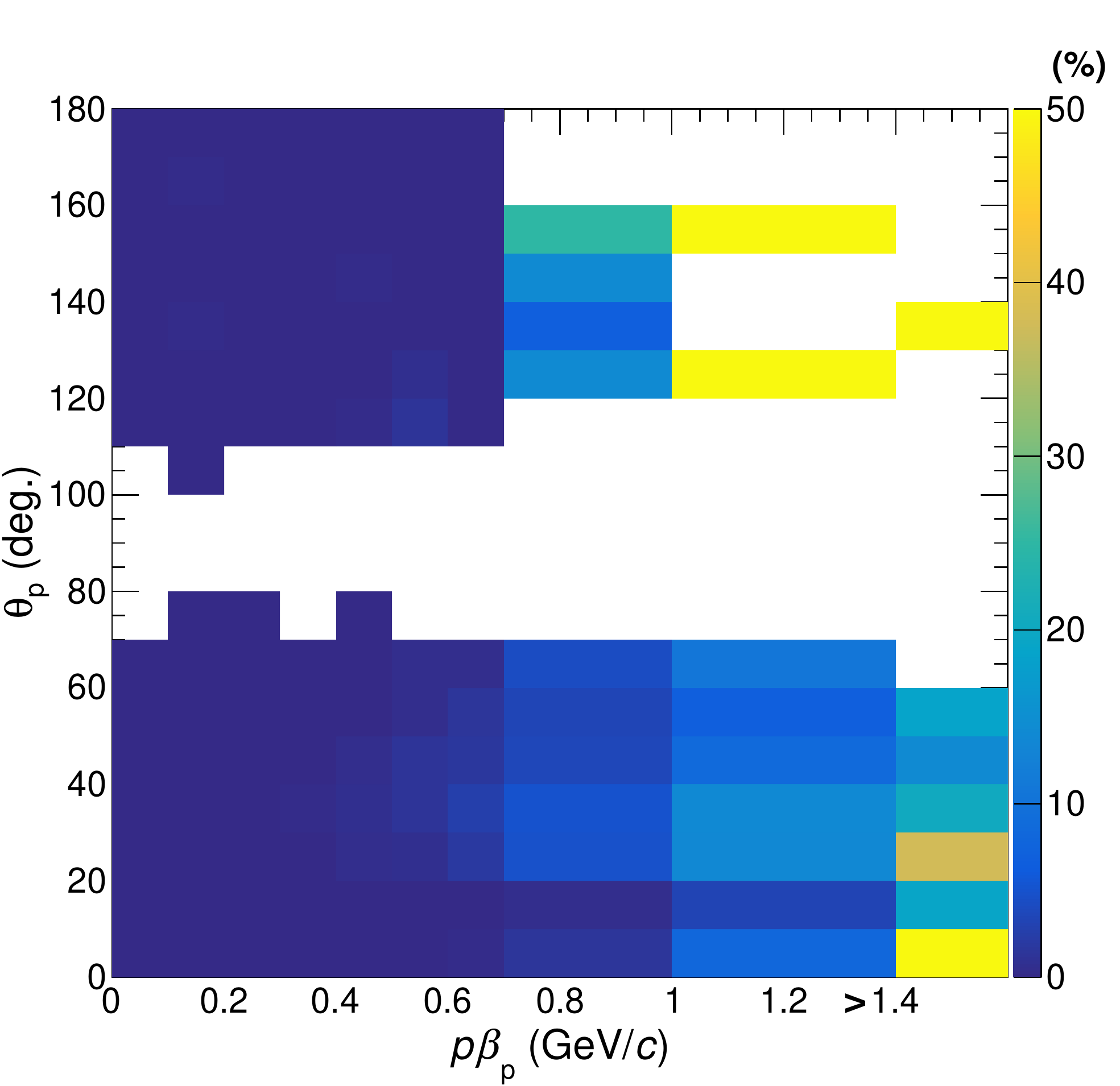}}
	\caption{Mis-PID rate distributions of charged pions and protons. The top figure shows the rate of pions, whereas the bottom figure shows the rate of protons. In these measurements, the momentum in the horizontal axis is expressed by $p\beta$, because the PID process uses $p\beta$ estimated by MCS measurements.}
	\label{fig:Mis-PID-rate}
\end{figure}

The PID of protons and charged pions was performed using the VPH and $p\beta$ of the tracks in the ECC bricks, whereas the muon ID was performed using track matching between the ECC, Shifter, and INGRID~\cite{ninja_run6_xsec}.
Figure~\ref{fig:VPH_dEdx_pbeta_correlation}~(top) shows correlations between the VPH and $p\beta$ of tracks recorded in the ECC bricks, while the figure ~\ref{fig:VPH_dEdx_pbeta_correlation}~(bottom) shows the theoretical curves of the mean rate of energy loss calculated using the Bethe equation~\cite{Rossi:99081,Bethe:1930ku,RevModPhys.9.245,doi:10.1146/annurev.ns.13.120163.000245}.
These figures clearly demonstrate that the MIPs and HIPs can be well separated using the VPH and $p\beta$ in the Fig.~\ref{fig:VPH_dEdx_pbeta_correlation}~(top) as well as using the mean rate of energy loss and $p\beta$ in the Fig.~\ref{fig:VPH_dEdx_pbeta_correlation}~(bottom).
The fractions of charged hadrons produced in $\nu_{\mu}$ CC interactions on iron were evaluated using NEUT, and 99.6\% of these particles were protons and charged pions.
Therefore in this analysis, charged hadrons from neutrino interactions were considered as protons or charged pions.
The likelihood functions for charged pions and protons are defined as
\begin{eqnarray}
L_{\rm pion} &\equiv \frac{1}{\sqrt{2\pi}\sigma_{\rm MIP}}{\rm exp}[\frac{-({\mathit v}-\mu_{\rm MIP})^{2}}{2\sigma_{\rm MIP}^{2}}],
\label{eq:LikelihoodFunction_pion}
\\
L_{\rm proton} &\equiv \frac{1}{\sqrt{2\pi}\sigma_{\rm HIP}}{\rm exp}[\frac{-({\mathit v}-\mu_{\rm HIP})^{2}}{2\sigma_{\rm HIP}^{2}}],
\label{eq:LikelihoodFunction_proton}
\end{eqnarray}
where $\mathit v$ denotes the VPH of the track, $\mu$ denotes the central value, and $\sigma$ denotes the standard deviation of the VPH distribution.
$\mu$ and $\sigma$ were obtained for each region of $p \beta$ in the range of 0.0 to 0.5\,GeV/$c$ by 0.1\,GeV/$c$, and tan$\theta$ in the range of 0.0 to 2.0 by 0.1.
Track angle $\theta$ is defined as the angle with respect to the direction perpendicular to the emulsion film.
In the region of $p \beta$ above 0.5\,GeV/$c$, and tan$\theta$ above 2.0, extrapolated values of the $\mu$ and $\sigma$ were used because the number of tracks available was limited.
The likelihood ratio $\mathcal R$ is defined by
\begin{equation}
{\mathcal R} \equiv \frac{L_{\rm pion}}{L_{\rm pion}+L_{\rm proton}}.
\label{eq:LikelihoodRatio}
\end{equation}
Charged hadron tracks with $\mathcal R$ less and greater than 0.5 were defined as protons and charged pions, respectively, in this study.
Figure~\ref{fig:VPH_pbeta} shows the correlation between VPH and $p\beta$ of the partner tracks after PID.
In the region $p\beta < 0.5\,{\rm GeV}/c$, the separation between protons and charged pions is well demonstrated. 
In addition, protons and pions above 0.5\,GeV/$c$ can be identified, because the PID parameters were tuned for each region of $p\beta$ and tan$\theta$.
Thus, based on the PID results, the multiplicities of the protons and charged pions were measured in neutrino interactions final state.
The performance of PID was estimated using the MC simulation.
The likelihood ratio distribution is shown in Fig.~\ref{fig:Evaluation_LikelihoodRatio}.
The proton ID was performed with 96.5$\%$ efficiency and 98.1$\%$ purity, whereas the charged pion ID was performed with 92.3$\%$ efficiency and 86.8$\%$ purity.
Figure~\ref{fig:Mis-PID-rate} shows mis-PID rates of charged pions and protons from neutrino interactions in the MC simulation.
In the region $p\beta$ below 0.5\,GeV/$c$, the average mis-PID rates were 0.5\% and 0.1\% for pions and protons, respectively.
The mis-PID rates for $p\beta$ above 1.0\,GeV/$c$ are 19.3\% and 15.7\% for pions and protons, respectively.

\begin{figure}[h]
		\includegraphics[width=8.6cm,pagebox=cropbox]{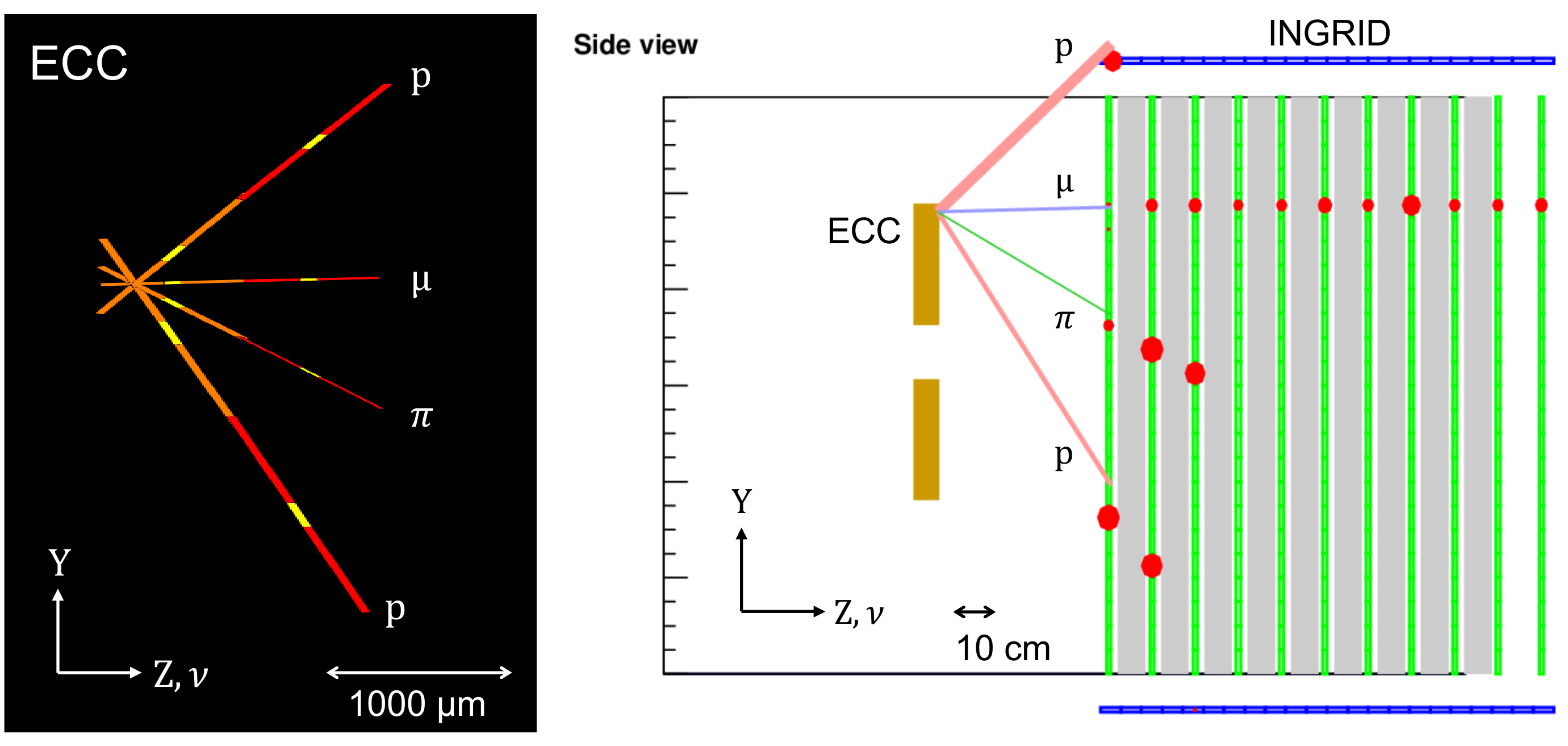}
		\caption{Event display of a $\nu_{\mu}$-iron CC interaction candidate. The left-hand side of the figure shows the event display in the ECC brick, while the right-hand side shows the event display both in the ECC bricks and in INGRID. On the left-hand side, with their colors representing the track segment and their width indicating the VPH. On the right-hand side, the lines from the ECC bricks are the ECC tracks extrapolated to INGRID, with the blue line representing a muon candidate and the pink and green lines representing a proton-like and pion-like tracks, respectively. The width of the lines indicates the VPH. The red markers represent hits and thier size represent deposited photoelectrons in INGRID.}
		\label{fig:eventdisplay}
\end{figure}

Figure~\ref{fig:eventdisplay} shows the event display of a $\nu_{\mu}$-iron CC interaction candidate.
The event contains a muon, a pion-like, and two proton-like tracks.

\section{Systematic uncertainties}\label{sec:systematics}
The systematic uncertainty sources in the measurements can be categorized into four groups: neutrino flux, neutrino interaction models, background estimation and detector response. 
The uncertainty from each source is evaluated from the data and MC simulation as follows.

\subsection{Neutrino flux}\label{subsec:sys_flux}
\begin{figure}[h]
		\includegraphics[width=8.6cm,pagebox=cropbox]{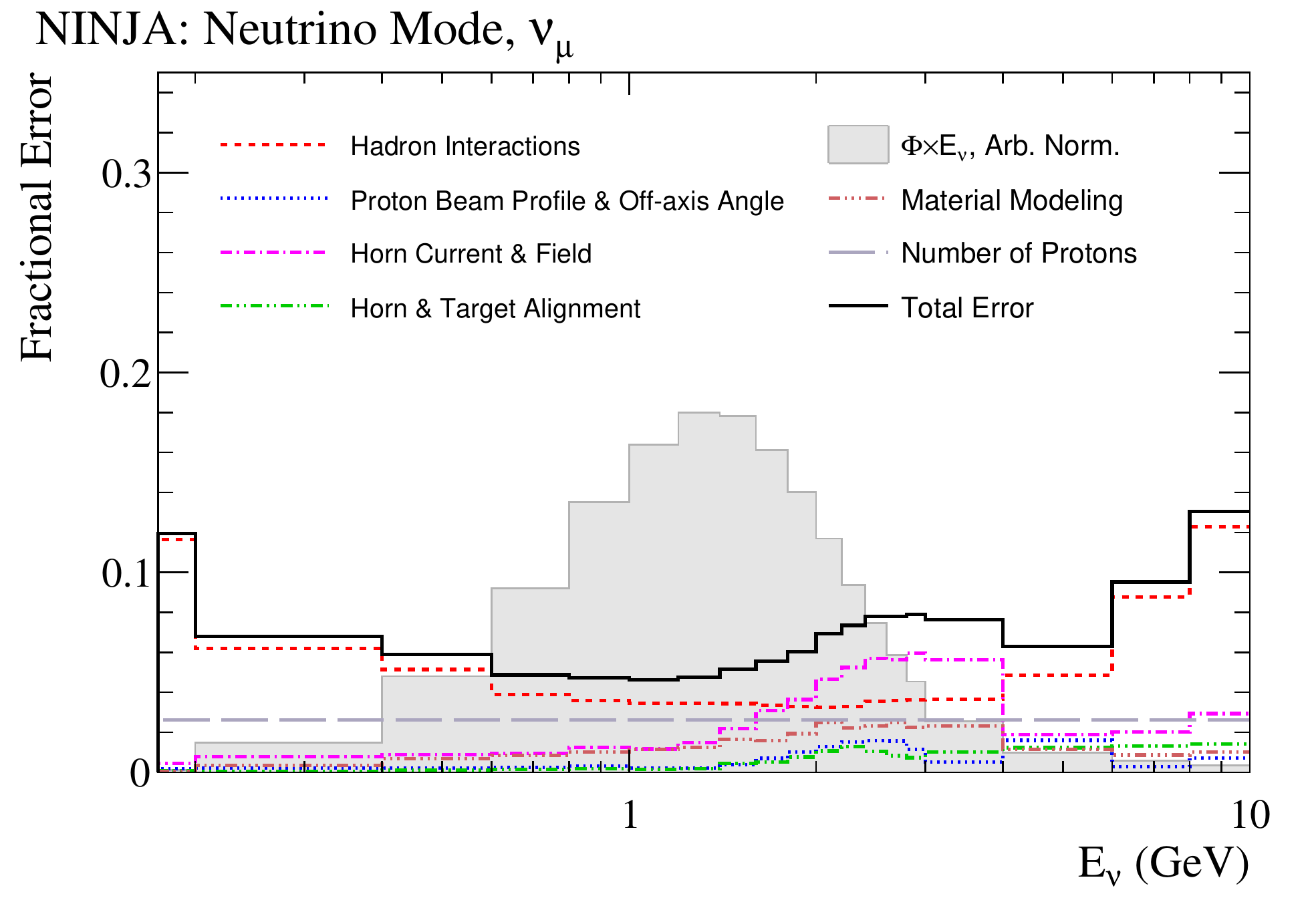}
		\caption{Neutrino flux uncertainties are due to hadron production and neutrino beamline optics uncertainties.}
		\label{fig:flux_error}
\end{figure}

The neutrino flux uncertainties arise from hadron production and neutrino beamline optics uncertainties.
A covariance matrix at the detector location was prepared following the same procedure as for the T2K experiment~\cite{Flux_JNUBEAM_2013}, with the relative errors in each energy bin shown in Fig.~\ref{fig:flux_error}.
The neutrino flux is made to fluctuate according to the covariance matrix and the $\pm$1$\sigma$ change of the number of predicted neutrino interactions at each bin is taken as the systematic uncertainty.

\subsection{Neutrino interaction}\label{subsec:sys_nuint}
The neutrino interaction and FSI models used in NEUT have several uncertainties.
Table~\ref{tab:nu_int_prm_value_uncertainty} shows parameters~\cite{PhysRevD.96.092006,10.1093/ptep/ptz070} used for modeling the neutrino interactions and the FSI in NEUT.
It lists the nominal values and the 1$\sigma$ uncertainties for the parameters.
Neutrino interaction uncertainties in NEUT are evaluated by changing the parameters and the change of the number of predicted neutrino interactions is taken as the systematic uncertainty for the MC prediction.
The dominant uncertainties of proton kinematics are the uncertainties of 2p2h interaction normalization and pion absorption.
These two uncertainties in the proton emission angles are about 9\% for forward emission and 17--19\% for backward emission, whereas the uncertainties in the momenta are about 6--14\%.
On the other hand, the dominant uncertainties of pion kinematics are the uncertainties of ${\textit{M}}^{\textrm{RES}}_{\textrm{A}}$, ${\textit{C}}^{\textrm{A}}_{\textrm{5}}$(0), CC other normalization, and CC coherent normalization.
These four uncertainties in the pion emission angles are about 6--8\%, whereas the uncertainties in the momenta are about 5--6\%.
\begin{table}[htbp]
	\caption{Nominal parameter values and their uncertainties in the neutrino interaction models. Detailed descriptions of the parameters are given in Refs.~\cite{PhysRevD.96.092006,10.1093/ptep/ptz070}.}	
	\label{tab:nu_int_prm_value_uncertainty}
	\begin{center}
	\begin{tabular}{lll}
		\hline
		\hline
		 Parameter & Nominal value & Uncertainty \\
		\hline
		${\textit{M}}^{\textrm{QE}}_{\textrm{A}}$   & 1.05\,GeV/$c^2$ & 0.20\,GeV/$c^2$\\
		${\textit{M}}^{\textrm{RES}}_{\textrm{A}}$  & 0.95\,GeV/$c^2$& 0.15\,GeV/$c^2$\\
		${\textit{C}}^{\textrm{A}}_{\textrm{5}}$(0)  & 1.01 & 0.12 \\
		Isospin $\frac{1}{2}$BG                           & 1.30 & 0.20 \\
		CC other shape                                            & 0 & 0.40 \\
		CC coherent normalization                       & 100$\%$ & 100$\%$ \\
		NC other normalization                             & 100$\%$ &   30$\%$ \\
		NC coherent normalization                       & 100$\%$ &   30$\%$ \\
		2p2h normalization                                     & 100$\%$ & 100$\%$ \\
		Fermi momentum ${\textit{P}}_{\textrm{F}}$                                      &  250\,MeV/$c$ & 30\,MeV/$c$ \\
		Binding energy ${\textit{E}}_{\textrm{b}}$                                             &     33\,MeV        &    9\,MeV \\
		Pion  absorption                                                                     & 1.1 & 50$\%$ \\
		Pion  charge exchange    & 1.0 & 50$\%$ \\
		($p_{\pi} <$ 500\,MeV/$c$) & & \\
		Pion  charge exchange    & 1.8 & 30$\%$ \\
		($p_{\pi} >$ 500\,MeV/$c$) & & \\
		Pion  quasi elastic           &  1.0 & 50$\%$ \\
		($p_{\pi} <$ 500\,MeV/$c$) & & \\
		Pion  quasi elastic           &  1.8 & 30$\%$ \\
		($p_{\pi} >$ 500\,MeV/$c$) & & \\
		Pion  inelastic                                                                          &  1.0 & 50$\%$ \\
		\hline
		\hline
	\end{tabular}
	\end{center}
\end{table}

\subsection{Background estimation}\label{subsec:sys_bkg}
Regarding the background estimation, the uncertainties of associated with the beam-induced particles from outside the ECC bricks and misconnected backgrounds are considered.
Most of the beam-induced backgrounds are caused by hadrons produced in neutrino interactions in the upstream wall of the detector hall.
The beam-induced background uncertainty mainly originates from the normalization of the sand muons, which are produced in neutrino interactions in the wall.
The number of sand muons in the MC simulation was found to be 30\% smaller than in the data and the difference is considered as a systematic uncertainty of the beam-induced backgrounds.
Misconnection background is a misconnection of the ECC track to INGRID or to cosmic muon coming from downstream and stopping in an ECC brick.
The uncertainty attributed to misconnected events was evaluated using mock data, which are the combination of the nominal and fake data in which the time information of the ECC track is shifted.
The total uncertainties of these backgrounds are sufficiently small compared to other uncertainties in most regions.

\subsection{Detector response}\label{subsec:sys_det}
Regarding the muon detection, the following uncertainties are considered: the muon track reconstruction, the track connection between detectors, the event selections, and the target mass and materials.
Further details on the muon detection uncertainties are described in Ref.~\cite{ninja_run6_xsec}.
Concerning the proton and charged pion detection, the uncertainties of the proton and pion track reconstruction, partner track search, and PID process were considered as the detector response uncertainties for proton and charged pion detection.

Regarding the track reconstruction, the uncertainties of the track reconstruction in the ECC bricks and the track connection between the ECC bricks were considered.
The efficiency of track reconstruction in the ECC bricks was evaluated using the data which is the sand muon tracks accumulated in the ECC bricks.
The track reconstruction uncertainty was evaluated with the reconstruction efficiency difference between the data and MC simulation, whereas the uncertainty of the black track reconnection was evaluated by the $\pm$1$\sigma$ change in the connection parameter errors using the MC simulation.
The efficiency of track connection among the ECC bricks was also evaluated using the sand muon data in the ECC bricks.
The track connection uncertainty was evaluated using the connection efficiency difference between the data and MC simulation.

Further, regarding partner track search, the uncertainties of the minimum distance tolerance and the rate of fake partner tracks were considered.
The maximum error of the minimum distance between the muon track and its partner track is 11\,${\rm \mu m}$, which originates from the track alignment uncertainty in the ECC bricks.
The uncertainty for the minimum distance was evaluated using the MC simulation by the $\pm$1$\sigma$ change in the minimum distance error.
Further, the uncertainty for fake partner tracks due to chance coincidence was estimated from the data.
Most fake partner tracks were low-momentum particles that passed near the interaction vertex and appeared as partner tracks because, due to a large scattering angle, the connection to the track segment beyond the vertex failed.
The rate of fake partner tracks was studied using mock data, which are muon track data with shifted positions.
The fake track rate of the charged pions was 4.3$\%$, whereas the rate of protons was negligible in this study.
Thus, the fake track rate for charged pions was considered as a systematic uncertainty.

In terms of PID, the difference in the VPH distributions between the data and MC simulation was considered.
The VPH uncertainty for the MIP tracks was 16\%, while the uncertainty for the HIP tracks was 15\%.
The uncertainty of the measurements was evaluated using the $\pm$1$\sigma$ change in VPH uncertainties.

\begin{figure*}
		\subfigure{
			\includegraphics[clip, width=1.0\columnwidth]{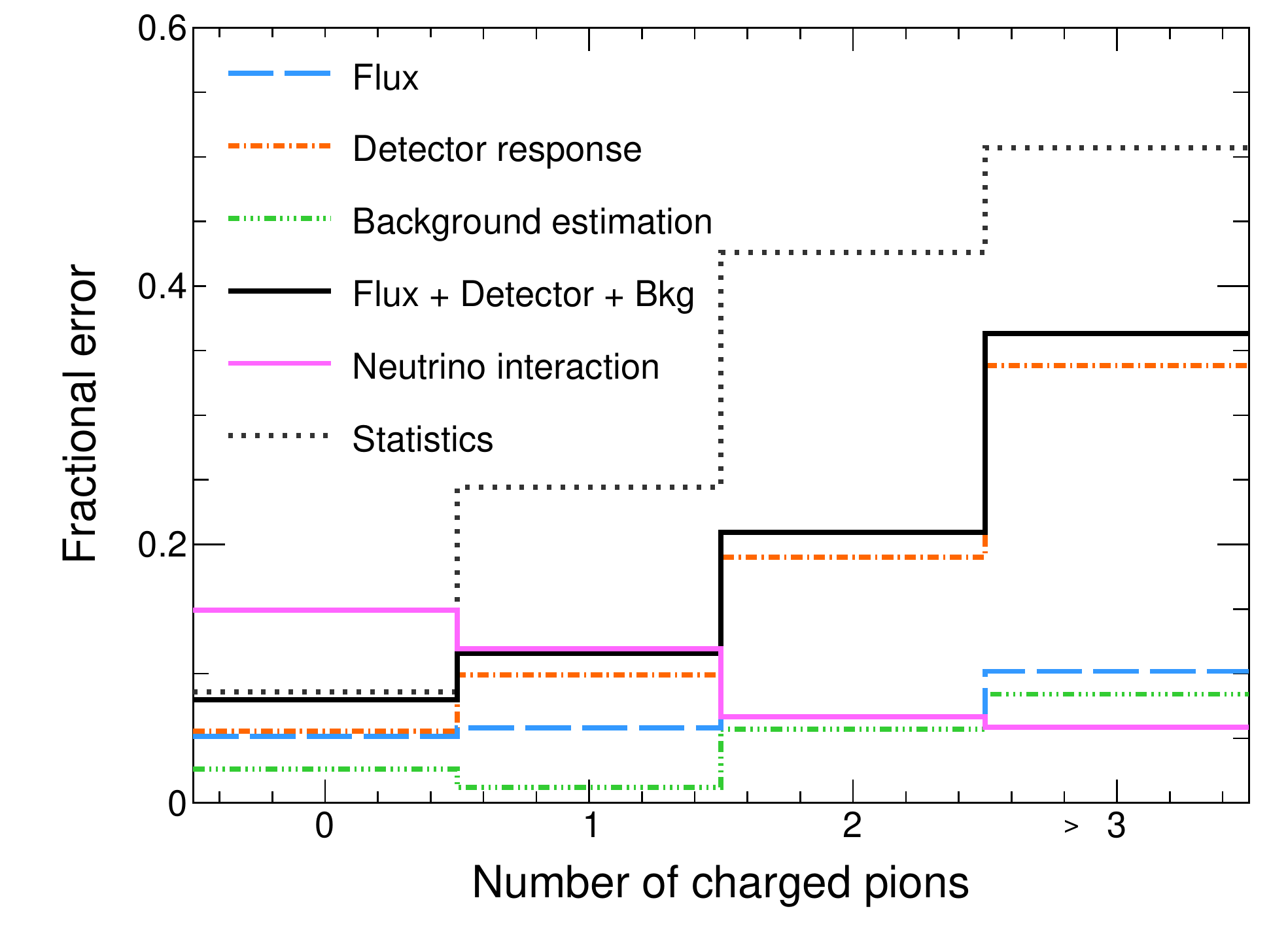}}
		\subfigure{
			\includegraphics[clip, width=1.0\columnwidth]{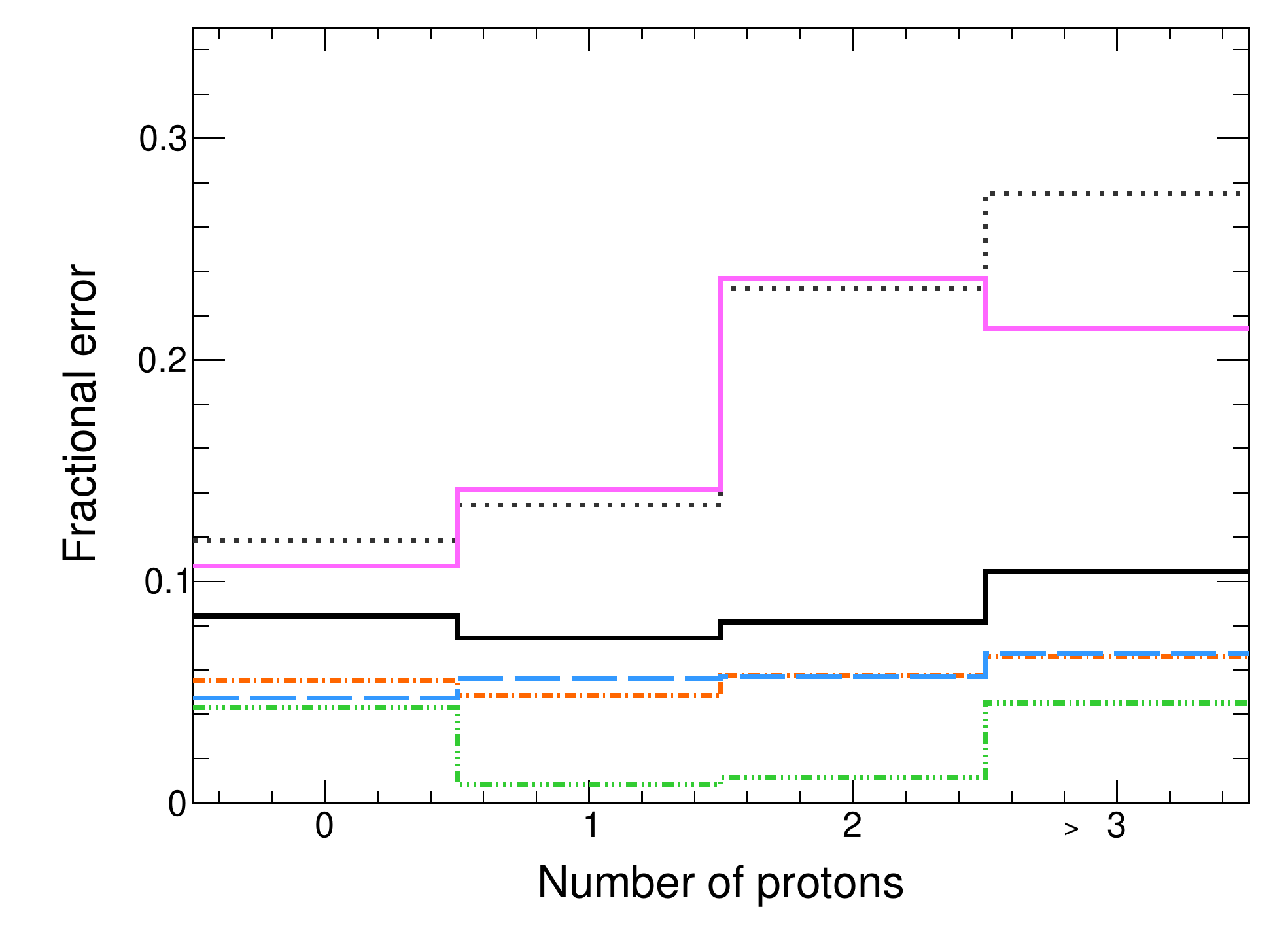}}
		\caption{Systematic uncertainties of the charged pion and the proton multiplicities. The left-hand side of the figures shows the uncertainties of the charged pion multiplicity, whereas the right-hand side shows the uncertainties of proton multiplicity. Statistical uncertainty of our data is also shown for comparison.}
		\label{fig:sys_mul}
\end{figure*}

\begin{figure*}
		\subfigure{
			\includegraphics[clip, width=1.0\columnwidth]{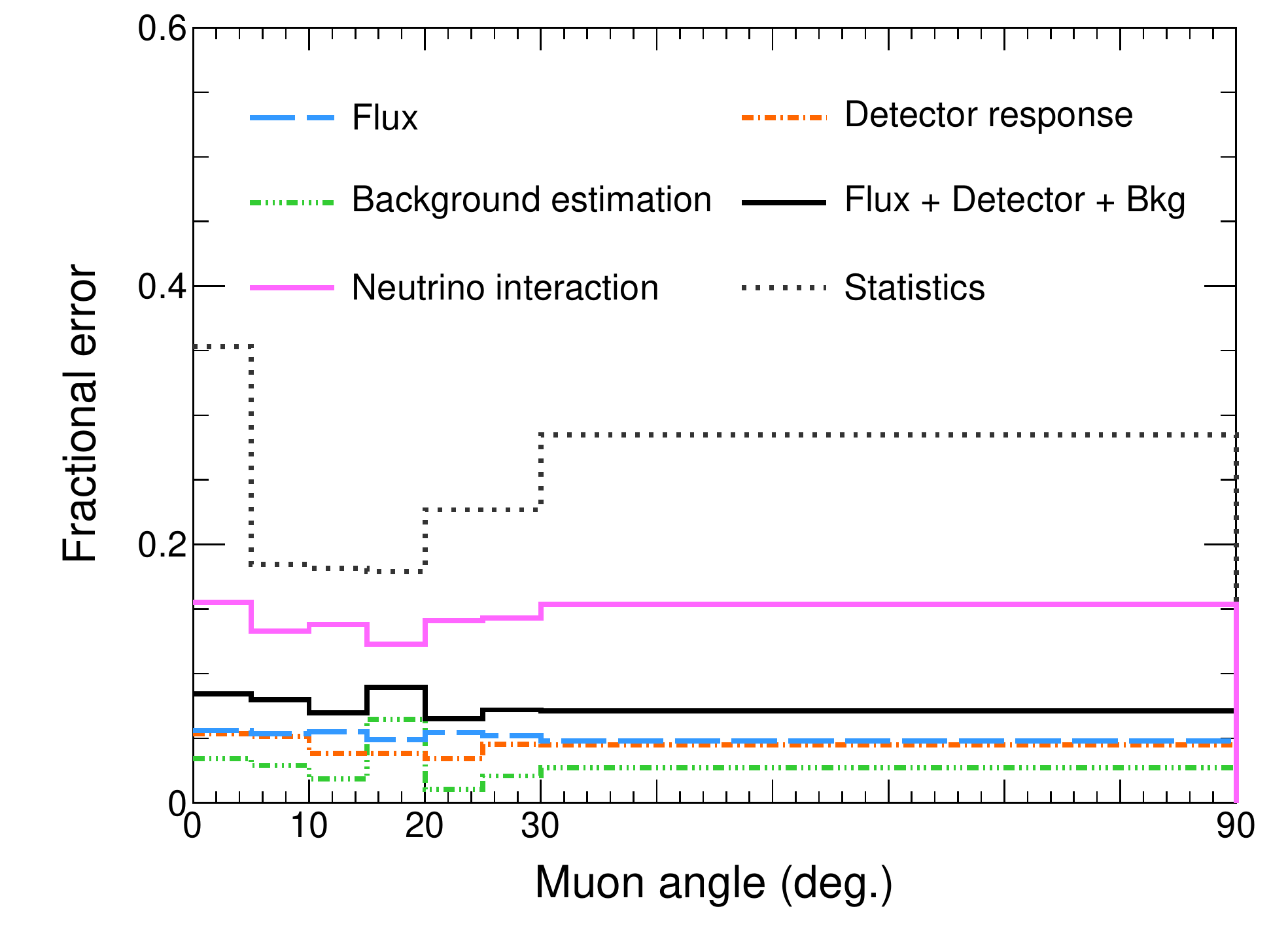}}
		\subfigure{
			\includegraphics[clip, width=1.0\columnwidth]{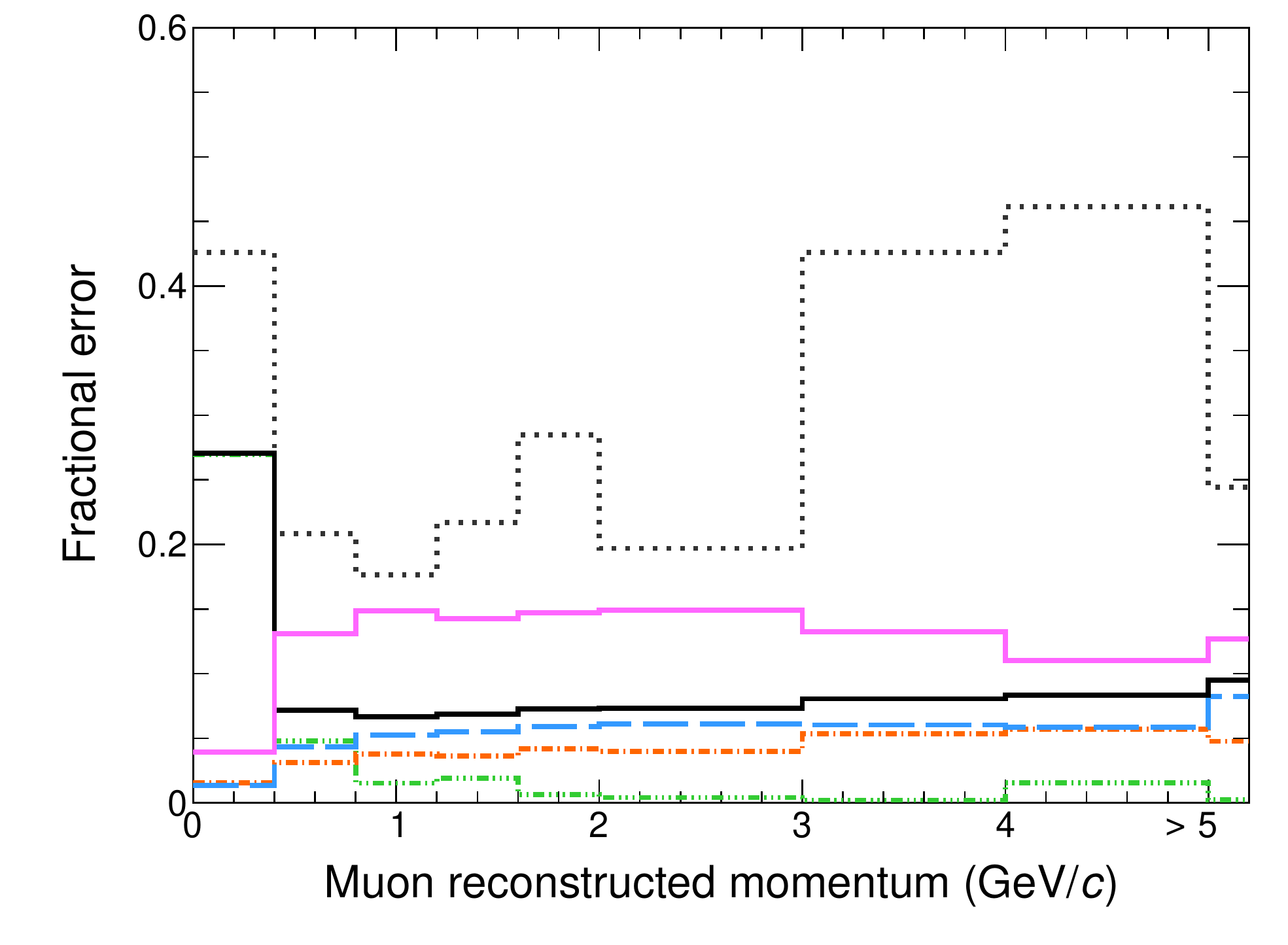}}
		\subfigure{
			\includegraphics[clip, width=1.0\columnwidth]{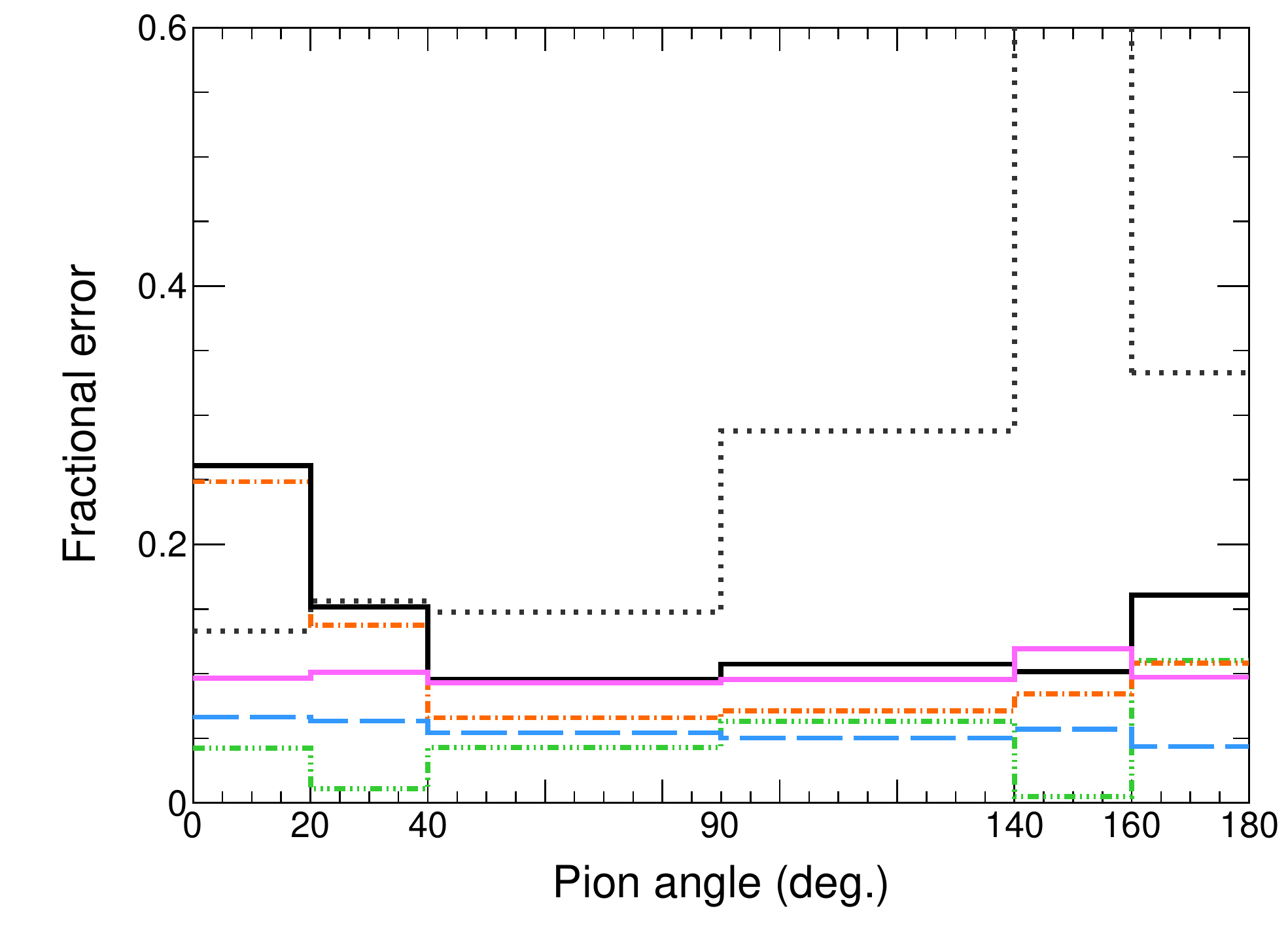}}
		\subfigure{
			\includegraphics[clip, width=1.0\columnwidth]{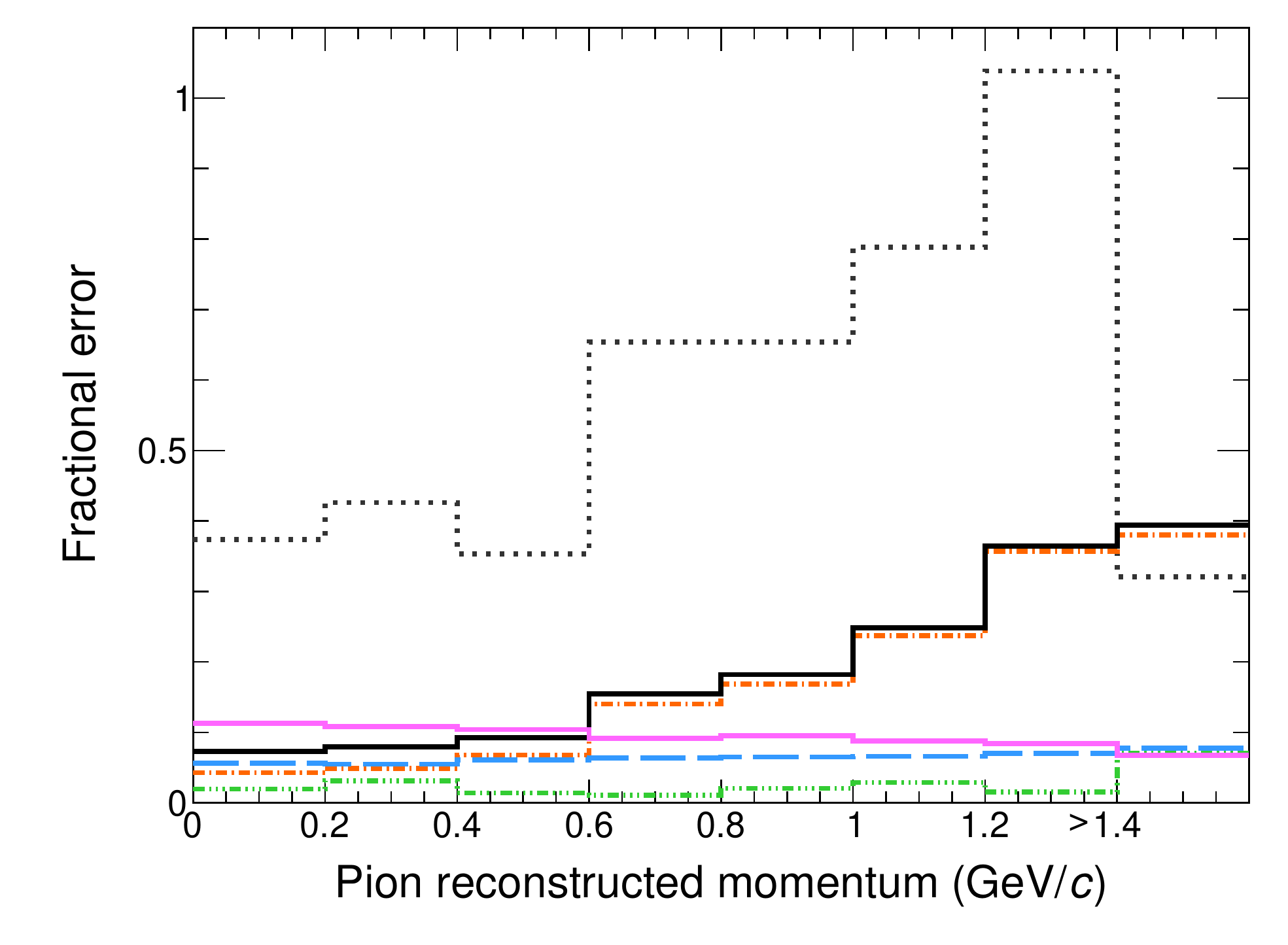}}
		\subfigure{
			\includegraphics[clip, width=1.0\columnwidth]{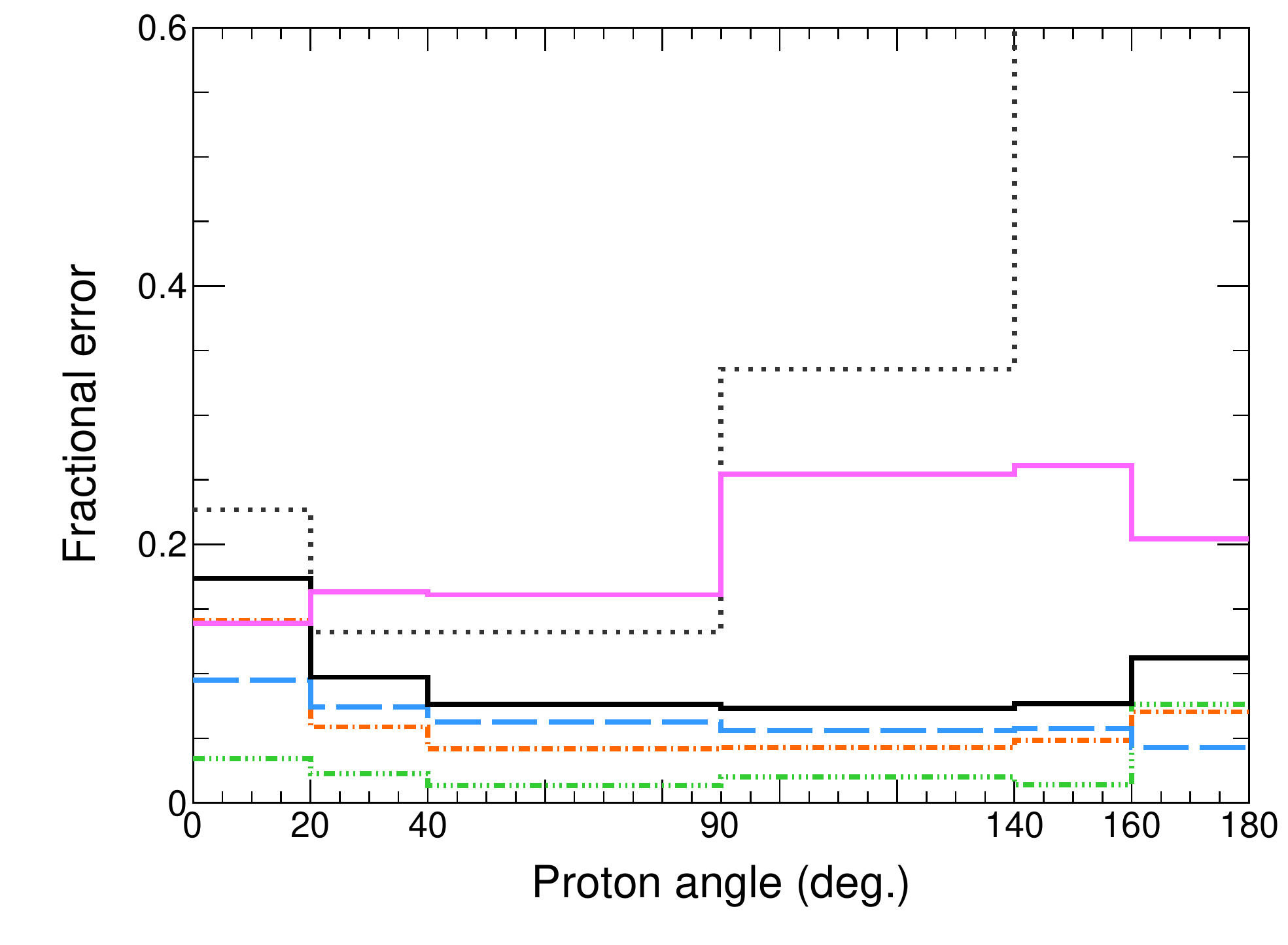}}
		\subfigure{
			\includegraphics[clip, width=1.0\columnwidth]{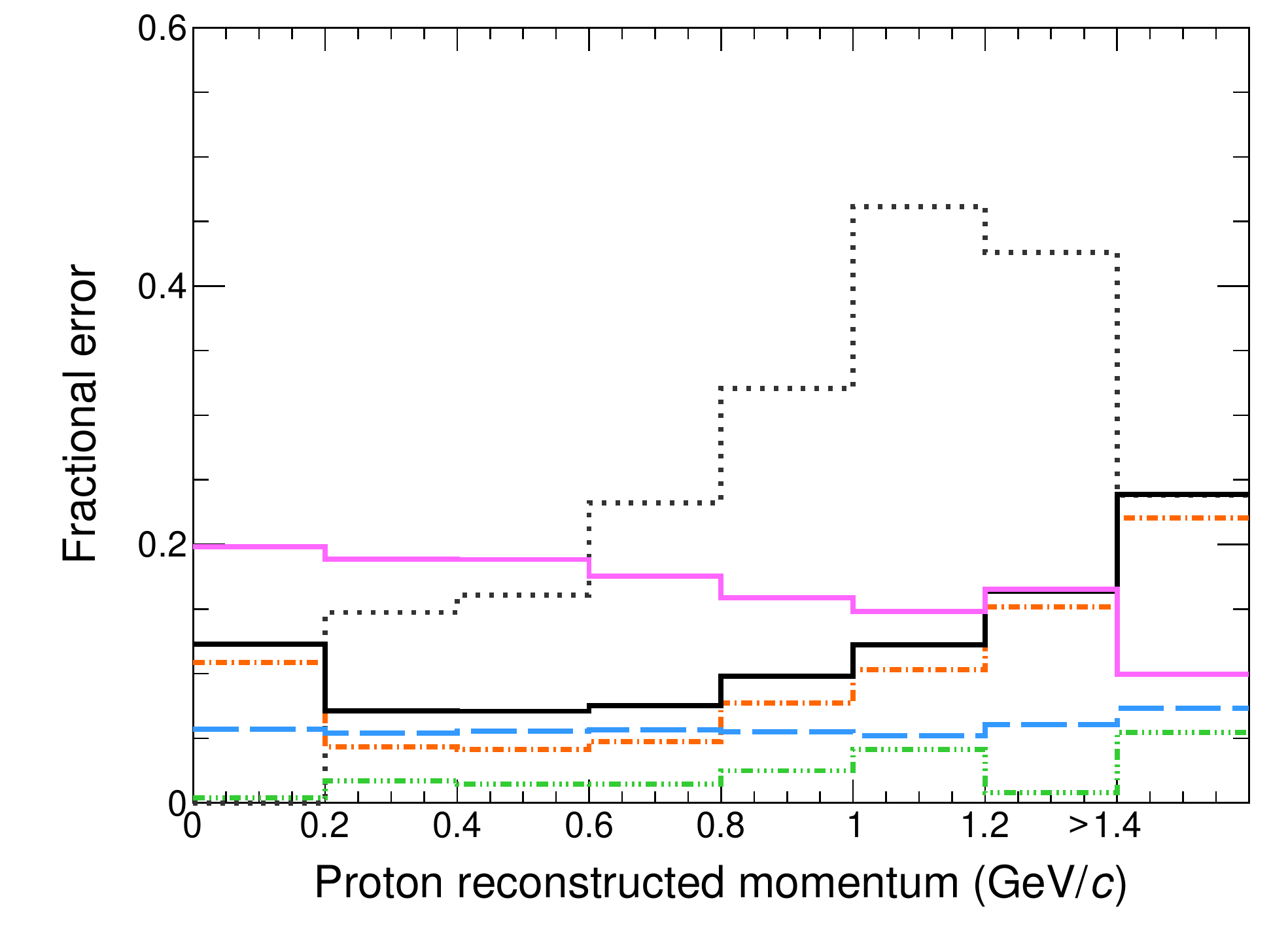}}
		\caption{Systematic uncertainties of the muon, charged pion, and proton kinematics. The figures in each column show the uncertainties in the muon, charged pion, and proton kinematics. The left-hand side of the figures shows the uncertainty of the emission angle distribution, whereas the right-hand side shows the uncertainty of the momentum distribution. Statistical uncertainty of our data is also shown for comparison.}
		\label{fig:sys_angle_momentum}
\end{figure*}

\begin{figure}
		\includegraphics[clip, width=1.0\columnwidth]{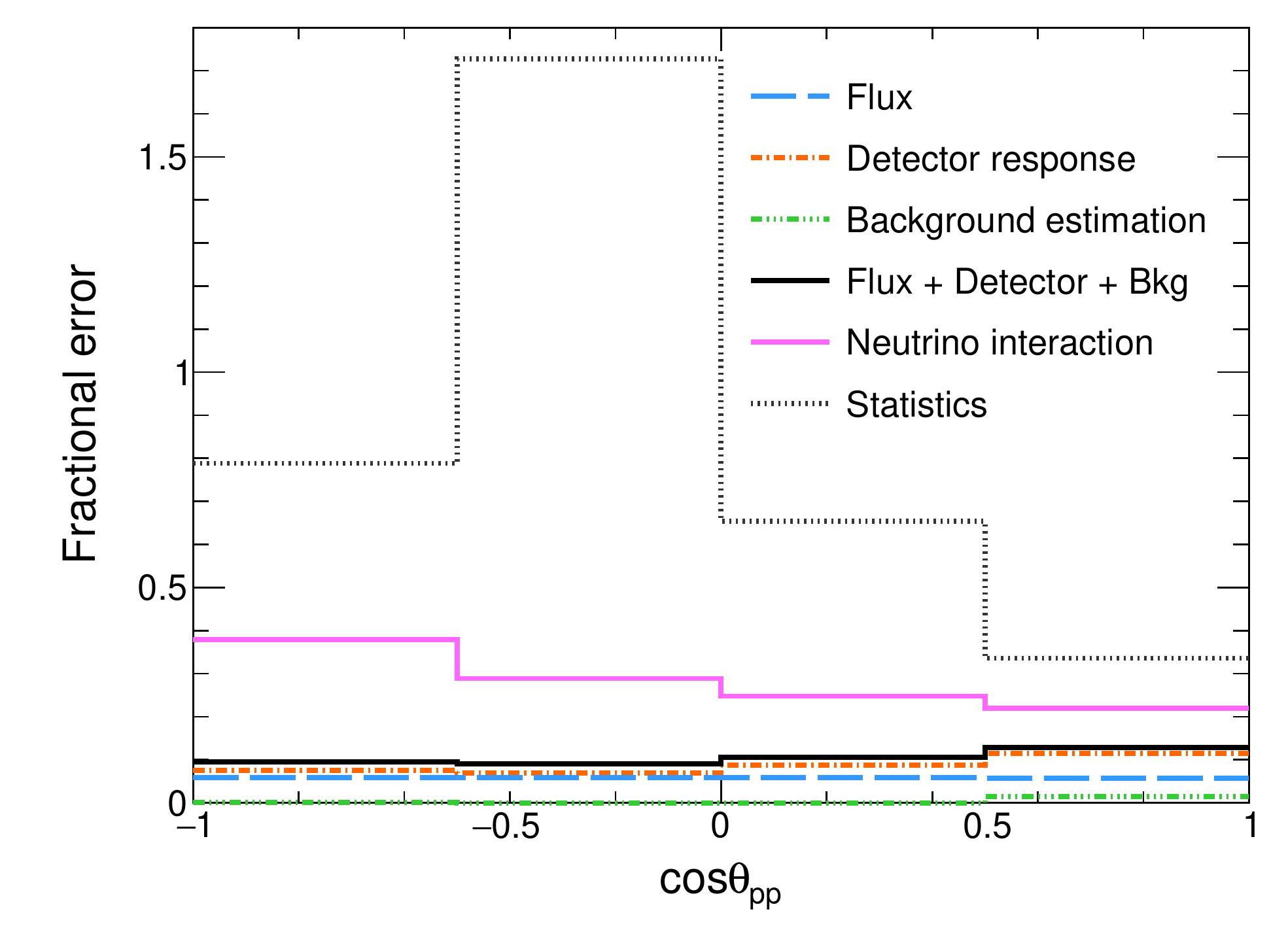}
		\caption{Systematic uncertainties of the opening angle between two protons of CC0$\pi$2$p$ events. Statistical uncertainty of our data is also shown for comparison.}
		\label{fig:sys_opang_2p}
\end{figure}

Figures~\ref{fig:sys_mul}--\ref{fig:sys_opang_2p} show the systematic and statistical uncertainties on the measurements, which are represented as the fractional error for each bin.
In this analysis, to study the neutrino interaction models via comparisons of the data and MC prediction, the flux, detector response, and background estimation uncertainties were included in the data as well as the statistical uncertainty, whereas the uncertainties of the neutrino interaction models were included in the MC prediction uncertainty.
Figure~\ref{fig:sys_mul} shows the systematic and statistical uncertainties for the multiplicities of charged pions and protons.
Figure~\ref{fig:sys_angle_momentum} shows the uncertainties for the kinematic measurements of muons, charged pions, and protons.
Figure~\ref{fig:sys_opang_2p} shows the uncertainty of the opening angle between two protons of CC pionless interactions with two protons in the final state~(CC0$\pi$2$p$).
In the analysis of this study for protons and pions, the uncertainty of the detector response was larger than that of the flux and background estimation, and the dominant uncertainty was the PID uncertainty, which originated from the VPH uncertainty.
Charged pions with momenta above 0.6\,GeV/$c$ exhibited a large PID uncertainty because the mis-PID rate of pions was also large, as shown in Fig.~\ref{fig:Mis-PID-rate}.
Further, large uncertainties in the pion and proton emission angles of less than 20$^{\circ}$ originate from these high-momentum pions, which are concentrated at small angles and have large mis-PID rate.
However, these uncertainties originated from the VPH can be reduced through further understanding of the detector response.
The total uncertainty will be smaller than that of the neutrino interaction uncertainties.
The uncertainties shown in Figs.~\ref{fig:sys_mul}--\ref{fig:sys_opang_2p} and the covariance matrixes representing the bin-to-bin correlations can be found in our data release~\cite{datarelease}.

\section{Results and discussion}\label{sec:results}
In this analysis, $\nu_{\mu}$ CC interactions on iron were defined as signal events, whereas $\nu_{\mu}$ NC, $\bar{\nu}_{\mu}$, $\nu_{e}$, and $\bar{\nu}_{e}$ interactions on iron were defined as background events.
ECC--Shifter--INGRID misconnection and external backgrounds that originate from the upstream wall of the detector hall and INGRID were also defined as background events.
However, the $\nu_{\mu}$ CC interactions on non-iron targets, such as emulsion films, in ECC bricks cannot be backgrounds because these interactions can be clearly rejected by the manual microscope check~\cite{ninja_run6_xsec}.

\begin{figure*}
		\subfigure{
			\includegraphics[clip, width=1.0\columnwidth]{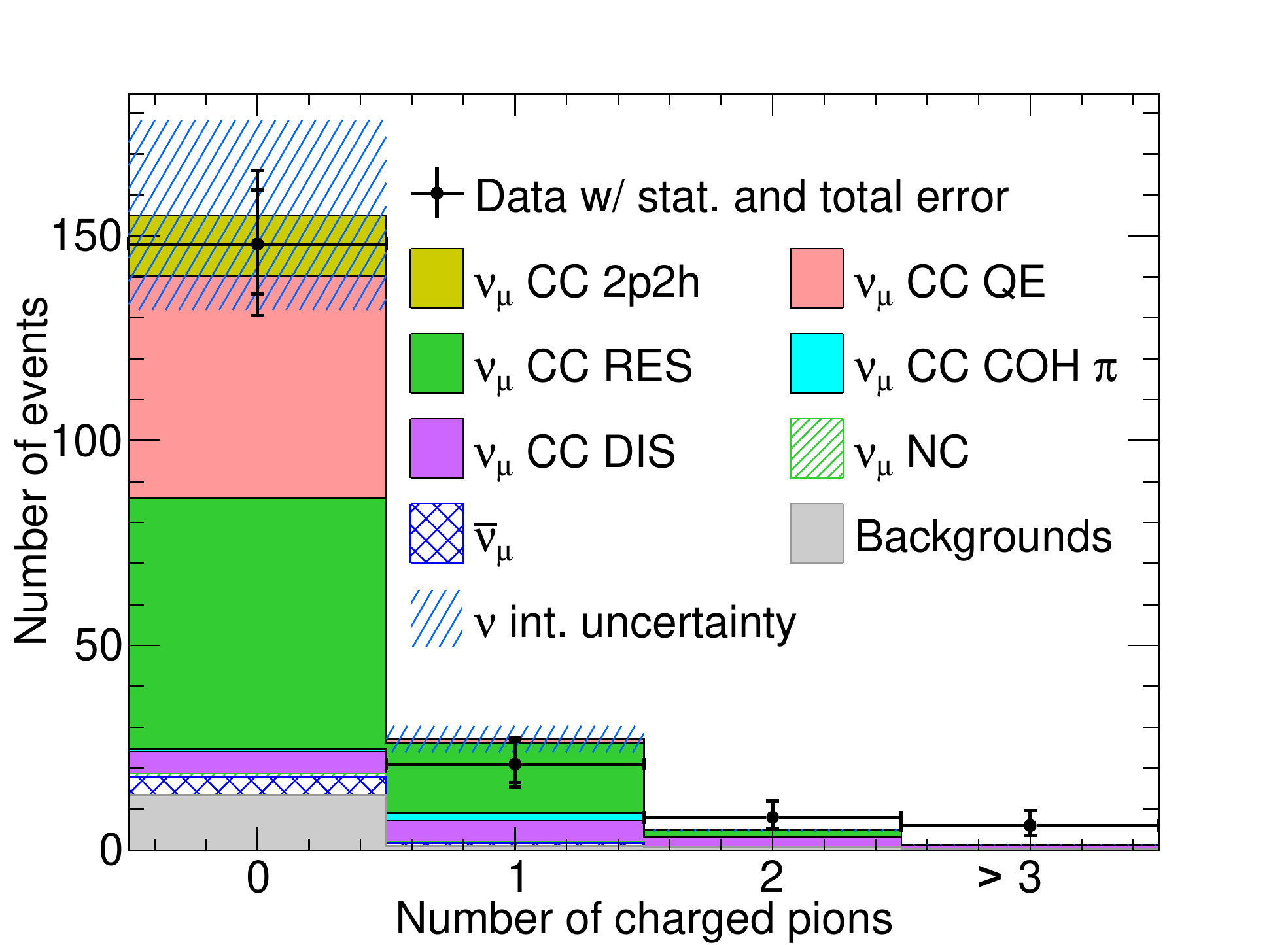}}
		\subfigure{
			\includegraphics[clip, width=1.0\columnwidth]{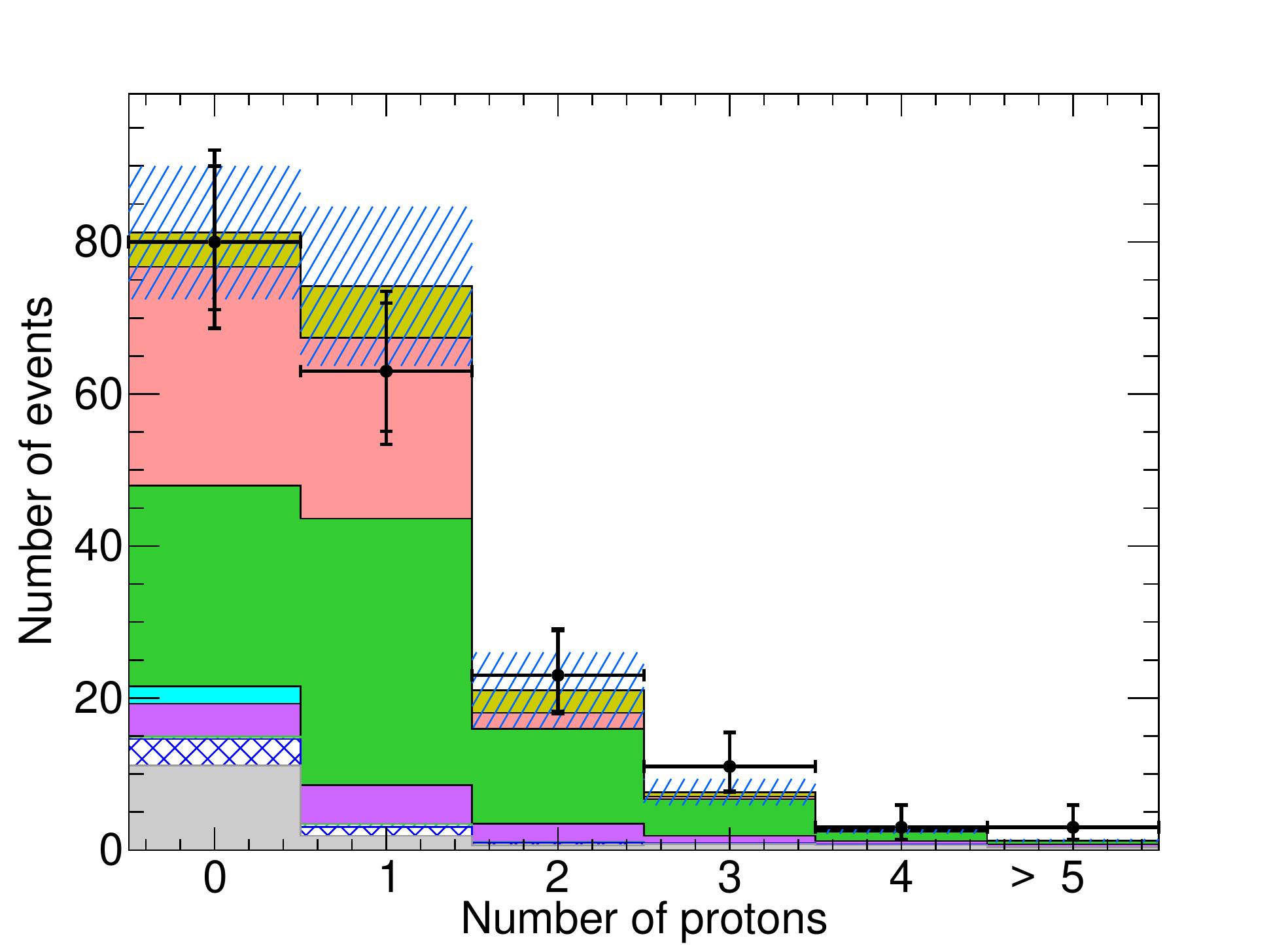}}
		\caption{Results of charged pion and proton multiplicities. The left-hand side of the figures shows the charged pion multiplicity, while the right-hand side shows the proton multiplicity. In the rightmost bin of the pion~(proton) distribution, all events with a multiplicity above three pions~(five protons) are contained.}
		\label{fig:numu_iron_num_pion}
\end{figure*}

The multiplicities of charged pions and protons produced in neutrino interactions on iron were measured.
Figure~\ref{fig:numu_iron_num_pion} shows the results.
In the figures shown in this section, the inside and outside error bars of the data represent the statistical and total errors, respectively.
The statistical error is a 68\% confidence interval of the Poisson distribution, while the total error shown on the data point is the quadrature sum of the statistical error and the uncertainties of the neutrino flux, detector response, and background estimation.
The hatched regions of the MC predictions represent the uncertainties in the neutrino interaction model.
In this analysis, comparisons between the data and the MC predictions are shown without unfolding the detector effects.
Although the statistical uncertainty is large, the number of events with more than two pions is significantly greater than the MC prediction.
The elementary processes of CCRES interactions that generate two or more pions are not implemented in the NEUT event generator, whereas the multi-pion events are expected as a result of the FSIs in the NEUT.
In addition, the prediction of pion multiplicity is significantly different among the neutrino interaction models because each model has a different pion production threshold depending on the value of the invariant mass of the hadronic system~\cite{pion_multiplicity_MC}.
Thus, our result on pion multiplicity provides a better understanding of pion production in neutrino interactions.
In contrast, good agreement between the data and MC prediction for proton multiplicity was observed. 
TABLE~\ref{tab:numu_iron_int_numubeam_num_of_p_pi_meas} shows the data of the correlation between the proton and the charged pion multiplicities, while TABLE~\ref{tab:numu_iron_int_numubeam_num_of_p_pi_mc} shows the MC prediction.
The values in each table show the number of events for exclusive channels CC$N\pi N^{\prime}p$~($N=0,1,2,..., N^{\prime}=0,1,2,...$): CC pionless interactions with no protons~(CC0$\pi$0p) and one proton~(CC0$\pi$1p).
The data agree well with the MC prediction within the error.
These results provide fundamental data to study exclusive interaction channels and a better understanding of neutrino interactions.
\begin{table*}[h]
	\caption{Data of the correlation between the number of protons and charged pions. The values in the table represent the number of neutrino interactions. The row and column show the number of pions and protons, respectively. The error is omitted because the statistical error is dominant compared with the systematic uncertainties.}
	\label{tab:numu_iron_int_numubeam_num_of_p_pi_meas}
	\vspace{2mm}
	\scalebox{1.3}{
	\begin{tabular}{cc|cccccc}
		 \multicolumn{2}{c}{} & \multicolumn{5}{c}{CC$N^{\prime}p$} & \\
		 & & 0 $p$ & 1 $p$ & 2 $p$ & 3 $p$ & $\geq$ 4 $p$ & Total \\
		\cline{2-8}
		 \multirow{5}{*}{\rotatebox[origin=c]{90}{CC$N\pi$} }
		 		 & 0 $\pi$ & 65  & 54   & 20  &   6  & 3 & 148\\
		 		 & 1 $\pi$ &   8  &   6   &   2   &   3  & 2 &   21\\
		   		 & 2 $\pi$ &   4  &   2   &   1   &   1  & 0 &     8\\
				 & 3 $\pi$ &   3  &   1   &   0   &   1  & 1 &     6\\
		   & $\geq$ 4 $\pi$ &   0  &   0   &   0   &   0  & 0 &     0\\
		 		     & Total & 80  & 63   & 23  & 11 &  6 & 183\\
	\end{tabular}
	}
\end{table*}
\begin{table*}[h]
	\caption{MC prediction of the correlation between the number of protons and charged pions. The values in the table represent the number of neutrino interactions. The row and column show the number of pions and protons, respectively. The error here is the systematic uncertainty of neutrino interaction models.}
	\label{tab:numu_iron_int_numubeam_num_of_p_pi_mc}
	\vspace{2mm}
	\scalebox{1.3}{
	\begin{tabular}{cc|cccccc}
		 \multicolumn{2}{c}{} & \multicolumn{5}{c}{CC$N^{\prime}p$} & \\
		 & & 0 $p$ & 1 $p$ & 2 $p$ & 3 $p$ & $\geq$ 4 $p$ & Total \\
		\cline{2-8}
		 \multirow{5}{*}{\rotatebox[origin=c]{90}{CC$N\pi$} }
		 		 & 0 $\pi$ &	$67.58 \pm 7.51$	&	$61.67 \pm 9.83$	&	$17.46 \pm 4.87$	&	$6.05 \pm 1.70$		&	$2.75 \pm 0.70$		&	$155.50 \pm 24.62 $\\
		 		 & 1 $\pi$ &	$13.00 \pm 2.23$	&	  $9.78 \pm 1.15$	&	  $2.86 \pm 0.23$	&	$0.95 \pm 0.10$		&	$0.70 \pm 0.07$		&	$  27.29 \pm 3.78 $\\
		 		 & 2 $\pi$ &	  $1.78 \pm 0.20$	&	  $1.83 \pm 0.13$	&	  $0.45 \pm 0.06$	&	$0.41 \pm 0.03$		&	$0.24 \pm 0.02$		&	$    4.71 \pm 0.43 $\\
		 		 & 3 $\pi$ &	  $0.42 \pm 0.05$	&	  $0.47 \pm 0.03$	&	  $0.11 \pm 0.02$	&	$0.04 \pm 0.01$		&	$0.04 \pm 0.02$		&	$    1.07 \pm 0.12 $\\
		 &   $\geq$ 4 $\pi$ &	  $0.09 \pm 0.02$	&	  $0.06 \pm 0.01$	&	  $0.02 \pm 0.01$	&	$0.012 \pm 0.002$	&	$0.012 \pm 0.003$	&	$    0.19 \pm 0.04 $\\
		 		     & Total &	$82.87 \pm 10.01$	&	$73.81 \pm 11.15$	&	$20.90 \pm 5.18$	&	$7.46 \pm 1.83$		&	$3.74 \pm 0.81$		&	$188.77 \pm 28.99$  \\
	\end{tabular}
	}
\end{table*}

\begin{figure*}
		\subfigure{
			\includegraphics[clip, width=1.0\columnwidth]{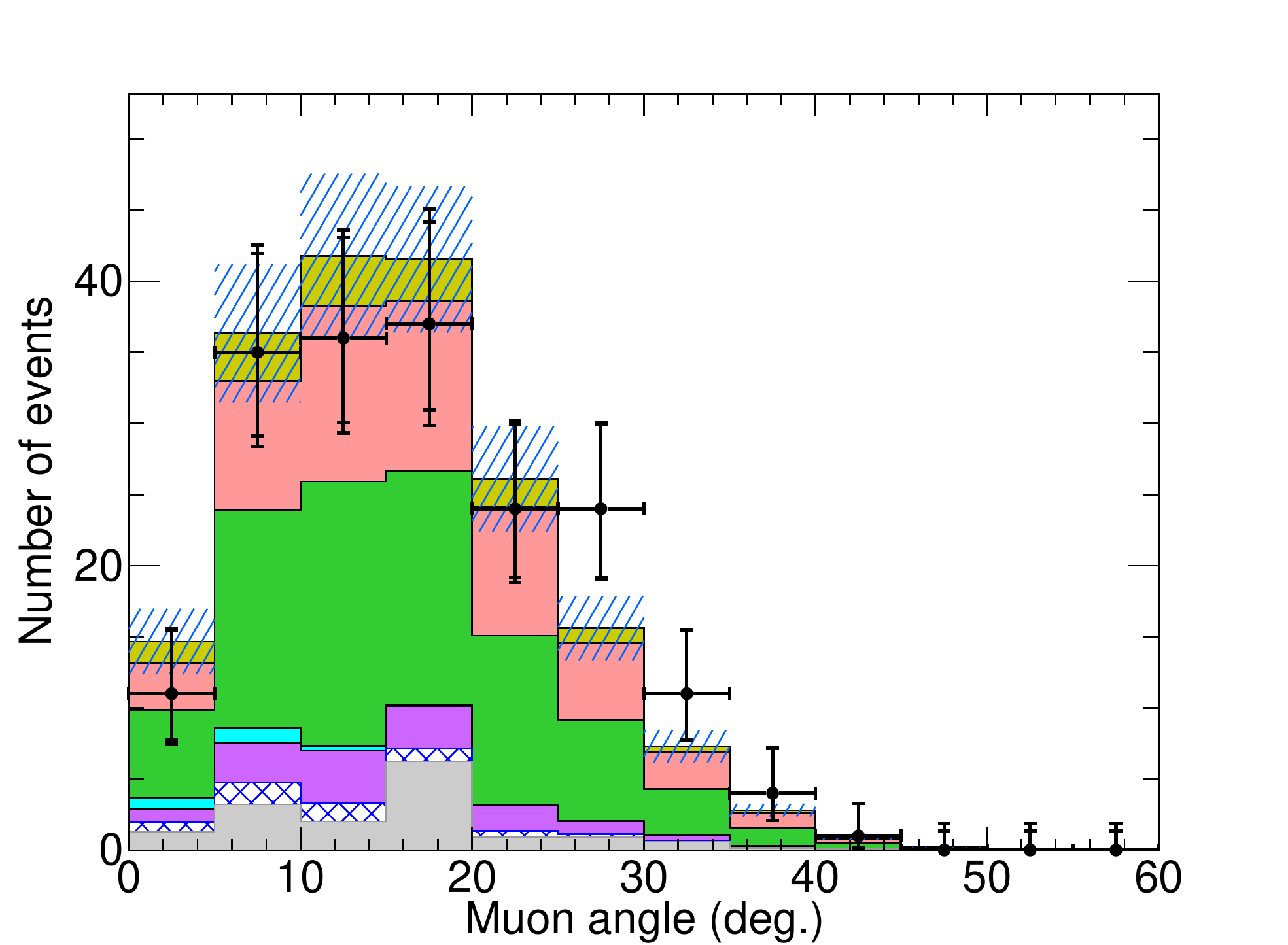}}
		\subfigure{
			\includegraphics[clip, width=1.0\columnwidth]{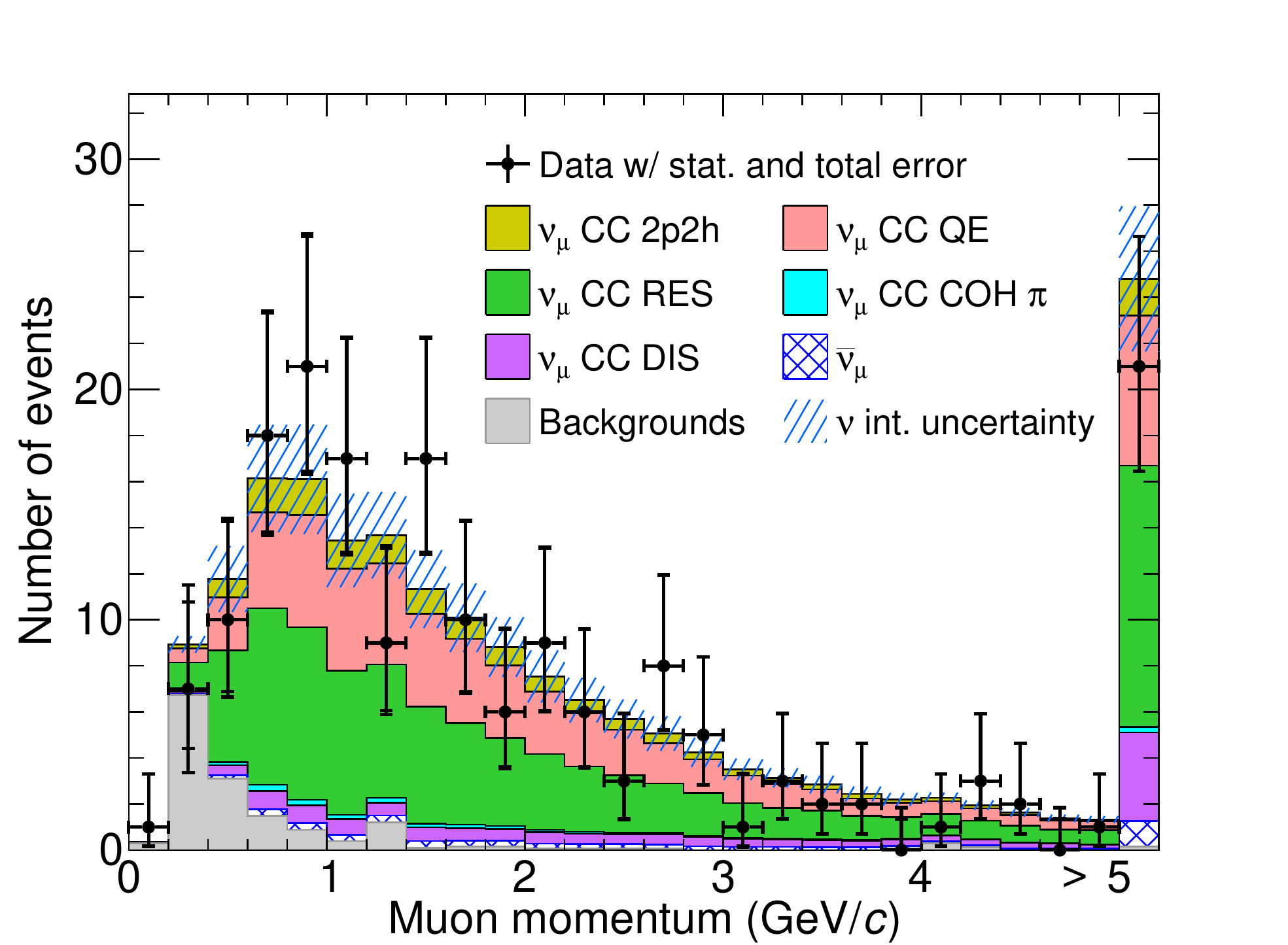}}
		\subfigure{
			\includegraphics[clip, width=1.0\columnwidth]{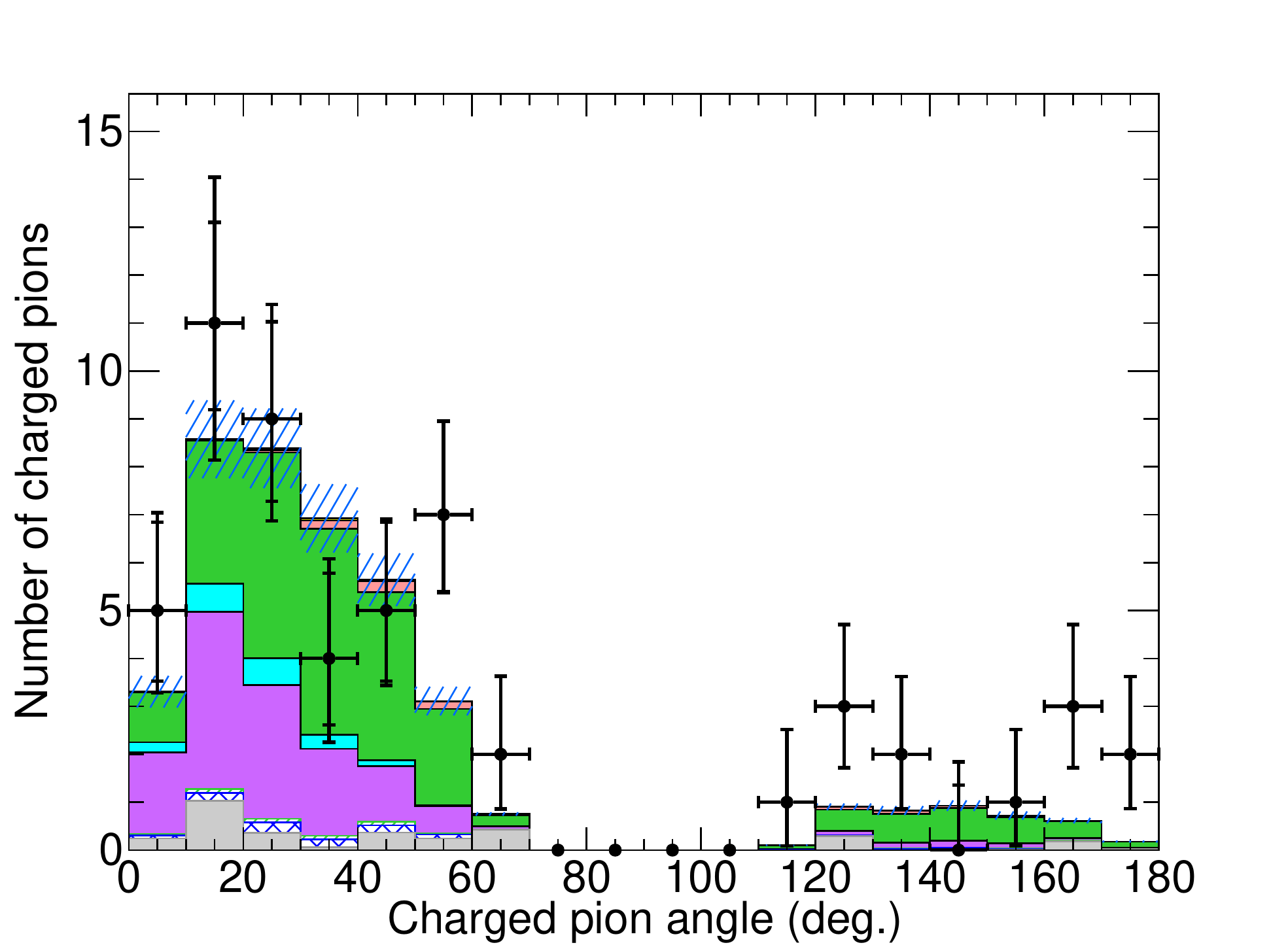}}
		\subfigure{
			\includegraphics[clip, width=1.0\columnwidth]{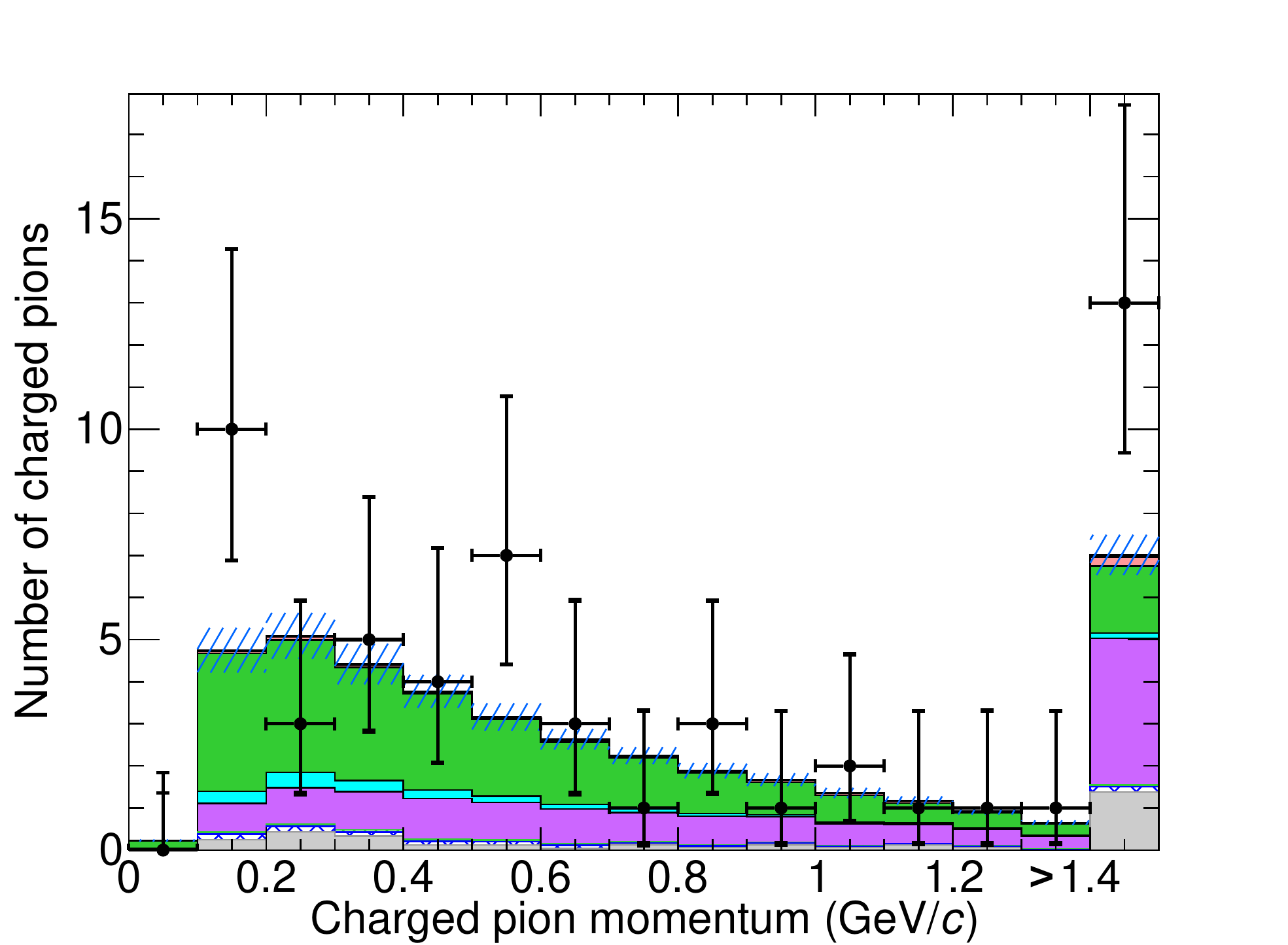}}
		\subfigure{
			\includegraphics[clip, width=1.0\columnwidth]{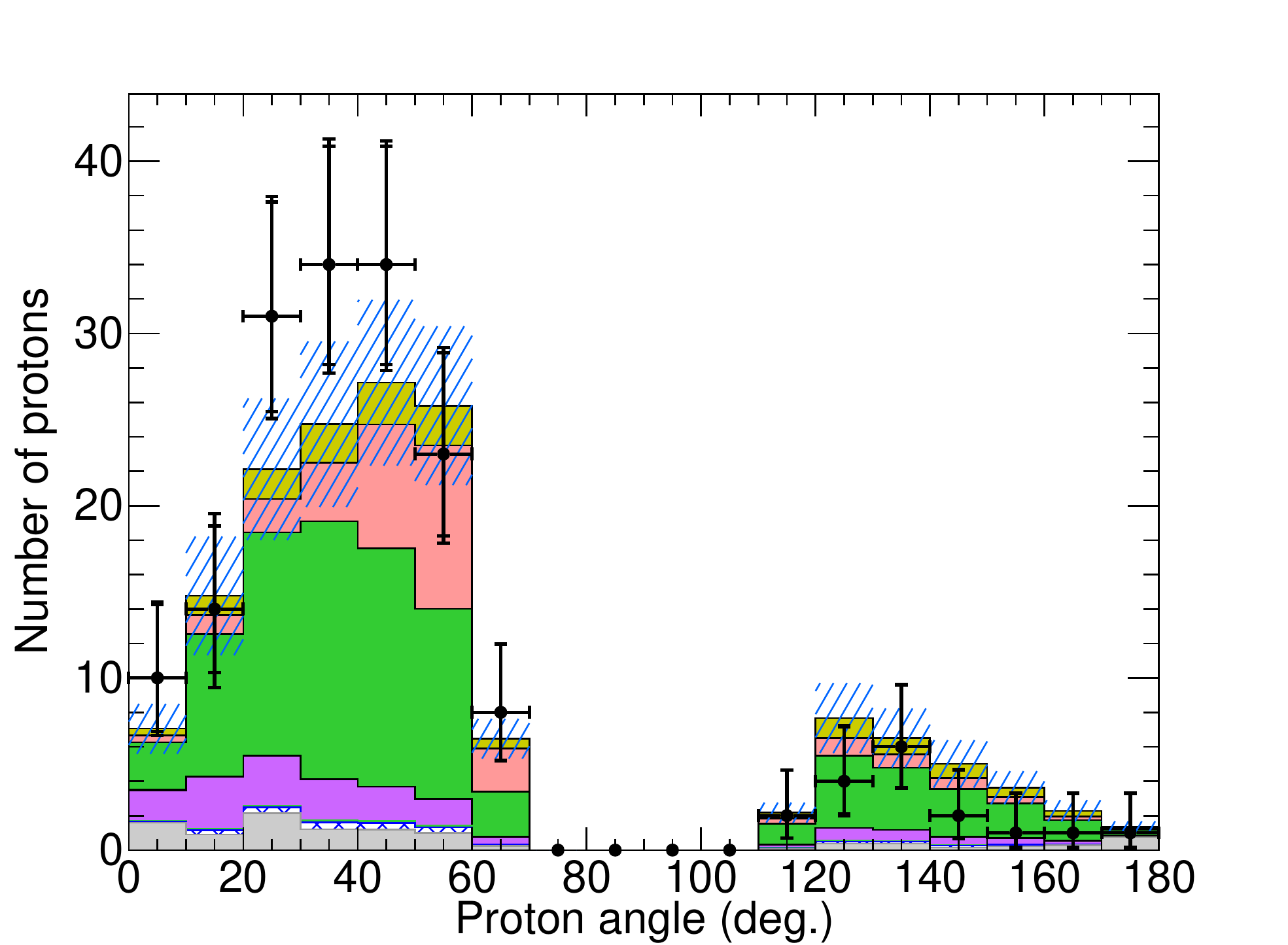}}
		\subfigure{
			\includegraphics[clip, width=1.0\columnwidth]{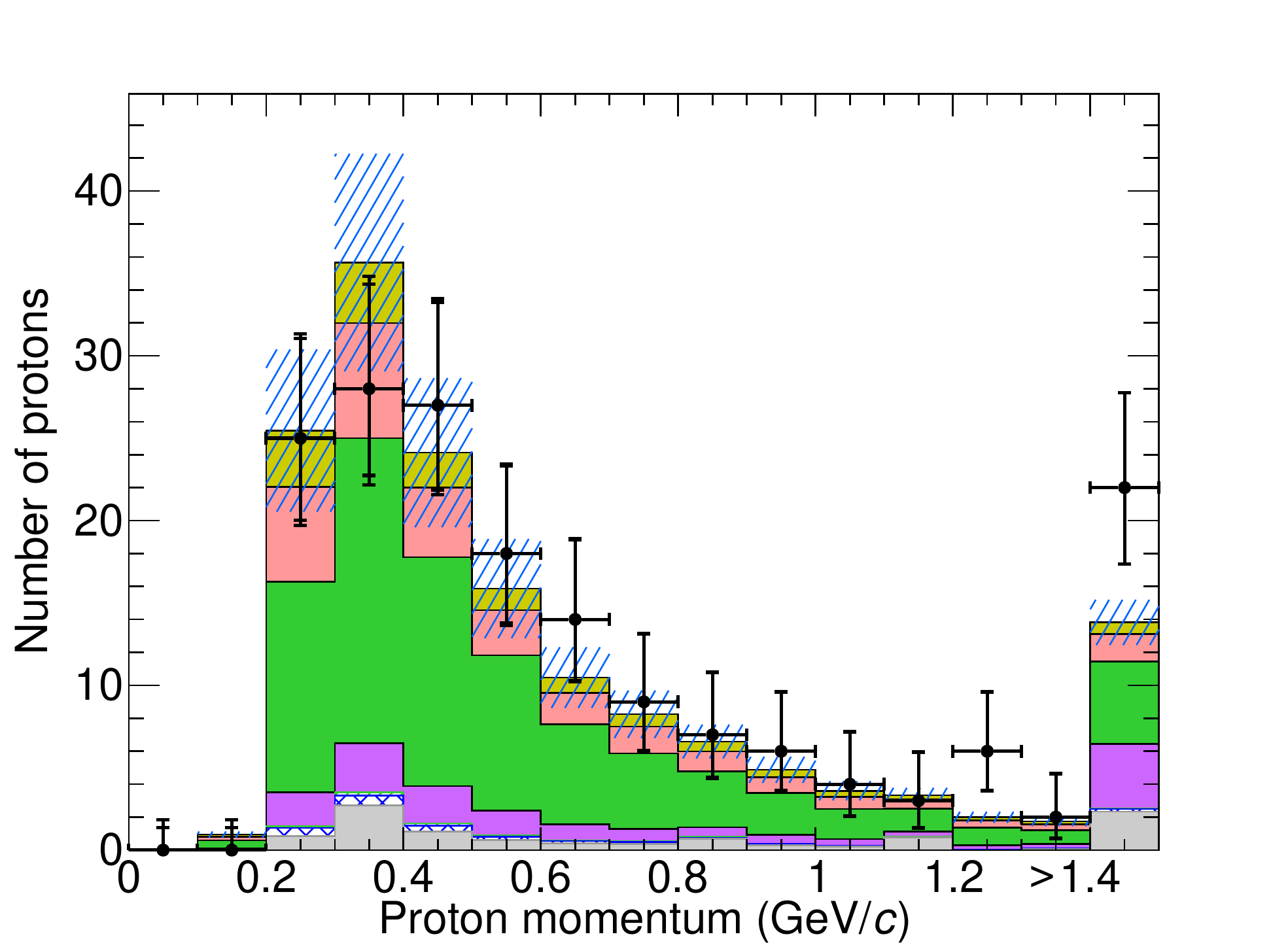}}
		\caption{Emission angles and momenta of muons, charged pions, and protons. The top of the figures shows the muon results, whereas the middle and bottom show the charged pion and proton results, respectively. The left-hand side of the figures shows the emission angle distributions and the right-hand side shows the momentum distributions. In the rightmost bin of the muon momentum distribution, all events with momenta greater than 5\,GeV/$c$ are contained. In the rightmost bins of the pion and proton momentum distributions, all events with momenta greater than 1.4\,GeV/$c$ are also contained. The range 70$^{\circ}$--110$^{\circ}$ in the pion and proton emission angle distributions cannot be reliably measured, and the errors are not shown.}
		\label{fig:angle_deg_momentum_muon_pion_proton}
\end{figure*}

\begin{figure*}
		\subfigure{
			\includegraphics[clip, width=1.0\columnwidth]{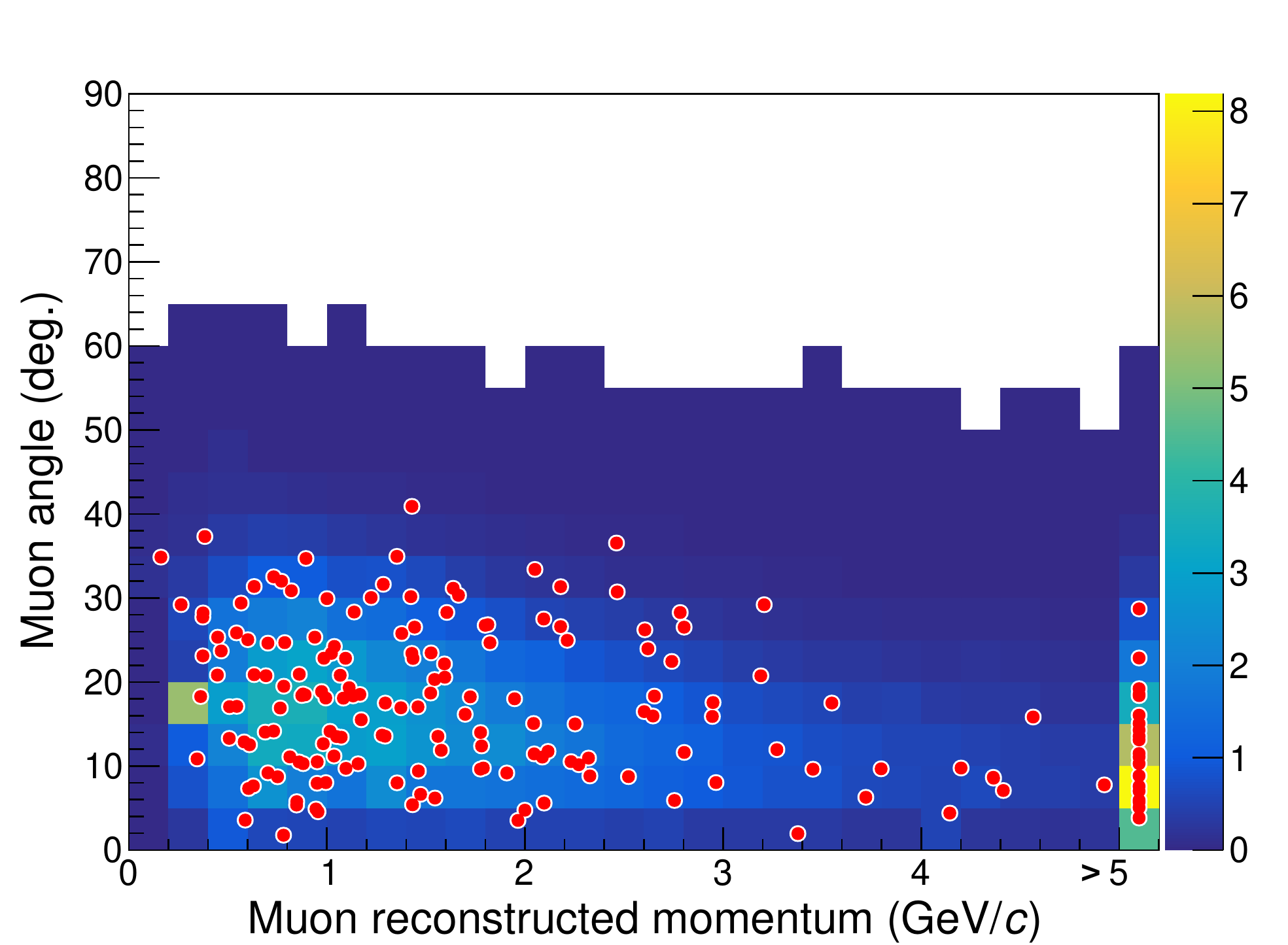}}\\
		\subfigure{
			\includegraphics[clip, width=1.0\columnwidth]{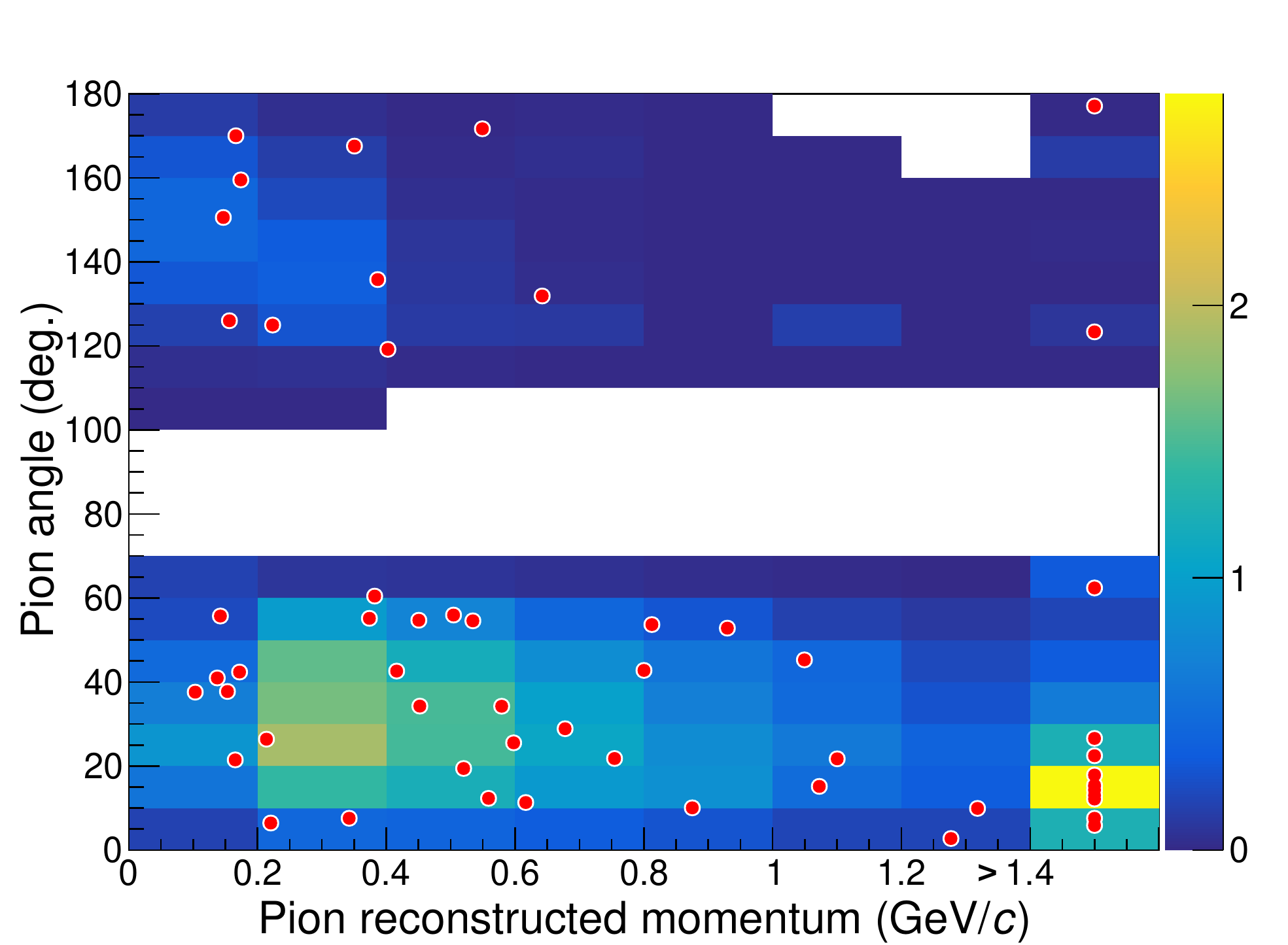}}
		\subfigure{
			\includegraphics[clip, width=1.0\columnwidth]{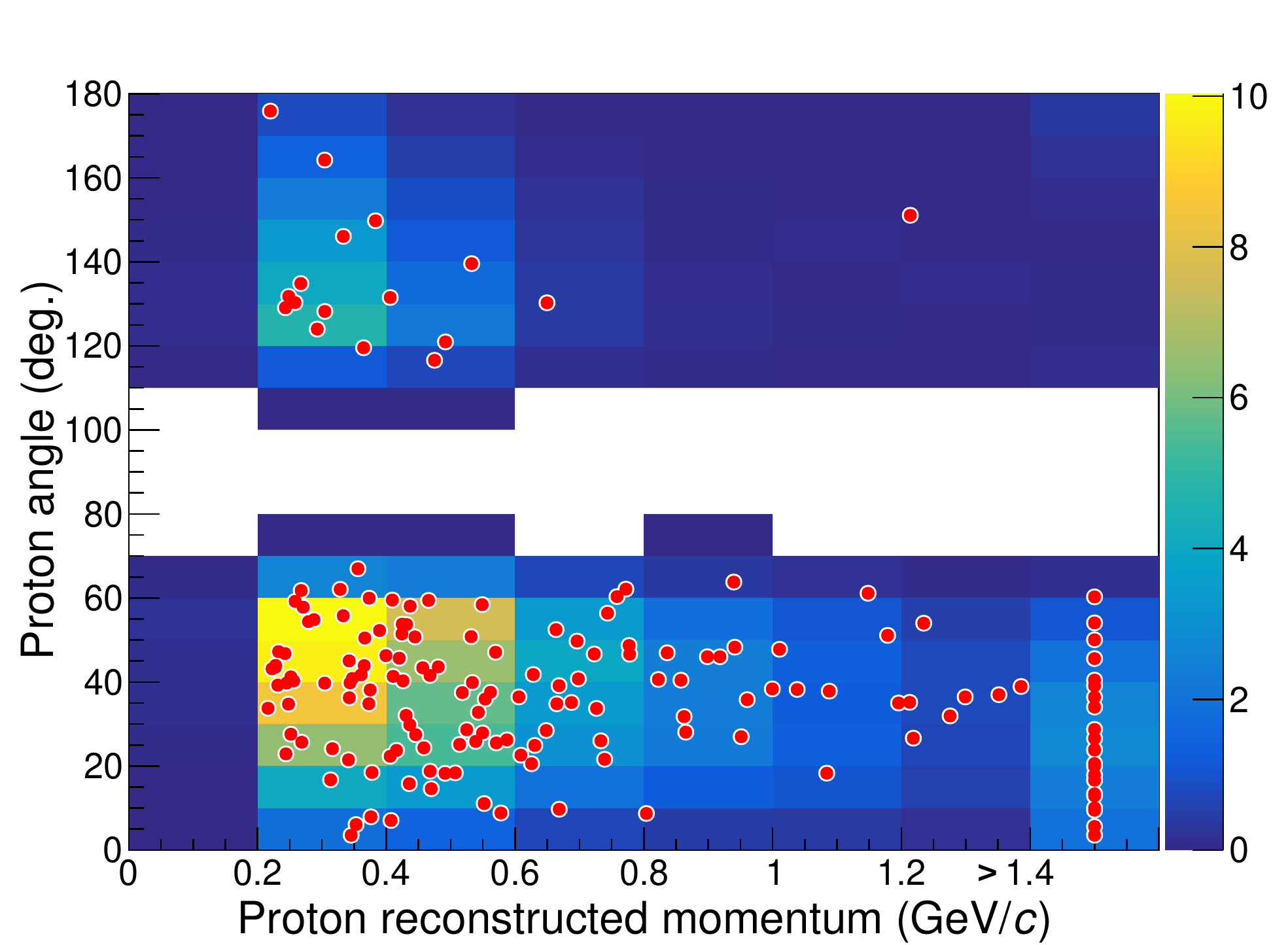}}
		\caption{Correlations between emission angle and momentum for muons, charged pions, and protons. The top of the figures shows that of the muons, while the bottom shows pions and protons, respectively. In the rightmost bin of the muon distribution, all events with momenta greater than 5\,GeV/$c$ are contained. In the rightmost bins of the pion and proton distributions, all events with momenta greater than 1.4\,GeV/$c$ are also contained. The data are represented by marker points, and the MC predictions are represented by a two-dimensional histogram. Colors in the histogram represent the number of events predicted by the MC simulation.}
		\label{fig:angle_deg_momentum_correlation_muon_pion_proton}
\end{figure*}

Further, the emission angles and momenta of muons, charged pions, and protons were also measured.
The results are shown in Fig.~\ref{fig:angle_deg_momentum_muon_pion_proton}.
Figure~\ref{fig:angle_deg_momentum_correlation_muon_pion_proton} shows the correlations between the emission angle and the momentum for muons, charged pions, and protons.
The results include low-momentum charged particles, particularly protons with momenta down to 200\,MeV/$c$, owing to the high granularity of the ECC.
The data and the MC predictions for the muon kinematic measurements were found to be consistent.
The muon results demonstrated the reliability of the detector and data analysis.

Regarding the pion emission angles, a tendency for the prediction to underestimate the back-scattered pions was observed.
The total number of back-scattered pions in the data was 12$^{+4.6}_{-3.4} {\rm (stat)} \pm 0.5 {\rm (syst)}$, whereas that of the MC prediction was 4.3$\pm 0.5 {\rm (syst)}$.
Most of the back-scattered pions shown in Fig.~\ref{fig:angle_deg_momentum_correlation_muon_pion_proton} have a momentum below 0.5\,GeV/$c$.
The PID in this low-momentum region performed well, as shown in Fig.~\ref{fig:Mis-PID-rate}.
In addition, regarding the pion momentum in the range of 0.1 to 0.2\,GeV/$c$, the number of predicted pions was underestimated, although the statistical uncertainty was large.
The momentum distributions of pions in other experiments were not well reproduced by the simulation programs~\cite{PhysRevD.101.012007,PhysRevD.100.072005}.
These back-scattered and low-momentum pions are expected to reflect rescattering in the nucleus by FSI.

In contrast, regarding the proton emission angles, a tendency for the prediction to overestimate the back-scattered protons was observed.
The total number of back-scattered protons in the data was 17$^{+5.2}_{-4.1} {\rm (stat)} \pm 2.3 {\rm (syst)}$, whereas that of the MC prediction was 28.6$\pm 7.2 {\rm (syst)}$.
In case of the proton momentum, a consistency between the data and MC prediction was observed.
The dominant interaction mode was CCRES according to the MC simulation.
It can be interpreted that the consistency between the data and the prediction is because the proton is generated via decay from a nucleon resonance state as CCRES, and the decay process is well understood.
However, although good agreement between the observed data and the MC prediction was observed, high statistics are required to confirm the conclusive result.
Regarding the study of nuclear effects, the measurements using an iron target complement the other experiments using the targets of bubble chambers~\cite{PhysRevD.16.3103,PhysRevD.23.2499,PhysRevD.28.436} and liquid argon time projection chambers~\cite{PhysRevD.102.112013} with proton low-momentum thresholds because the iron nucleus is the largest of these target nuclei.
Further, the results can be compared with various changeable targets in ECC, such as carbon, water, and iron, which is a big advantage in studying neutrino-nucleus interactions, including nuclear effects.

\begin{figure*}
		\subfigure{
			\includegraphics[clip, width=1.0\columnwidth]{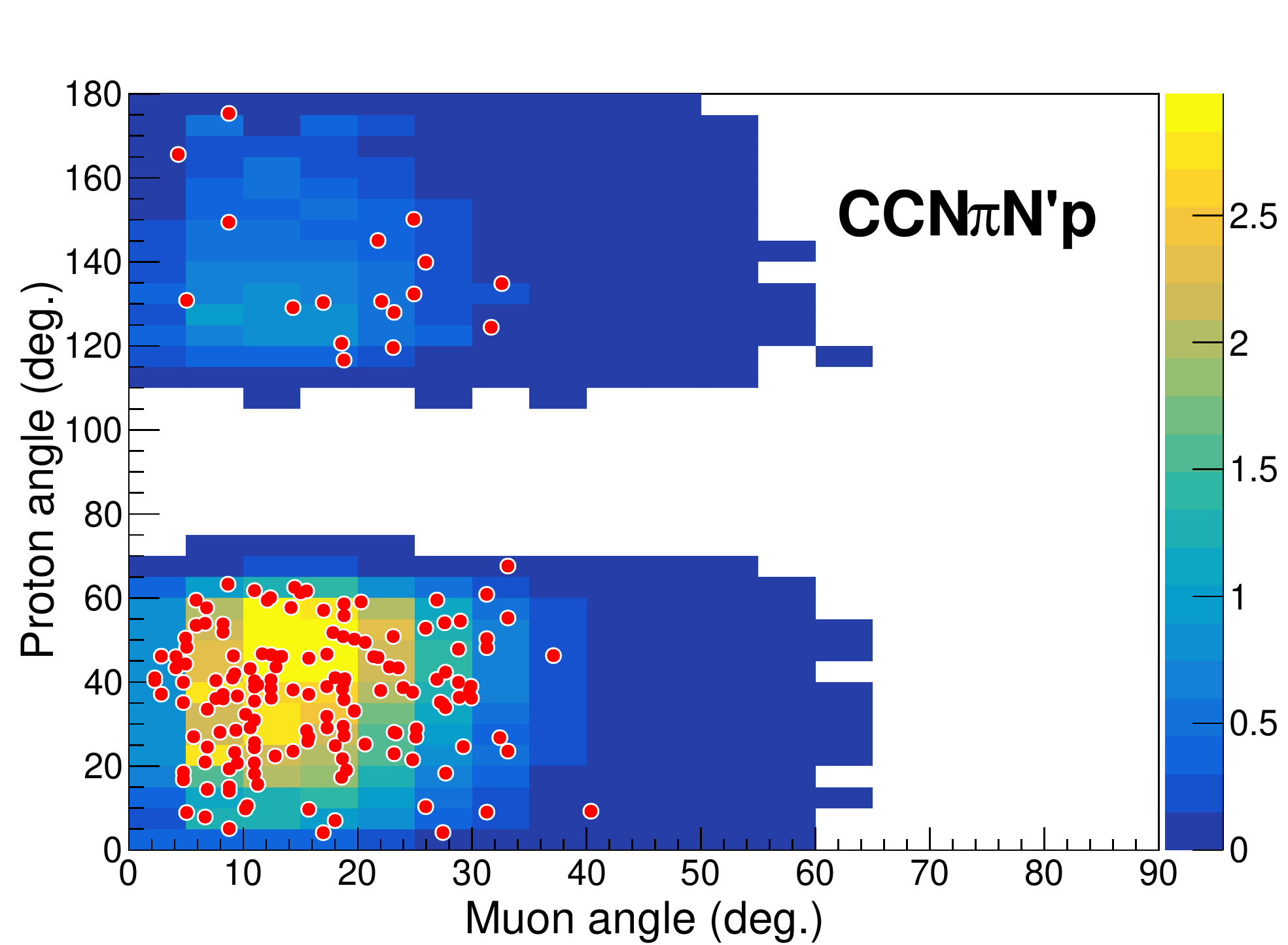}}\\
		\subfigure{
			\includegraphics[clip, width=1.0\columnwidth]{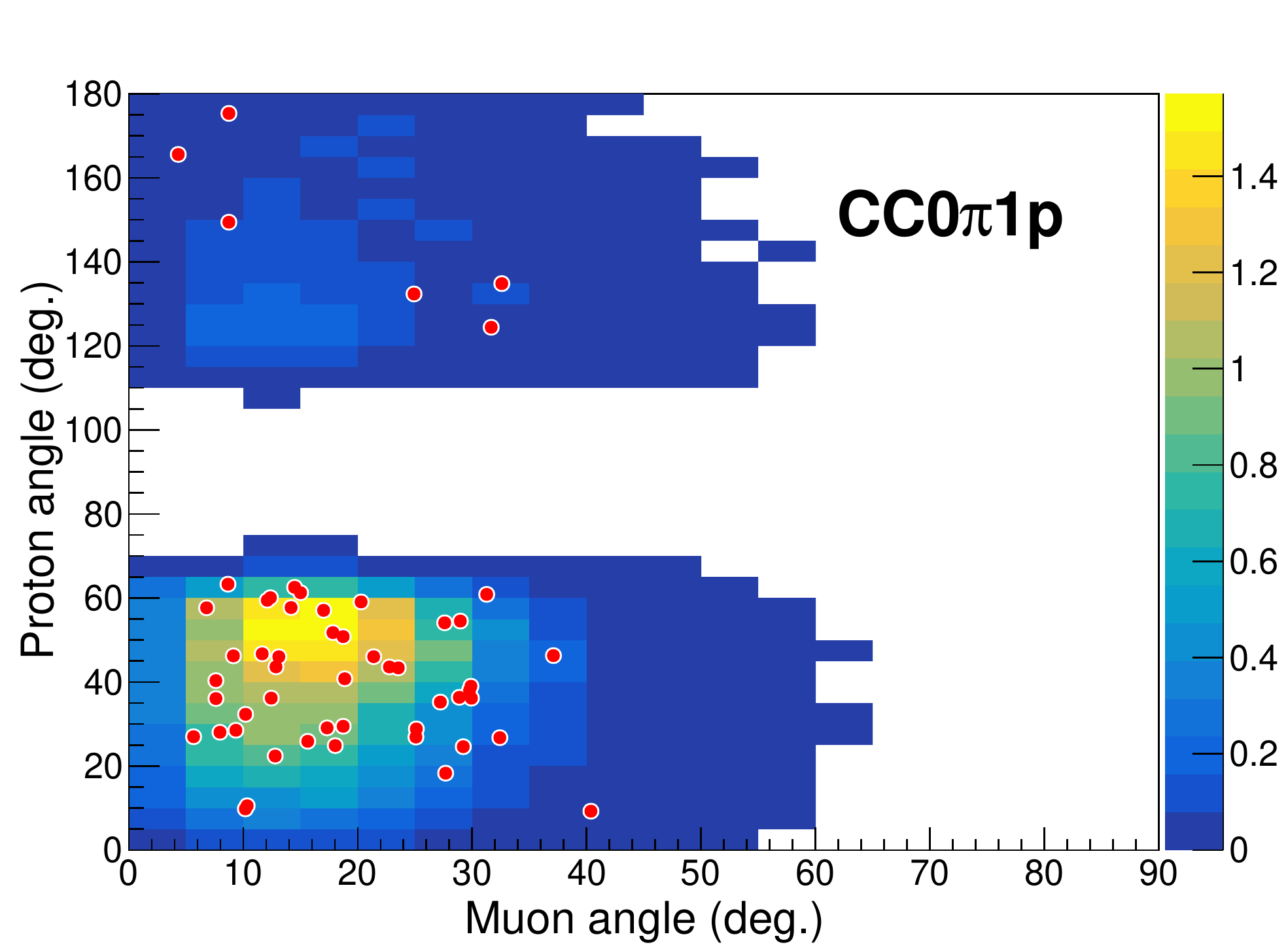}}
		\subfigure{
			\includegraphics[clip, width=1.0\columnwidth]{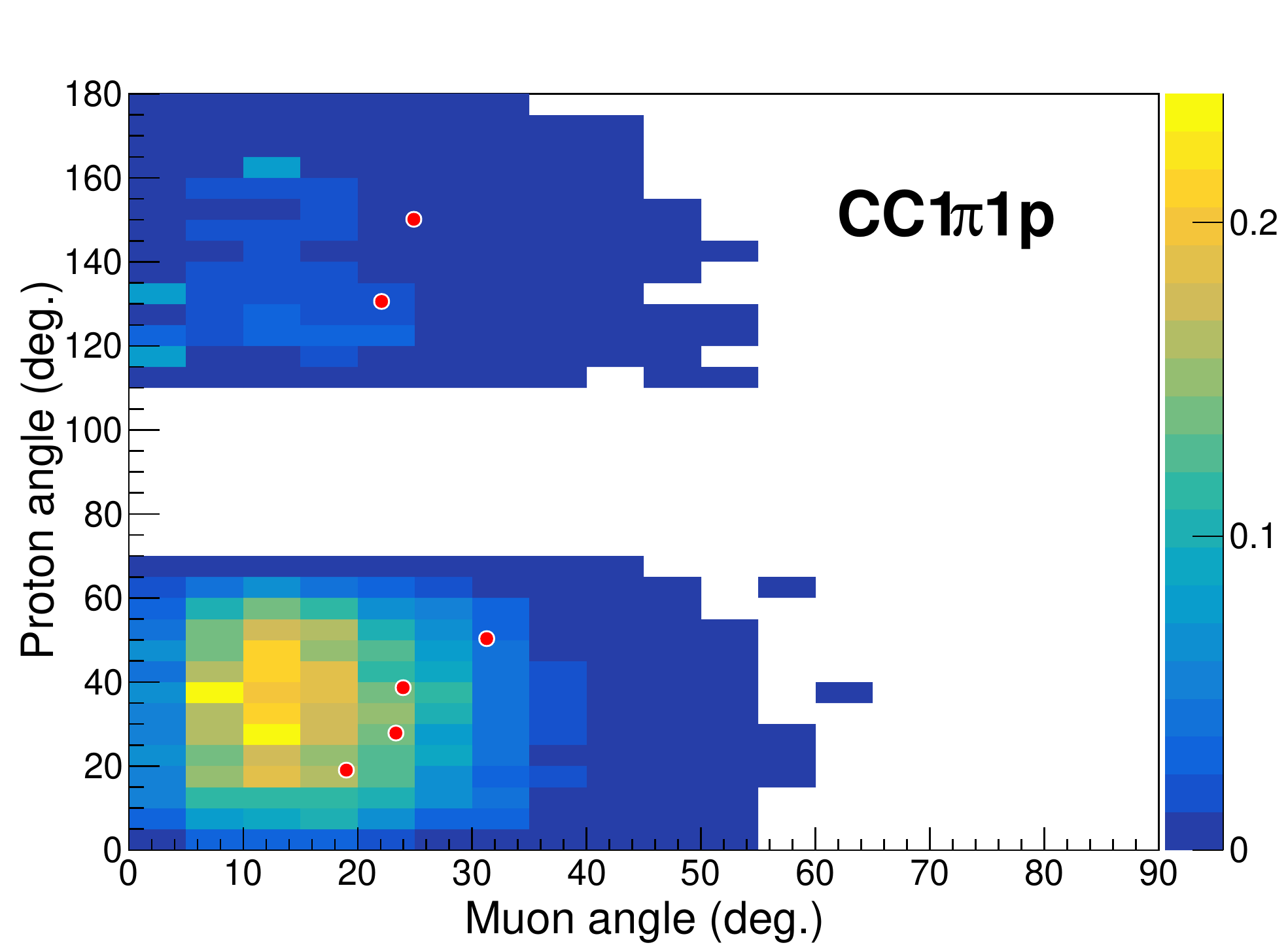}}
		\caption{Correlations between emission angles of muon and protons. The top half of the figures shows the correlation of CC$N\pi N^{\prime}p$ events, while the bottom half shows those of CC$0\pi 1p$ and CC$1\pi 1p$ events, respectively. The data are represented by marker points, and the MC predictions are represented by a two-dimensional histogram. Colors in the histogram represent the number of events predicted by the MC simulation.}
		\label{fig:angle_muon_proton}
\end{figure*}

Figure~\ref{fig:angle_muon_proton} shows the emission angle correlations between muons and protons.
Regarding the CC$N\pi N^{\prime}p$ and CC$0\pi 1p$ events, the data were in good agreement with the MC prediction.
In terms of CC$1\pi 1p$ events, in CC interactions with one pion and one proton, although the number of events is small, the muon angles tended to be larger than that of the MC prediction. However, the proton angles were consistent with the MC prediction because the dominant interaction mode of CC$1\pi 1p$ events was CCRES and the proton was generated from decay process.
It is possible that the momentum transfer predicted using the MC simulation is smaller than that of the data.
A similar tendency has been observed in other experiments, such as MiniBooNE~\cite{PhysRevD.83.052007}, MINERvA~\cite{PhysRevD.94.052005,PhysRevD.100.052008,PhysRevD.100.072005}, and T2K~\cite{PhysRevD.101.012007}.

\begin{figure}
		\includegraphics[clip, width=1.0\columnwidth]{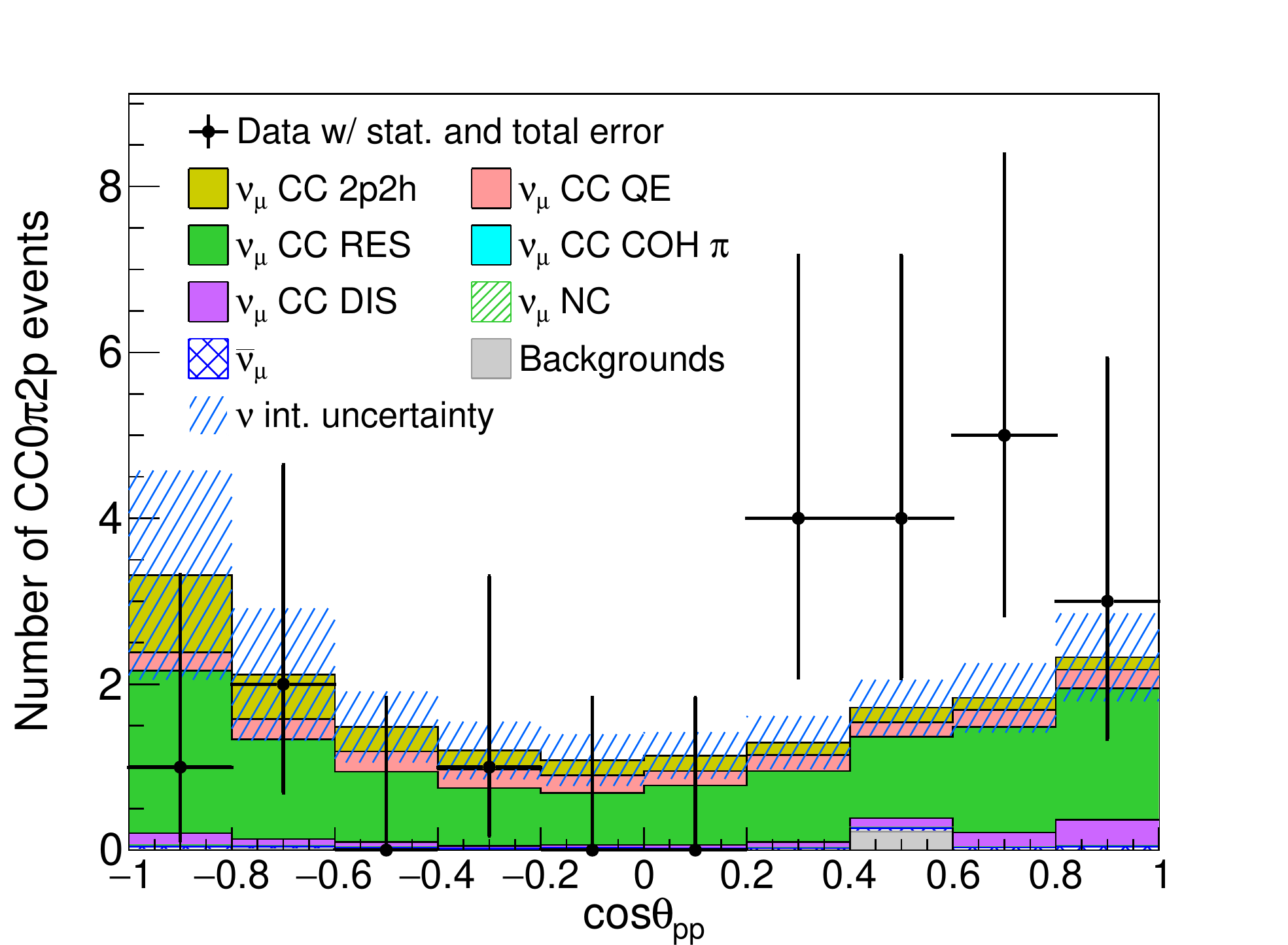}
		\caption{Opening angle between two protons of CC0$\pi$2$p$ events. Back-to-back protons are represented by cos$\theta = -1$, whereas protons emitted in the same direction are represented by cos$\theta = 1$.}
		\label{fig:angle_deg_momentum_correlation_pion_proton}
\end{figure}

The opening angle between the protons of CC$0\pi 2p$ events were also measured, as shown in Fig.~\ref{fig:angle_deg_momentum_correlation_pion_proton}.
Back-to-back protons are represented by cos$\theta = -1$, whereas protons emitted in the same direction are represented by cos$\theta = 1$.
The two protons generated by the 2p2h interactions are expected to exhibit back-to-back emissions~\cite{PhysRevD.90.012008}.
However, there were fewer back-to-back protons in the data than in the MC prediction, while there were more protons in the same direction in the data than in the MC prediction, although the statistical uncertainty was large.
It can be considered that the back-to-back protons may be induced incidentally by FSIs rather than the physical processes of the 2p2h interactions.

These results are the first step towards a detailed understanding of neutrino-nucleus interactions and are important data for building reliable neutrino interaction models.
The results and related data can be found in Ref.~\cite{datarelease}.
In addition to these data, transverse kinematic imbalance~\cite{PhysRevC.94.015503,PhysRevLett.121.022504,PhysRevD.98.032003}, inferred proton kinematics~\cite{PhysRevD.98.032003}, and certain results of the measurements have been provided as supplemental material~\cite{supplemental_material} for this study.
In the future, these interactions, including FSIs, are expected to be understood using more statistics.

\section{Conclusion}\label{sec:conclusion}
An accurate understanding of neutrino interactions around the 1\,GeV energy region is important for current and future long-baseline neutrino-oscillation experiments.
This study reported the measurements of protons and charged pions emitted from $\nu_{\mu}$ CC interactions on iron using a nuclear emulsion detector.
The measurement was performed by exposing a 65\,kg iron target to a neutrino beam with 4.0$\times$10$^{19}$\,POT at a mean neutrino energy of 1.49\,GeV at J\nobreakdash-PARC.
The multiplicities, emission angles, and momenta of protons and charged pions from the $\nu_{\mu}$ CC interactions on iron were measured.
The measurements show the first results of particles with momentum thresholds of 200 and 50\,MeV/$c$ for protons and pions, respectively.
Although the statistical uncertainty was large, the data and MC predictions were compared to study the differences in this study.
Regarding pion measurements, predictions tended to underestimate the number of events involving more than two pions, back-scattered pions, and pions with a momentum in the range of 0.1 to 0.2\,GeV/$c$, respectively.
In contrast, concerning the proton measurements, the multiplicity and momentum distributions were consistent between the data and the MC simulation, while for the emission angle distribution, a tendency for the prediction to overestimate the back-scattered protons was observed.
Further, in terms of the emission angle correlations between muons and protons, the muon angles tended to be larger than those predicted by MC, whereas the proton angles were consistent with the MC prediction in the correlations of CC$1 \pi 1 p$ events.
In addition, with regard to the opening angle between protons of CC$0 \pi 2 p$ events, fewer back-to-back protons in the data than those in the MC prediction were observed, whereas more protons in the same direction in the data than those in the MC prediction were found.
The presented measurements are the first step toward improving the understanding of neutrino-nucleus interactions in the transition region from CCQE to CCRES interactions using an emulsion detector.
The results are expected to serve as the foundation for building reliable neutrino interaction models.
The related data shown in this paper can be found in Ref.~\cite{datarelease}.
However, further detailed studies are required to understand the inconsistencies using high-statistics data, which were recently collected by exposing 75\,kg water and 130\,kg iron targets.

\clearpage
\begin{acknowledgments}
The authors would like to acknowledge the support of the T2K Collaboration in performing the experiments.
Furthermore, the assistance of the T2K neutrino beam group in providing high-quality beams and MC simulations is appreciated.
The authors thank the T2K INGRID group for providing access to their data.
Further, they acknowledge the work of the J\nobreakdash-PARC staff in facilitating the superb accelerator performance.
In addition, the authors would like to thank P. Vilain~(Brussels University, Belgium) for his careful reading of the manuscript and for his valuable comments.
The authors acknowledge the financial support of the Ministry of Education, Culture, Sports, Science, and Technology in Japan~(MEXT) and the joint research program of the Institute of Materials and Systems for Sustainability, Nagoya University.
This work was financially supported by the Japan Society for the Promotion of Science~(JSPS) KAKENHI Grant Numbers JP25105001, JP25105006, JP26105516, JP26287049, JP25707019, JP20244031, JP26800138, JP16H00873, JP18K03680, JP17H02888, JP18H03701, JP18H05537, and JP18H05541.
\end{acknowledgments}

\bibliography{./references}

\end{document}